\documentclass[preprint]{aastex}

\usepackage{epsf}
\usepackage{graphicx}
\usepackage{ae}
\usepackage[T1]{fontenc}

\newcommand{\sm}[1]{\mbox{{\scriptsize #1}}}

\newcommand{\simge} {\,{}^>_{\sim}\,}

\newcommand{\be}{\begin{equation}}
\newcommand{\ee}{\end{equation}}
\newcommand{\bea}{\begin{eqnarray}}
\newcommand{\eea}{\end{eqnarray}}
\newcommand{\bdm}{\begin{displaymath}}
\newcommand{\edm}{\end{displaymath}}
\newcommand{\bef}{\begin{figure}}
\newcommand{\eef}{\end{figure}}
\newcommand{\befs}{\begin{figure*}}
\newcommand{\eefs}{\end{figure*}}
\newcommand{\befone}{
  \begin{figure*}
  \centering
  \begin{minipage}{\textwidth}
  }
\newcommand{\eefone}{\end{minipage}\end{figure*}}

\newcommand{\cm}{\mbox{cm}}

\newcommand{\km}{\mbox{km}}

\renewcommand{\sec}{\mbox{s}}
\newcommand{\g}{\mbox{g}}

\newcommand{\pc}{\mbox{pc}}

\newcommand{\K}{\mbox{K}}

\newcommand{\ys}{\mbox{years}}
\newcommand{\Htwo}{\mbox{H$_2$}}

\newcommand{\Ma}{{\cal M}}
\newcommand{\Vol}{{\cal V}}

\newcommand{\di}{\mbox{d}}

\def\eps@scaling{.98}

\def\showone#1{
  \centering
  \leavevmode
  \includegraphics[width={\eps@scaling\linewidth}]{#1.pdf}
}

\def\epstwo@scaling{0.49}

\def\showtwo#1#2{
  \centering
  \leavevmode
  \includegraphics[width={\epstwo@scaling\linewidth}]{#1.pdf} \hfil
  \includegraphics[width={\epstwo@scaling\linewidth}]{#2.pdf}
}

\def\showtwover#1#2{
  \centering
  \leavevmode
  \epsfxsize=\eps@scaling\linewidth
  \epsfbox{#1.pdf} \hfil
  \epsfxsize=\eps@scaling\linewidth
  \epsfbox{#2.pdf}
}

\def\epstwover@scaling{0.4}

\def\plottwover#1#2{
  \centering
  \leavevmode
  \includegraphics[height={\epstwover@scaling\textheight}]{#1} \hfil \,
  \includegraphics[height={\epstwover@scaling\textheight}]{#2} \hfil \,
}

\def\showfour#1#2#3#4{
  \centering
  \leavevmode
  \includegraphics[width={\epstwo@scaling\linewidth}]{#1.pdf} 
  \includegraphics[width={\epstwo@scaling\linewidth}]{#2.pdf}
  \hfil \,
  \includegraphics[width={\epstwo@scaling\linewidth}]{#3.pdf} 
  \includegraphics[width={\epstwo@scaling\linewidth}]{#4.pdf} 
}

\def\showsix#1#2#3#4#5#6{
  \centering
  \leavevmode
  \includegraphics[width={\epstwo@scaling\linewidth}]{#1.pdf} \hfil
  \includegraphics[width={\epstwo@scaling\linewidth}]{#2.pdf} \hfil
  \includegraphics[width={\epstwo@scaling\linewidth}]{#3.pdf} \hfil
  \includegraphics[width={\epstwo@scaling\linewidth}]{#4.pdf} \hfil
  \includegraphics[width={\epstwo@scaling\linewidth}]{#5.pdf} \hfil
  \includegraphics[width={\epstwo@scaling\linewidth}]{#6.pdf} \hfil
}

\begin{document}

\title{Can Protostellar Jets Drive Supersonic Turbulence in Molecular Clouds?}

\author{Robi Banerjee\altaffilmark{1}, Ralf S.~Klessen\altaffilmark{1},
  Christian Fendt\altaffilmark{2}}
\affil{$^1$Institute of Theoretical Astrophysics, University of
Heidelberg, Albert-Ueberle-Str. 2, 69120 Heidelberg, Germany \\
$^2$Max Planck Institute for Astronomy, K\"onigsstuhl 17, 69117 Heidelberg}

\begin{abstract}
Jets and outflows from young stellar objects are proposed candidates
to drive supersonic turbulence in molecular clouds. Here, we present
the results from multi-dimensional jet simulations where we
investigate in detail the energy and momentum deposition from jets
into their surrounding environment and quantify the character of the
excited turbulence with velocity probability density functions. Our
study include jet--clump interaction, transient jets, and magnetised
jets. We find that collimated supersonic jets do not excite supersonic
motions far from the vicinity of the jet. Supersonic fluctuations are
damped quickly and do not spread into the parent cloud. Instead
subsonic, non-compressional modes occupy most of the excited
volume. This is a generic feature which can not be fully circumvented
by overdense jets or magnetic fields. Nevertheless, jets are able to
leave strong imprints in their cloud structure and can disrupt dense
clumps. Our results question the ability of collimated jets to sustain
supersonic turbulence in molecular clouds.
\end{abstract}

\keywords{magneto-hydrodynamics, ISM: evolution, methods: numerical}

\section{Introduction}

The interstellar medium (ISM) and star forming molecular clouds
are permeated by turbulent, supersonic gas motions~\citep[e.g.~see
recent reviews][and references therein]{Elmegreen04, MacLow04,
Ballesteros07}. Supersonic turbulence is a main ingredient in the
process of star formation. It can, on the one hand, compress material
that might become Jeans unstable and collapse to form stars. On the
other hand, supersonic motions can disperse clumps and cores in
radiative shocks. Although supersonic turbulence is a major player in
forming stars, its origin is still obscure. A difficulty of supersonic
turbulence is that it decays quickly and has to be continuously driven
to be maintained~\citep{MacLow98a, Stone98, Padoan99}. The energy
input can in principle be provided by sources within the molecular
cloud (e.g.~radiation from massive stars, outflows) or from outside
(e.g.~supernovae, galactic spiral arms).  In particular, the birth of
stars is in most cases accompanied by outflows and high velocity
jets. Protostellar jets propagate with velocities of about $300 \, \km
\, \sec^{-1}$ as seen in the radial velocity shift of forbidden
emission lines and proper motions of jet knots. Many of these jets
remain highly collimated with opening angles less than $5^{\circ}$
over a distance up to several $\pc$~\citep{Mundt90, Raga01}. As
proposed first by~\citet{Norman80}, these Herbig-Haro (HH) outflows
could sustain the energetics in molecular clouds. This is an
attractive idea as this process could be self-regulated. The amount of
turbulent energy controls the strength of gravitational collapse and
subsequent star formation activity: an increased energy input from protostellar
outflows results in higher levels of turbulence, which reduces the star
formation activity. This in turn lowers the energy injection by
outflows, increasing again the star formation rate.

In order to study the effects of jet-driven turbulence, it is
useful to estimate the kinetic energy associated with protostellar jets,
and compare it to the energy needed to drive supersonic turbulence in a
typical star forming region. It should be mentioned that the simple
analysis of the energies involved can at best be an order-of-magnitude
approach. Real values may easily differ by factors of a few. However,
it demonstrates that protostellar jets and outflows may indeed play an
important role in determining density and velocity structure in star
forming regions~\citep[see, e.g.,][]{Stanke00, Stanke02}.

We begin with an estimate of the protostellar jet kinetic
luminosity. It can be described as
\begin{equation}
L_{\rm jet} =  \frac{1}{2}\dot{M}_{\rm jet} \,
v^2_{\rm jet} \approx 2.9\times10^{32}\,{\rm erg\,s}^{-1}
\left(\frac{\dot{M}_{\rm jet}}{10^{-8}\,{\rm M}_{\odot} {\rm
yr}^{-1}}\right)
\left(\frac{v_{\rm jet}}{300\,{\rm km\,s}^{-1}}\right)^2\;,
\end{equation}
with $\dot{M}_{\rm jet}\approx 10^{-8}\,{\rm M}_{\odot} {\rm yr}^{-1}$
being the mass loss associated with the jet material that departs from
the protostellar disk system at typical velocities of ${v_{\rm
jet}}\approx 300\,{\rm km\,s}^{-1}$. A simple estimate of the jet
lifetime in this phase is $\tau_{\rm jet} \approx (3\,{\rm
pc}/{300\,{\rm km\,s}^{-1}}) \approx 10^4\,{\rm yr}$ where we take a
spatial extent of $3\,{\rm pc}$. This coincides to within factors of a
few with the typical duration of the class 0 and early class I phases
of protostellar evolution. In these phases we expect the strongest
outflow activity~\citep[see the review by][]{Andre00}.  This kinetic
luminosity is less than but comparable to the radiative luminosity of
protostars.  The outflow-ISM coupling is more direct and, thus,
supposedly more efficient than the energy exchange between the
protostellar radiation and the ISM. However, the detailed coupling
strength is not known and will be investigated here. The total
amount of energy provided by the jet is
\begin{equation}
E_{\rm jet} =  L_{\rm jet}\,\tau_{\rm jet} \approx 10^{44} \,
{\rm erg}\;.
\end{equation}

If we assume a typical cluster-forming region of molecular cloud
material, say with mass $M=1000 \,{\rm M}_{\odot} = 2\times 10^{36}
\,{\rm g}$ and a rms velocity dispersion of $v_{\rm rms} = 1
\,\km\,\sec^{-1}$ (corresponding to a Mach 5 flow at the canonical
cloud temperature of $10\,$K), then the total turbulent kinetic energy
in the region is
\begin{equation}
E_{\rm kin} =  \frac{1}{2} M \, v^2_{\rm rms} \approx  10^{46}\,{\rm erg}\;.
\end{equation}
In principle, this means that
\begin{equation}
N_{\rm jet} = f\,E_{\rm kin} /  E_{\rm jet} \approx 100 \, f \;,
\end{equation}
jets could carry enough energy to drive the turbulence if the coupling
to the ambient medium, $f$, is reasonable efficient.  Given that a
molecular cloud region of $M=1000 \,{\rm M}_{\odot}$ will form a
cluster of several thousand stars \citep[see e.g., the Orion Nebula
cluster,][]{Hillenbrand97} protostellar jets are serious candidates to
drive turbulence in star-forming regions.

This analysis, however, is incomplete without a comparison of
timescales. We have estimated the lifetime of individual jets being of
order of $10^4\,$years. This has to be compared with the decay
timescale of turbulence in the star-forming region. The dissipation of
supersonic turbulence is roughly 
\begin{equation}
\dot{E}_{\rm kin} = - \eta M\,k\,v^3_{\rm rms}\;,
\end{equation}
with $\eta \approx 0.07$ \citep{MacLow99}.  $M$ is the total mass,
$k$ the wavenumber of the most dominant velocity mode, and again
$v_{\rm rms}$ the rms velocity dispersion. We know from
observations~\citep[e.g.,][]{Ossenkopf02} that molecular cloud
turbulence always is dominated by the largest-scale modes, we thus
take $k=1/L$ with $L\approx 1\,$pc being the size of the star-forming
region.  The timescale for turbulent decay then becomes
\begin{equation}
\tau_{\rm decay} = E_{\rm kin} / \dot{E}_{\rm kin} = \frac{1}{2\eta}
\frac{L}{v_{\rm rms}} \approx \, 6.8\times
10^6\,{\rm yr}\,\left(\frac{L}{1\,{\rm pc}}\right) \left(\frac{v_{\rm
rms}}{1\,{\rm km\,s^{-1}}}\right)^{-1}\;.
\end{equation}
Turbulence decays on several rms crossing times $L/v_{\rm rms}$, which
is considerably larger than the lifetime of individual jets. The
energy inserted by individual jets and outflows therefore will remain
in the cloud region during most of its star-formation period, which
typically lasts a few $10^5\,$years~\citep{Klessen03, MacLow04}.
This simple estimate indicates that protostellar jets and outflows
indeed carry sufficient energy to drive turbulence. The question that
remains, is whether this energy can be transferred to the ambient cloud
material with high-enough efficiency and with the right spatial and
temporal characteristics. 

Recently, the idea of jet-driven turbulence has been reconsidered in
phenomenological estimates~\citep{Matzner07, Quillen05} and several
numerical simulations~\citep[e.g.,][]{Chernin94, Gouveia99, MacLow00,
Micono00, Li06, Cunningham06a, Cunningham06b}. In particular, direct
numerical simulations of this issue have the possibility to answer the
question whether outflows and jets can drive supersonic turbulence in
molecular clouds. One of the first numerical study of outflow driven
turbulence was done by~\citet{MacLow00}. This work showed (with
randomly placed spherical and collimated outflow sources in a
molecular cloud) that the excited turbulence decays quickly and that
such point source driven turbulence is highly dissipative. A more
recent numerical investigation with the same resolution was done
by~\citet{Li06}. There the outflow sources were launched in regions
where protostars formed. Therefore, back-reactions from multiple
outflows in star forming regions could be studied
self-consistently. These authors claim that the outflows generated by
their protostellar objects maintain the turbulence in the molecular
cloud, although it is difficult to distinguish between the driving
forces -- gravity or outflows -- from their simulation.

Detailed numerical studies of isolated jets by \citet{Chernin94}
showed that low Mach number jets entrain gas along their edges via
Kelvin-Helmholtz instabilities and high Mach number jets transfer
energy and momentum mainly through the bow shock region at the head of
the jets. A similar study but for supersonic jets showed that a large
fraction of the jet momentum can be transfered to the ambient
medium~\citep{Micono00}. However, this investigation concentrated on
global properties of the energy and momentum transfer but did not
quantify the amount and structure of the excited turbulence. Numerical
investigations by \citet{Cunningham06a} showed that momentum
entrainment of the surrounding gas actually is reduced in jet-jet
interactions. The redirected radiative jets are not spread out widely
in this process, limiting the ability to strongly impact the jet
environment. On the other hand, ~\citet{Cunningham06b} found that
cavities produced by decaying jets stir up the gas inside the cavities
while they are back-filled. The idea of cavity driven turbulence was
recently put forward by~\citet{Quillen05} who compared the energetics
of expanding cavities by wind blown bubbles and outflows NCG 1333 with
the energy necessary to power turbulence in this molecular cloud. The
authors conclude that the energy from outflows seen in $^{12}$CO and
the observed cavities is sufficient to maintain turbulence in the
nebula~\citep[see also][for outflow activities in NGC1333]{Warin96,
Knee00}.

Another important aspect of jet entrainment, namely the influence of
radiative cooling, was recently presented by~\citet{Moraghan06}. For
high velocity jets it is the Mach number that determines the shape and
size of the jet and its bow shock, whereas for lower velocity jets the
cooling efficiency plays a crucial role in shaping the jet entrained
environment. If the gas can cool efficiently, lower velocity jets stay
more collimated and are less likely to develop large instabilities
compared the non-radiative jets. Here, we are mainly interested in
high velocity jets as a potential source of supersonic turbulence. For
those, radiative effects are of minor influence. We therefore use an
isothermal equation of state (EOS) throughout the presented study.

We note there is a literature concerning the question of possible
jet-driven molecular outflows, in particular to explain the
mass-velocity relation (usually a power law $\di m(v)/\di v \sim
(v/v_{\sm{jet}})^{-\alpha}$) and/or the outflow velocity-distance
relation (a linear "Hubble law" increase with distance) observed in
molecular outflows \citep[e.g.][]{Stahler94, Lada96}. Numerical
simulations which try to explain these findings, however, concentrate
on the momentum exchange from jets to ambient medium and and do not
investigate the turbulence pattern induced by this interaction
\citep[see e.g.][]{Downes99, OSullivan00, Downes03, Rosen04}.

Jets can travel over a long distance and are likely to interact with
density lumps on their way. One of the first numerical studies of
jet--cloud core interaction was carried out by~\citet{Raga95}. This
work showed that jets are highly deflected and de-collimated if the
density contrast between the jet and the dense clump is very
high. Whether jets can penetrate cloud cores and disrupt them, or are
just deflected, will have an impact on the cloud structure. To address
this point we present one example of jet--clump interaction in this
work.

If jets are the main driver of supersonic turbulence in molecular
clouds they have to impart high velocities to a large fraction of the
cloud material with high velocities. On the one hand, such jets
have large momenta, but on the other hand, highly supersonic jets
are well collimated and do not entrain much gas. These general jet
properties are also seen in the SPH simulations
by~\cite{Chernin94}. In \citet{Cunningham06a} the authors concluded
that the most efficient coupling of outflows with their surrounding
cloud will be by low-velocity fossil outflows. But, such low-velocity
jets are unlikely sources of {\em supersonic} turbulence.

In this study we present the results from numerical simulations of
individual jets interacting with their surrounding gas. We focus on
the impact of collimated jets on their environment, in particular, on
the jet-excited velocity structure in the surrounding gas. To quantify
the impact of jets on the cloud we present velocity probability
density functions (PDFs) for a number of different setups. These
include two and three dimensional jets, transient jets (jets whose
powering engines are shut off during the simulation), jet--clump
interactions, and jets that run into a magnetised environment. The
presented probability density functions can be interpreted as volume
filling factors of fluctuations with certain amplitudes. These
diagrams show that the fraction of fluctuations excited by jets that
reach supersonic velocities is negligibly small. Additionally, any
supersonic excitations do not propagate far into the cloud and decay
rapidly.

\section{Numerical Method}

We performed the numerical investigation with the adaptive mesh
refinement (AMR) code FLASH~\citep{FLASH00} in which we model the jet
as a kinetic energy injection from the box boundary. We parametrise
the jet speed by its Mach number, $\Ma$. In our simulations, the
temperature of the jet medium and ambient medium are the same, which
means that the internal and external jet Mach numbers are equal. For
most of our simulations the density of the jet material,
$\rho_{\sm{jet}}$, and the ambient medium, $\rho_{\sm{amb}}$, are also
equal but the jet density can be varied with the contrast parameter
$\delta = \rho_{\sm{jet}}/\rho_{\sm{amb}}$. The energy injection can
be switched on and off after a certain amount of time. We denote a jet
that is not continuously driven a transient jet. Typically our jet
runs into a homogeneous density distribution, but we also present a
study of jet--clump interaction where the jet runs into a spherical
over-density, $\rho_{\sm{cl}}$, with a density contrast given by
$\delta_{\sm{cl}} = \rho_{\sm{cl}}/\rho_{\sm{amb}}$. Furthermore, we
present studies where the jet runs into a magnetised medium where the
homogeneous magnetic field is either parallel or perpendicular to the
jet axis. In all cases we use the ideal HD and MHD treatment of the
FLASH code and neglect physical viscous and resistive effects. For the
MHD runs we use the standard diffusive cleaning of magnetic field
divergence. In Table~\ref{tab:runs} we summarise the parameters for
our different simulation runs.

We use outflow boundary conditions on the non-injection side of the
simulation box and constant boundary conditions on the side where the
wind is injected. For all but one simulations we use an isothermal
equation of state approximated by an adiabatic index of $\gamma =
1.0001$. Note that the FLASH code always evolves the entire energy
(i.e. kinetic, magnetic and thermal energy) within its standard
advection scheme and does not include a treatment for a pure
isothermal gas. For comparison we also show the results of one
barotropic run (i.e. $p \propto \rho^{\gamma}$) with a
barotropic index of $\gamma = 1.4$.

As we do not include radiative processes in our simulations (for
simplicity and to limit the parameter space) we present our results in
simulation units and give some examples to convert them to physical
units. We choose units where the density of the ambient medium is
unity ($\rho_{\sm{amb}} = 1$) and so is the pressure. Therefore, the
sound speed is unity throughout the simulation ($c_s = 1$) and the unit
of the gas velocity corresponds also to the sonic Mach number. For
example if we assume a mean number density of $10^3 \, \cm^{-3}$, a
mean molecular weight of $\mu = 2.1$ (i.e.~$\rho = 3.51\times 10^{-21}
\g \, \cm^{-3}$), and a gas temperature of $T = 10 \, \K$ (i.e.~$c_s =
0.198 \, \km \, \sec^{-1}$) a Mach 5 jet has a flow speed of $\sim 1
\, \km \, \sec^{-1}$. If we furthermore choose the length of the
simulation box to be $1 \, \pc$ ($L_{\sm{Box}} = 24 \, \rm{length \,
units}$; $1 \, \rm{length \, unit} = 1.286\times 10^{17} \cm$), one
time unit corresponds to $2.05\times 10^5 \, \ys$.

Our simulation boxes have dimensions of $24\times 8$ and $24\times
8\times 8$ length units in the 2D and 3D cases. Note that the images
presented show only subareas of the full simulation box.

The initial grid is resolved with up to 5 refinement levels which
corresponds to an effective resolution of $128$ grid points in the
$y$-direction (and $z$-direction in the 3D cases) and $3\times 128$
grid points in the $x$-direction. Furthermore we allow for maximal 8
refinement levels (7 in the 3D cases) during runtime. We use the
standard second-derivative refinement criterion based on the gas
density. This refinement criterion is particular useful to capture
density contrasts which allows us to track propagating shock
fronts at high resolution. The effective resolution corresponding to 8
refinement levels is 1024 grid points in the $y$-direction and
$3\times 1024$ along the jet axis ($x$-direction).

\begin{table*}
\begin{tabular}{l|c|r|c|c|c|c} \hline
run & dim & Mach & duration & $\delta$ & clump & MHD \\ \hline \hline
M5c & 2D & 5 & $\infty$ & 1 & no & no \\ \hline
M5t [g1.4] & 2D & 5 & 1.3 & 1 & no & no \\ \hline
M10c & 2D & 10 & $\infty$ & 1 & no & no \\ \hline
M20tCl & 2D & 20 & 0.6 & 1 & yes, $\delta_{\sm{cl}} = 10$ & no \\ \hline
M5t3D & 3D & 5 & 1.3 & 1 & no & no \\ \hline
M10tOd3D & 3D & 10 & 1.3 & 10 & no & no \\ \hline
M10tMpll3D & 3D & 10 & 1.3 & 1 & no & yes, parallel field \\ \hline
M10tMpe3D & 3D & 10 & 1.3 & 1 & no & yes, perpendicular field \\ \hline
\end{tabular}
\caption{Summarises the parameters of the different simulations. We
  refer to a certain simulation in the text by the name {\it
  run}. {\em Mach} is the Mach number of the jet, where the sound
  speeds in the jet medium and ambient medium are the same. The jet is
  driven for the time {\it duration} in simulation units after which
  the energy injection is switched off. The density contrast of the
  jet and possible a clump are given by $\delta =
  \rho_{\sm{jet}}/\rho_{\sm{amb}}$ and $\delta_{\sm{cl}} =
  \rho_{\sm{cl}}/\rho_{\sm{amb}}$, respectively. Magnetic field runs
  are denoted by {\it MHD}, where the homogeneous magnetic field is
  either parallel or perpendicular to the jet axis. For comparison we
  show the results of simulation M5t also with a barotropic EOS
  ($\gamma = 1.4$, M5tg1.4) in all other cases we use an isothermal
  EOS.}
\label{tab:runs}
\end{table*}

\section{Analysis and Results}

Our main focus is the question whether jets from young stellar objects
can excite {\em supersonic} turbulent motions in their parent
cloud. For this purpose we inject jets with different properties into
an ambient medium that is either homogeneous or clumpy. A well known
general trend is that high velocity flows ($\Ma \gg 1$) generate less
pronounced instabilities. For instance, Kelvin-Helmholtz instabilities
at the edge of the jet will appear only at very large wavenumbers,
i.e.~small wavelengths, which will decay quickly. The width of the bow
shock also shrinks with increasing jet speed. Furthermore, fast jets
are less prone to deflections on clumps and are only slightly
redirected in jet-jet interactions~\citep{Cunningham06a}. The most
distinctive impact on the environment by means of efficient gas
entrainment comes from flows that are transonic or only mildly
supersonic. Transient jets naturally reach such a transonic
state. 'Dying' jets leave fossil cavities that couple
quite well to the molecular cloud~\citep{Cunningham06b}. We 
discuss the effects of such transient jets in this section.

To study the impact of jets on their surrounding media it is quite
useful to quantify explicitly the jet-excited motions of the gas.  A good
way of doing this is to calculate the relative strength of particular
velocity excitations and follow their time evolution. For this purpose
we quantify the strength of the (turbulent) fluctuations produced by
the jet with probability density function (PDF) for velocity
fluctuations, $v$
\bea
P(f_i)     & = & \sum_j w_j(f_i) \\
w_j(f_i)   & = & \left\{ \Delta V_j/\Vol \, \,
                 | \, \, f_i \le f_j < f_{i+1} \right\}  ,
\eea
where $\Delta V_j / \Vol$ is the relative volume occupied by the
quantity $f_j$ at the mesh point $j$, and $[f_i,f_{i+1}]$ is the bin
range for the $i$'th bin in which $f_j$ is sampled.
With this definition the PDF is normalised, i.e.
\be
\sum_i P_i = 1 \; ,
\ee
so the $P_i$'s can be interpreted as volume filling factors of a
particular velocity interval. For the calculation of the presented PDFs
we only take into account the volume $\Vol$ which is affected by
the jet at each time, i.e.
\be
\Vol = \sum_i V_i \quad \mbox{with} \quad 
  \left\{i \; | \; v_i \ne 0\right\}
\ee
which is the volume with non-zero velocity fluctuations $v_i$.
With the help of the above described PDFs one can accurately measure
the fraction of sub- and supersonic gas motions.

The quantification of turbulent structure with the help of density or velocity PDFs is
often used in numerical simulations~\citep[e.g.][]{Vazquez94,
Chernin94, Falgarone94, Lis96, Klessen00c, LiKlessenMacLow03, MacLow05, Padoan07}. 
For supersonic turbulence
the generated density fluctuations can often be fitted by a log-normal
distribution which width is mainly determined by the {\it rms} Mach
number of the flow and similar holds for the velocity PDF. Here we use 
velocity PDFs (or volume weighted histograms) to distinguish
between sub- and supersonic fluctuations. For all of presented cases
this distinction comes about naturally: the supersonic regime is
strongly suppressed compared to the subsonic regime. We also tried to
fit log-normal distribution to the PDFs in the subsonic regime, but
their application are of limited use because of the complex structure
in velocity space (e.g.~douple or multiple peaks).

Furthermore, we compute contributions to the kinetic energy from the
subsonic and supersonic regimes separately and show their time
evolution. From these calculations one can derive decay laws for the
two different regimes and compare their importance to the overall
energy with time.

\subsection{General Jet Properties, Continuous vs. Transient jets}

\befs
\showfour{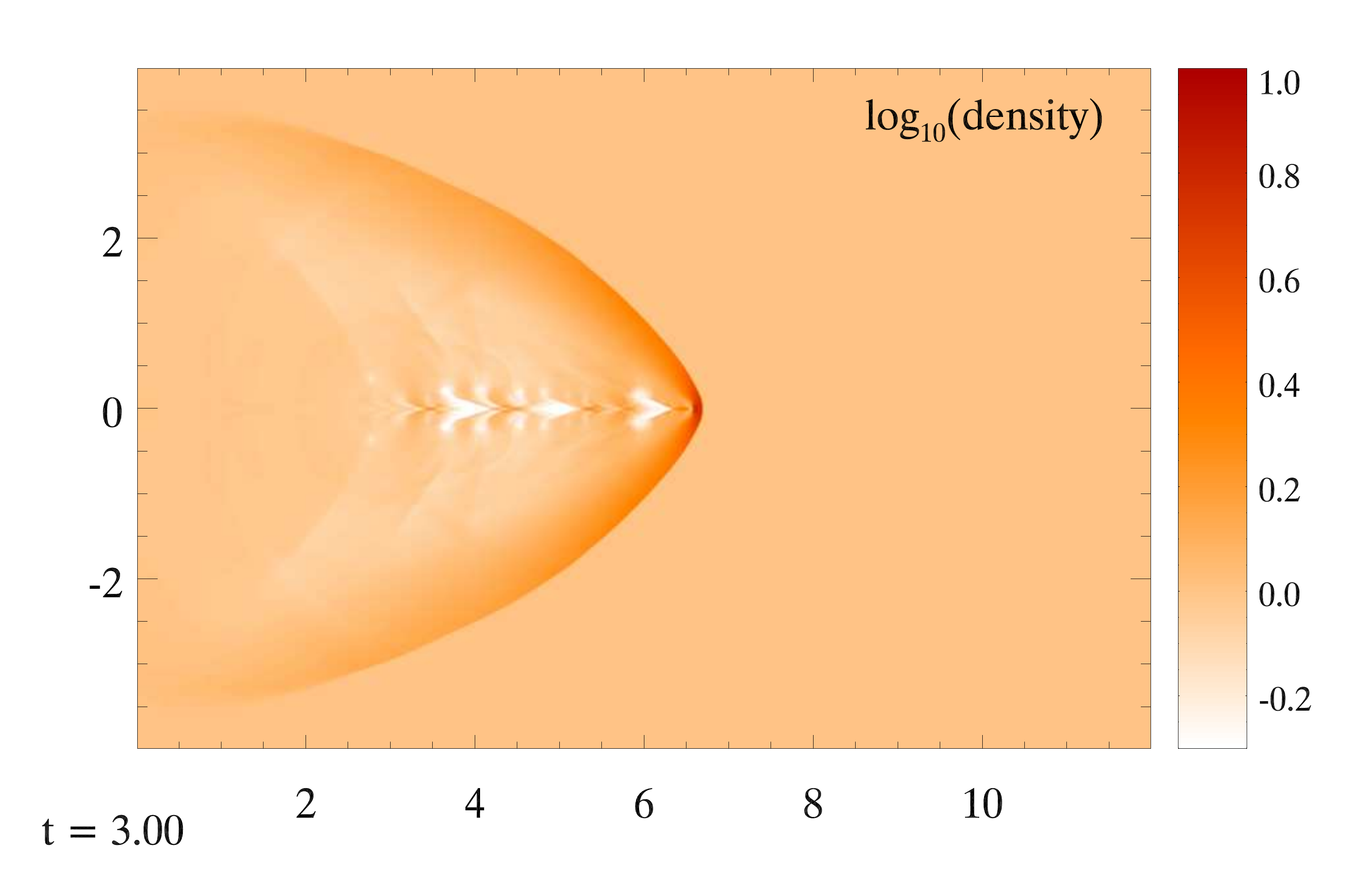}
	 {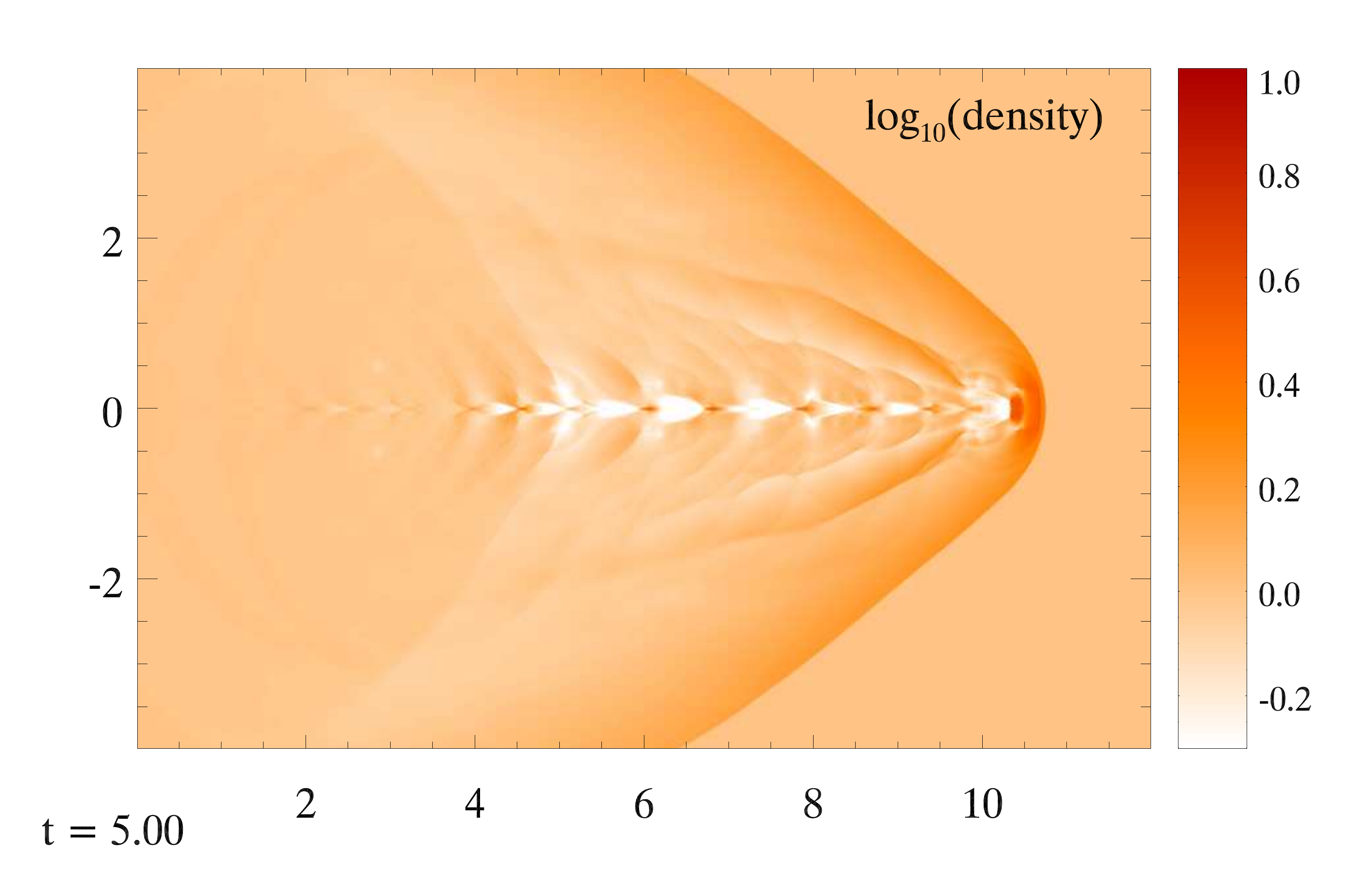}
	 {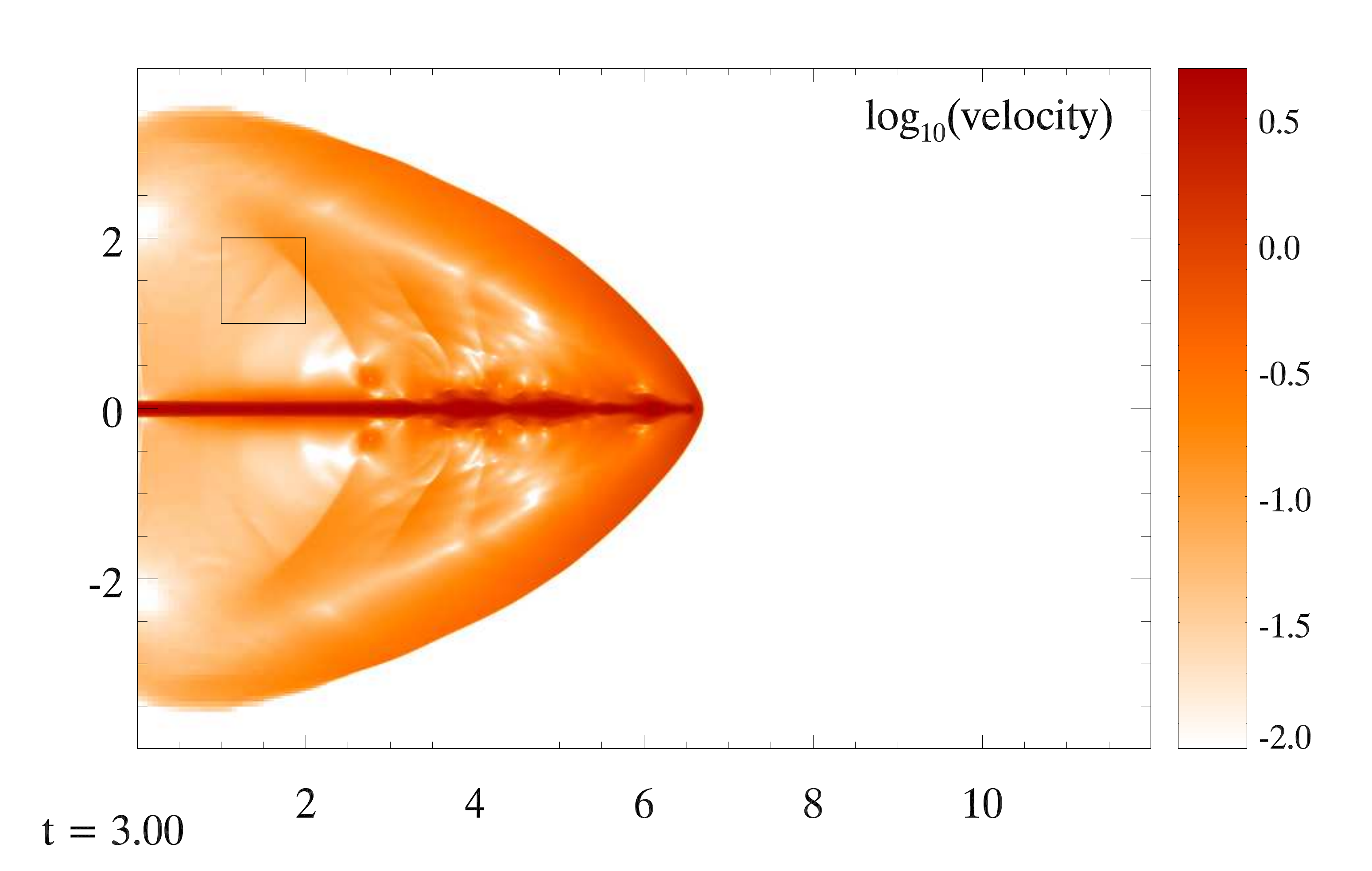}
	 {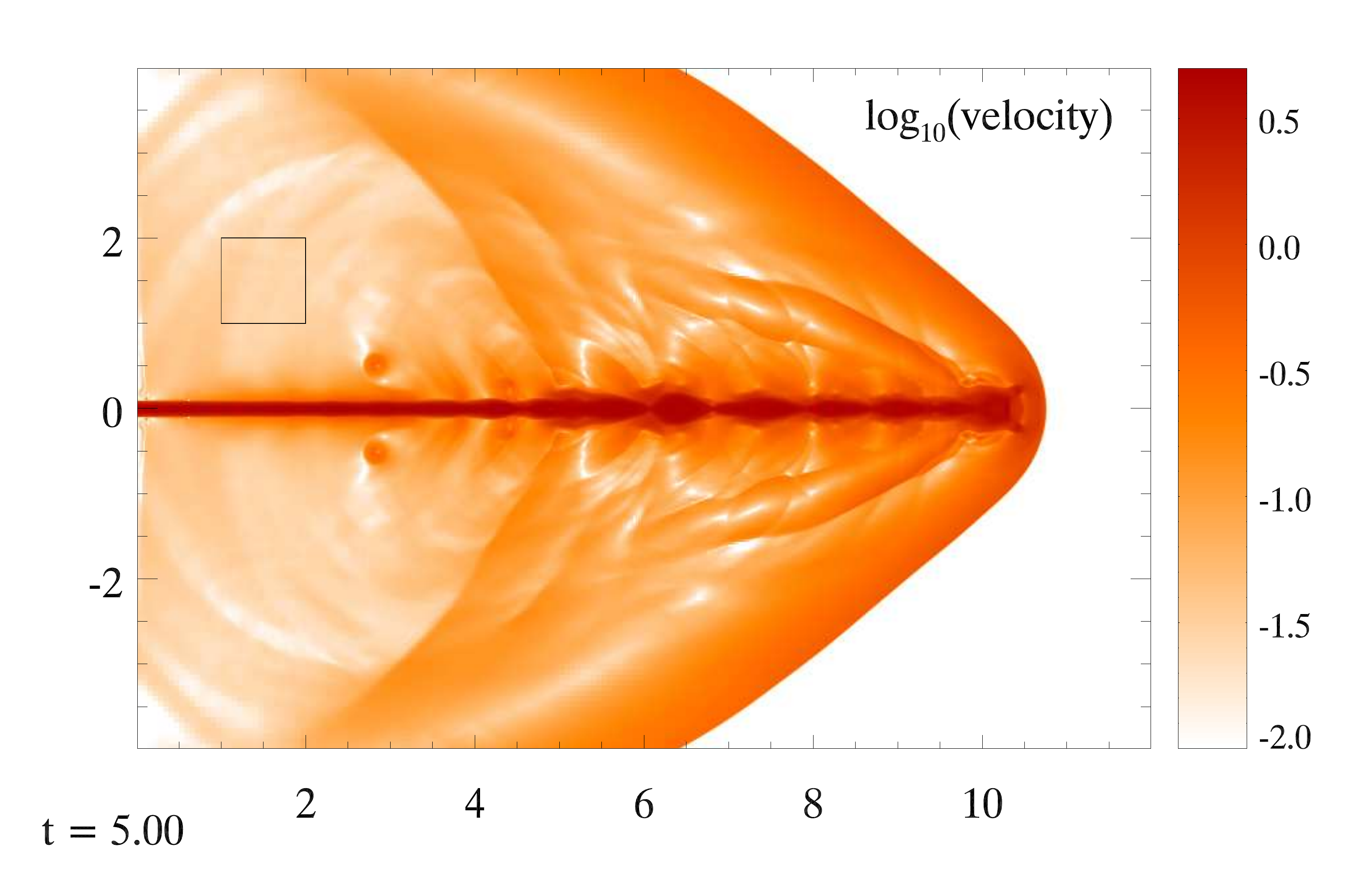}
\caption{Density (top) and velocity (bottom) evolution of a Mach 5 jet
(run M5c) at two different times, $t = 3.0$ (left) and $t = 5.0$
(right). The jet is continuously powered and runs into a homogeneous
medium. The jet develops knots from reflections off the jet edge. This
structure propagates also into the ambient media. Additionally,
Kelvin-Helmholtz instabilities develop at the edge of the jet. The
turbulent flow outside the jet is mainly subsonic.}
\label{fig:jet_M5c}
\eefs

\befs
\showfour{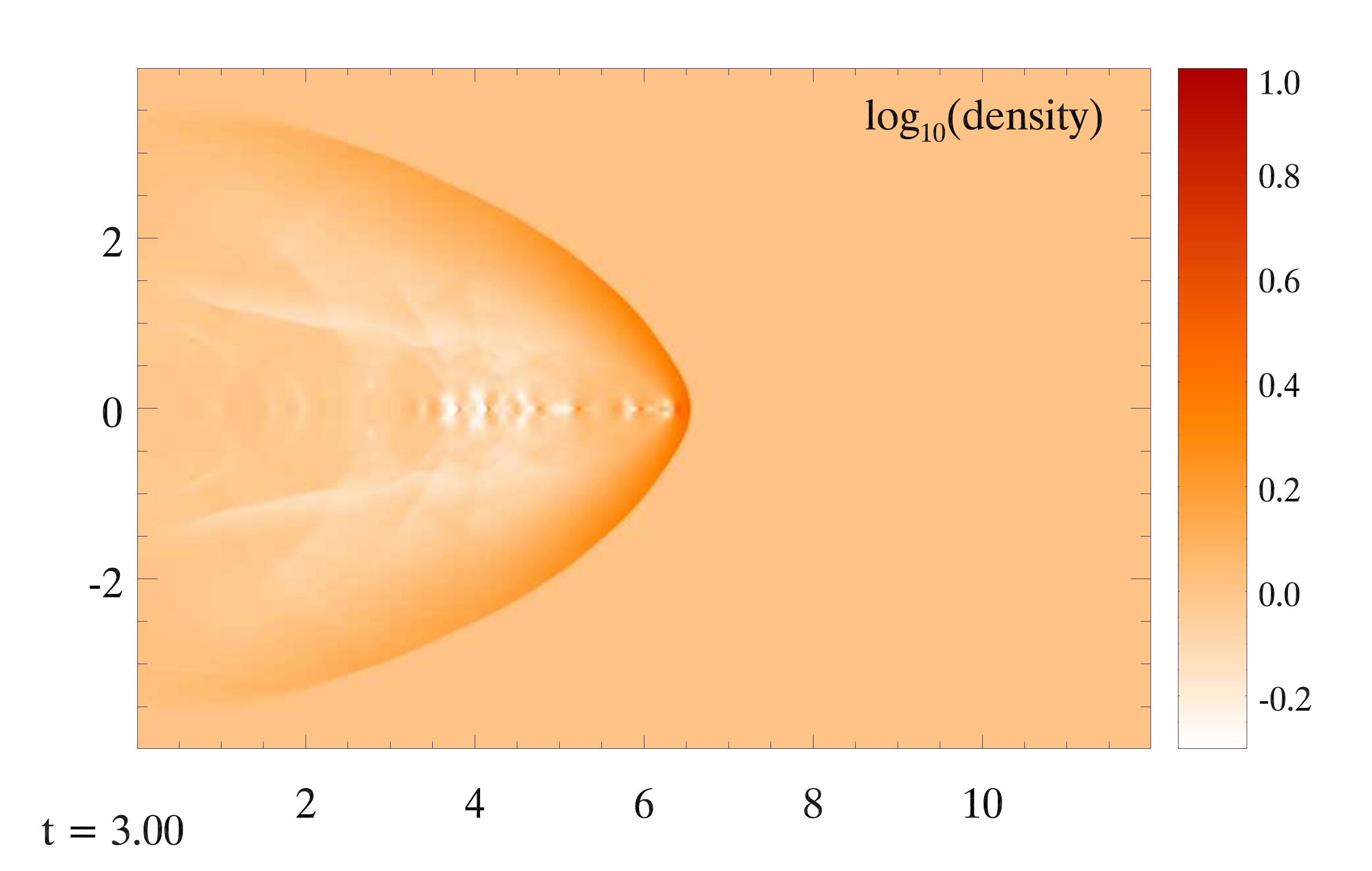}
	 {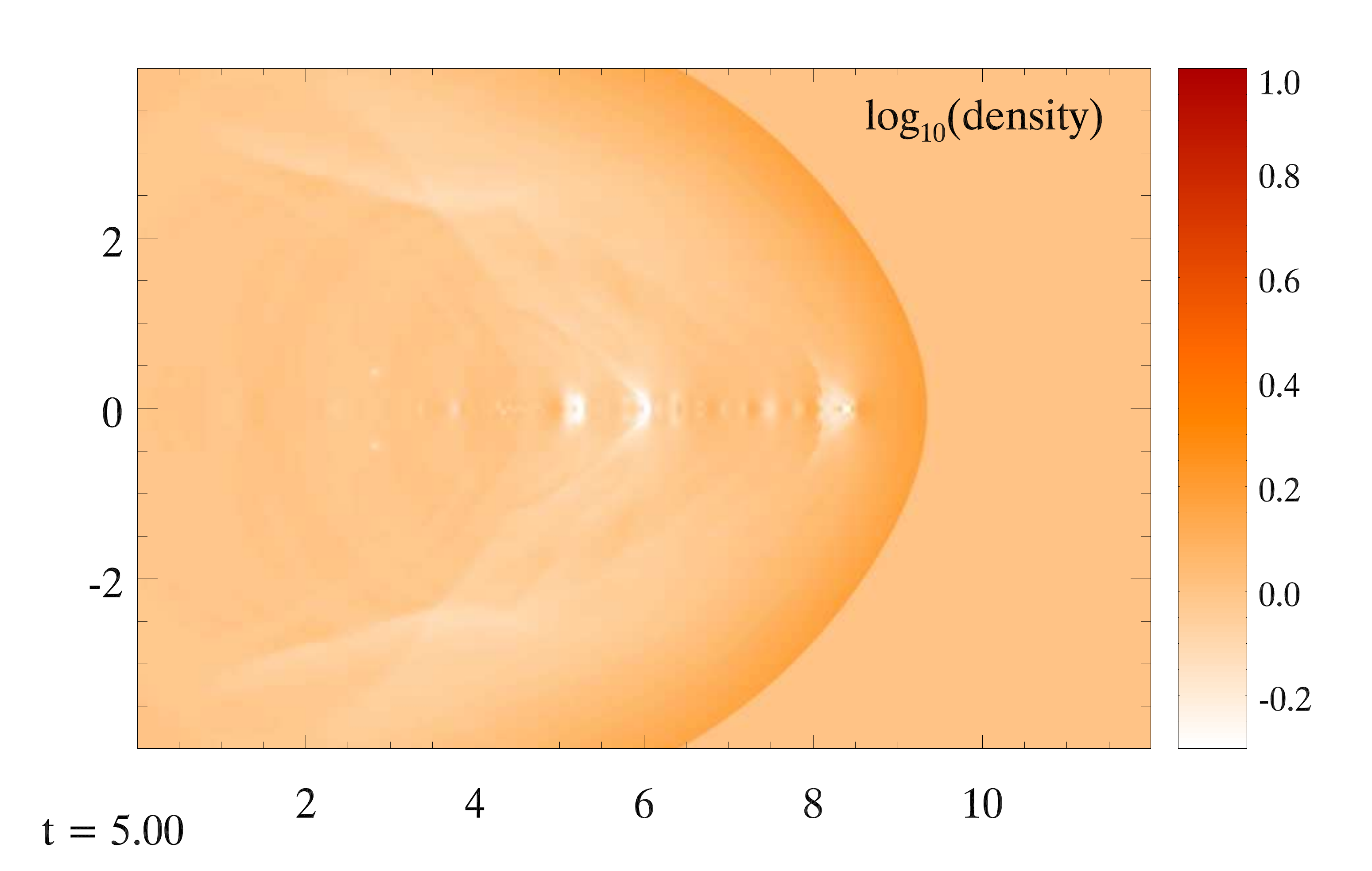}
	 {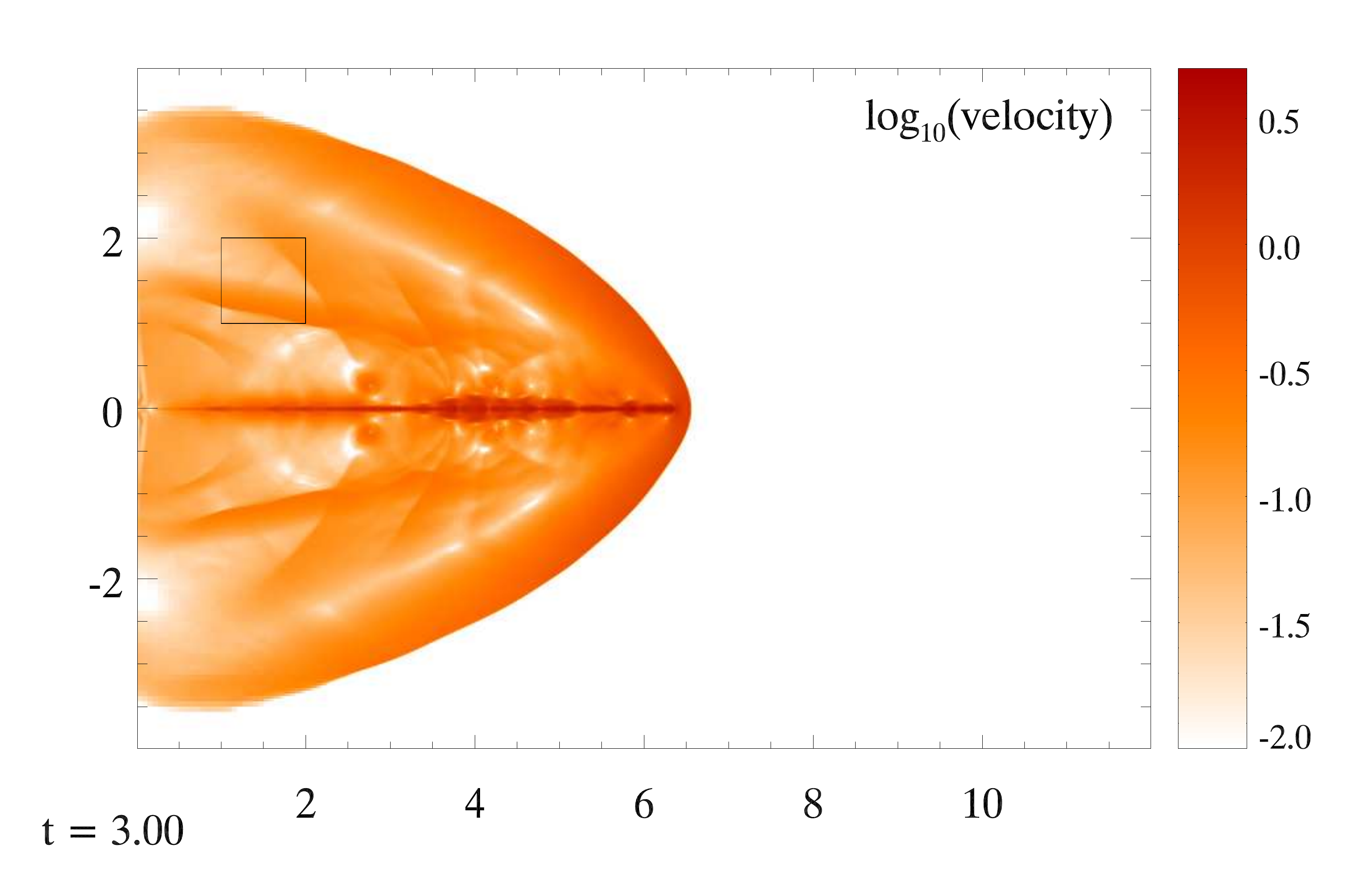}
	 {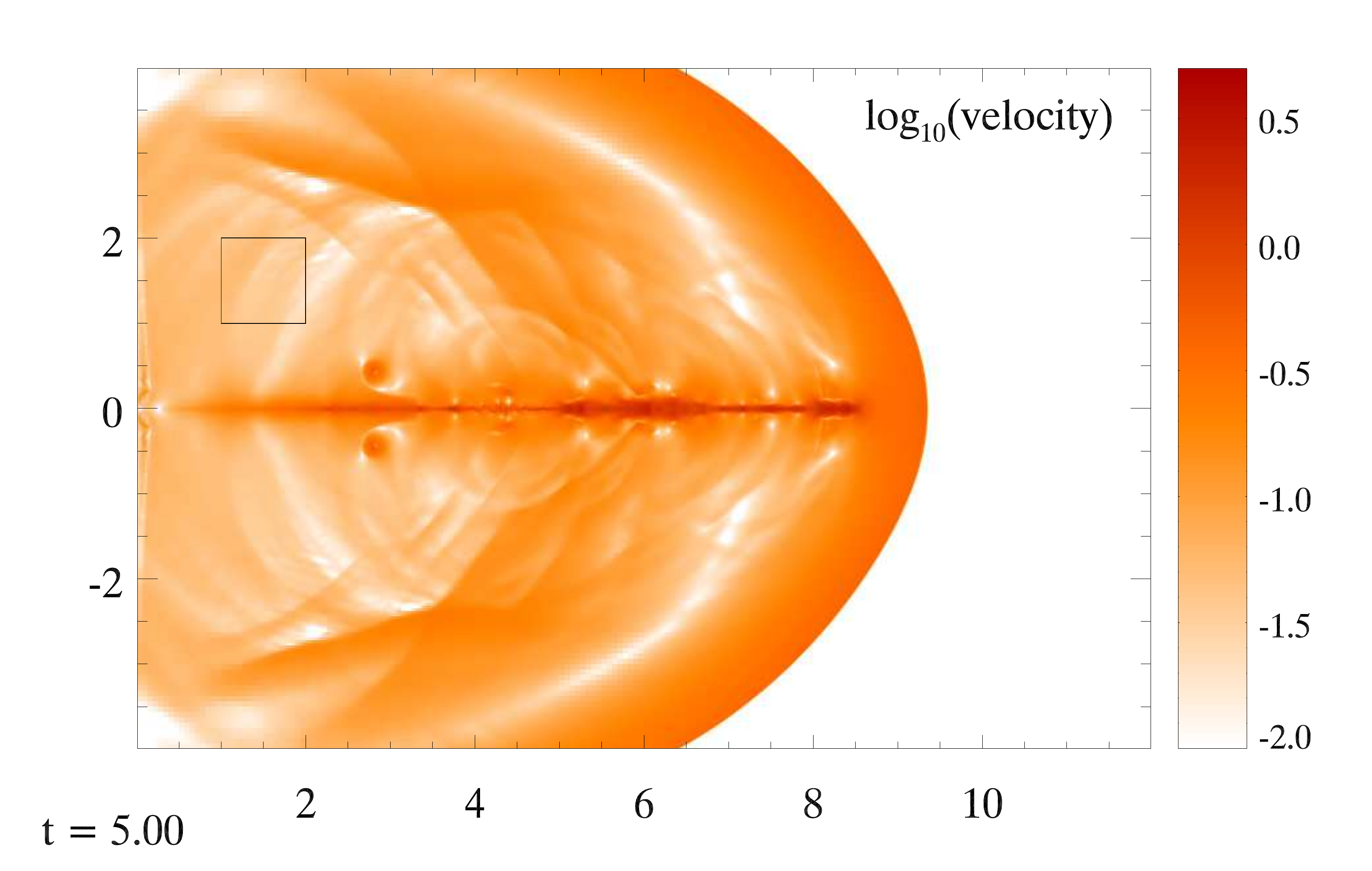}
\caption{Density (top) and velocity (bottom) evolution of a transient
Mach 5 jet (run M5t) at two different times, $t = 3.0$ (left) and $t =
5.0$ (right). The jet is shut off at $t = 1.3$ and runs into a
homogeneous medium. Compared to the continuously driven jet
(cf.~Fig~\ref{fig:jet_M5c}) the structure at late times ({\em right
panels}) entrained by the jet has already homogenised, an indication
of decaying large amplitude fluctuations.}
\label{fig:jet_M5t}
\eefs

\bef
\showtwo{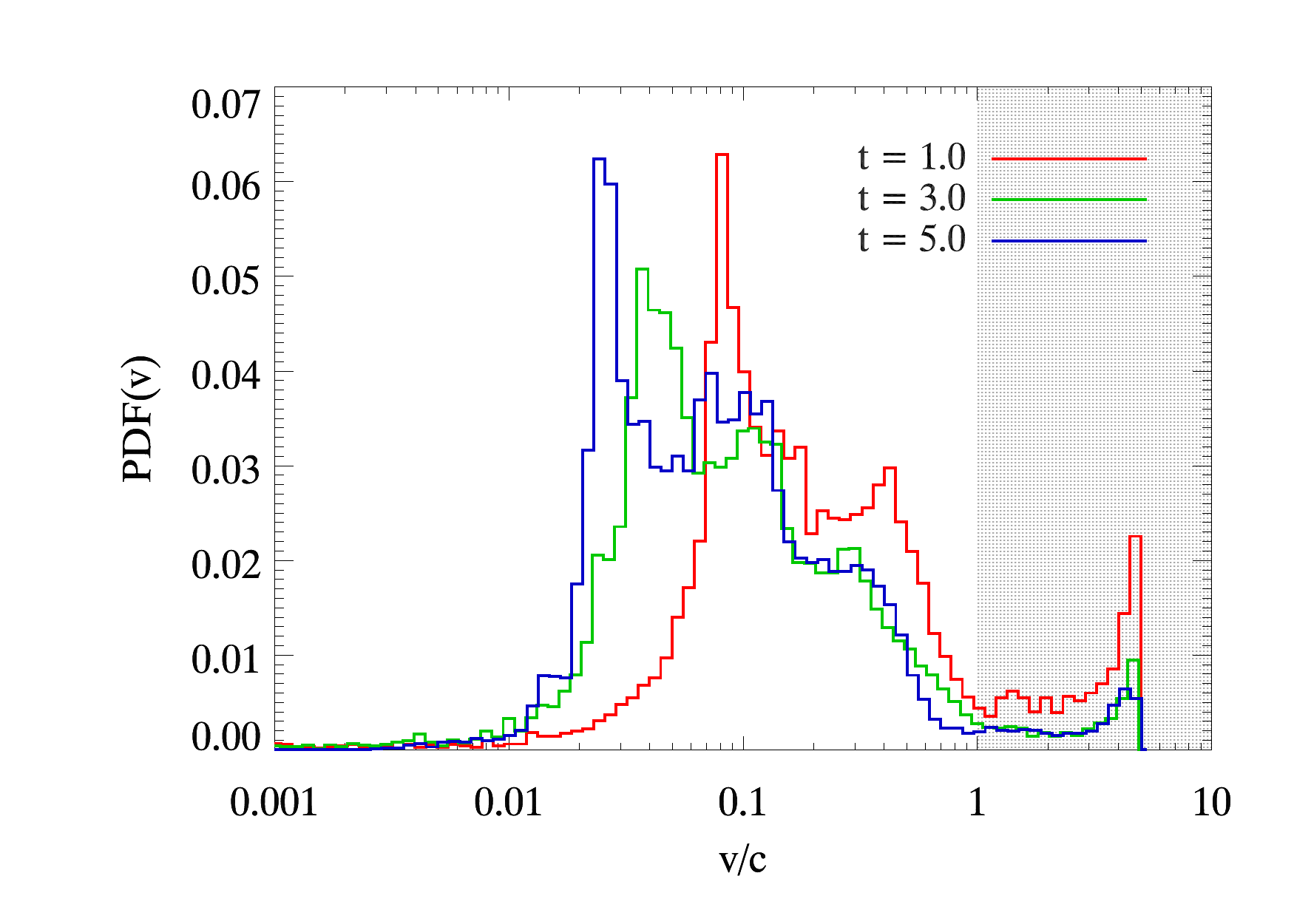}
        {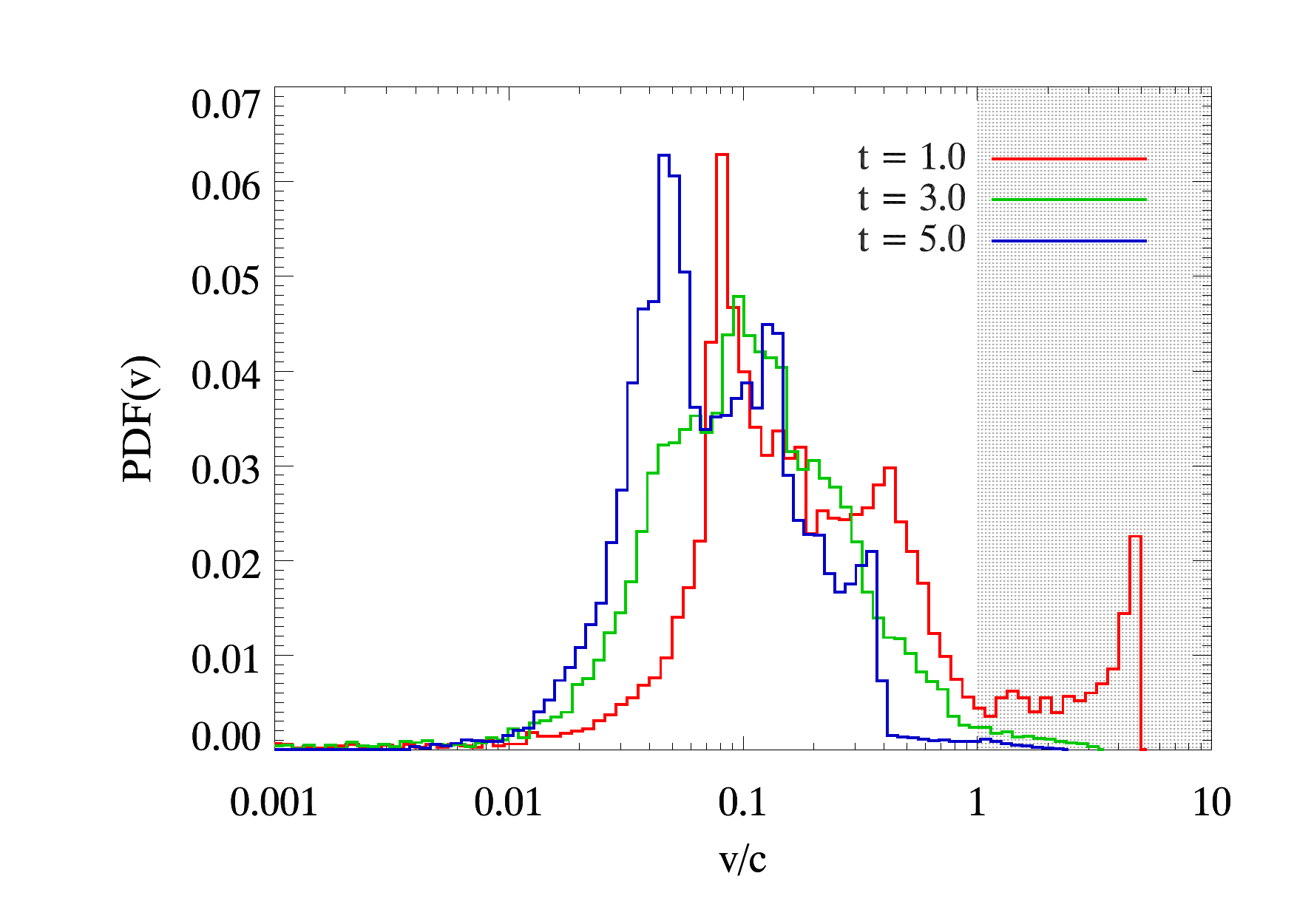}
\caption{Shows the time evolution of the probability density function of 
  velocity fluctuations in the case of an continuously driven jet ({\em
  left panel}) and a jet whose engine ceased at $t = 1.3$ ({\em right
  panel}). See also Figs.~\ref{fig:jet_M5c} and \ref{fig:jet_M5t} and
  text. The amplitudes of the velocity fluctuations are given in units
  of the sound speed (i.e.~as Mach number). The transient jet does not
  show any significant supersonic fluctuations after the driving has
  been stopped. Even in the continuously driven case most of the
  velocity fluctuations are subsonic (less than 2\% of the
  fluctuations are supersonic). Obviously, the peak at $v/c = 5$ comes
  from the Mach 5 jet itself.}
\label{fig:vel_PDFs_M5ct}
\eef

\bef
\showtwo{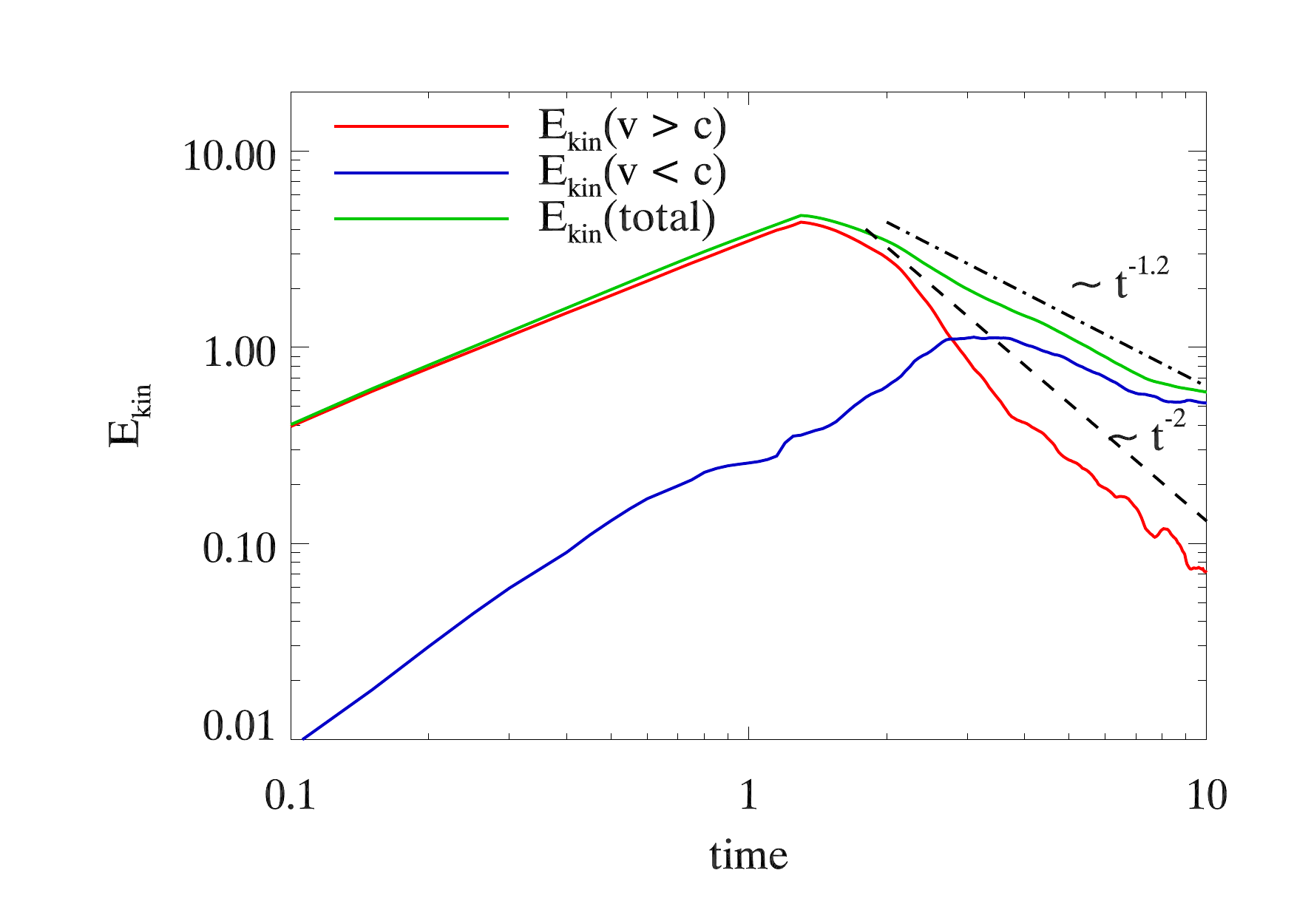}
	{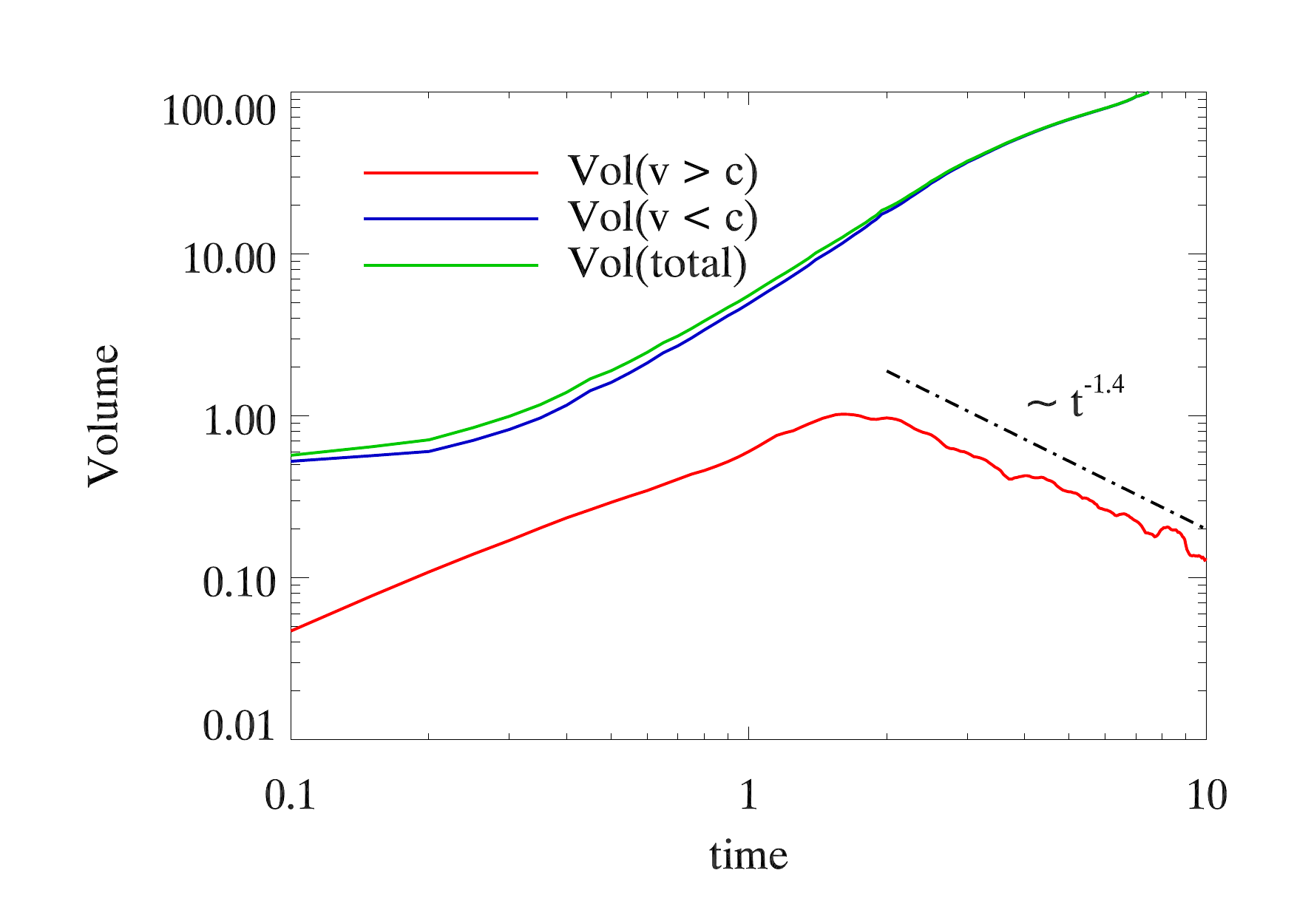}	
\caption{Shows the time evolution of the kinetic energies
  ($E_{\sm{kin}}$, {\it left panel}), and corresponding affected
  volume ($V$, {\it right panel}) for the transient jet M5t. The
  quantities are divided into a supersonic regime, $v > c$, and a
  subsonic regime, $v < c$. The decay of supersonic energy
  contributions is much faster (faster than $\propto t^{-2}$) than the
  subsonic one. The time shift between the peaks in the kinetic
  energies shows that the subsonic fluctuations are powered by the
  decay of the supersonic motions. Unlike the subsonic fluctuations,
  supersonic motions do not spread after the jet is shut off, as can
  be seen in the right panel.}
\label{fig:M5t_Ekins}
\eef

\bef
\showtwo{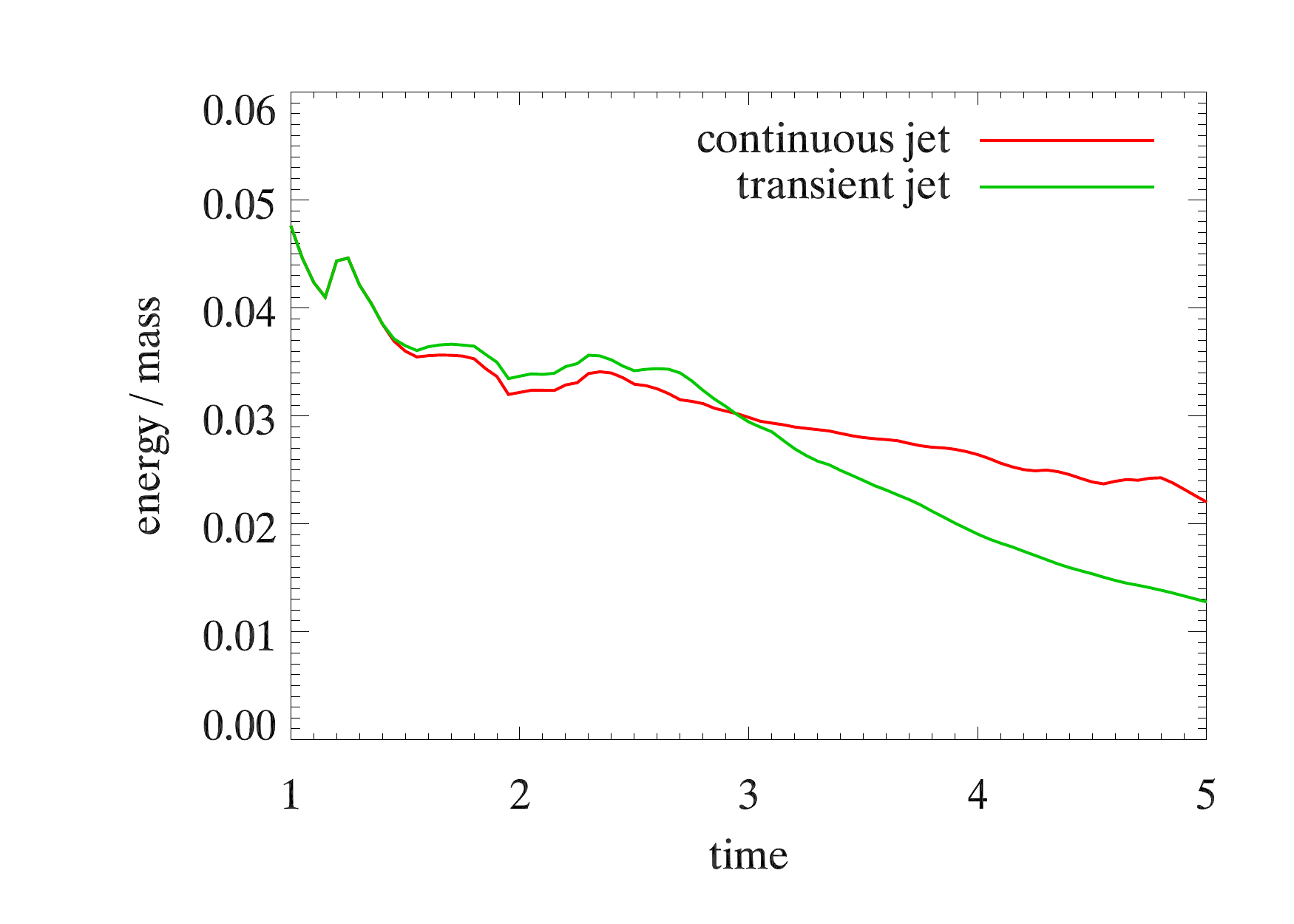}
	{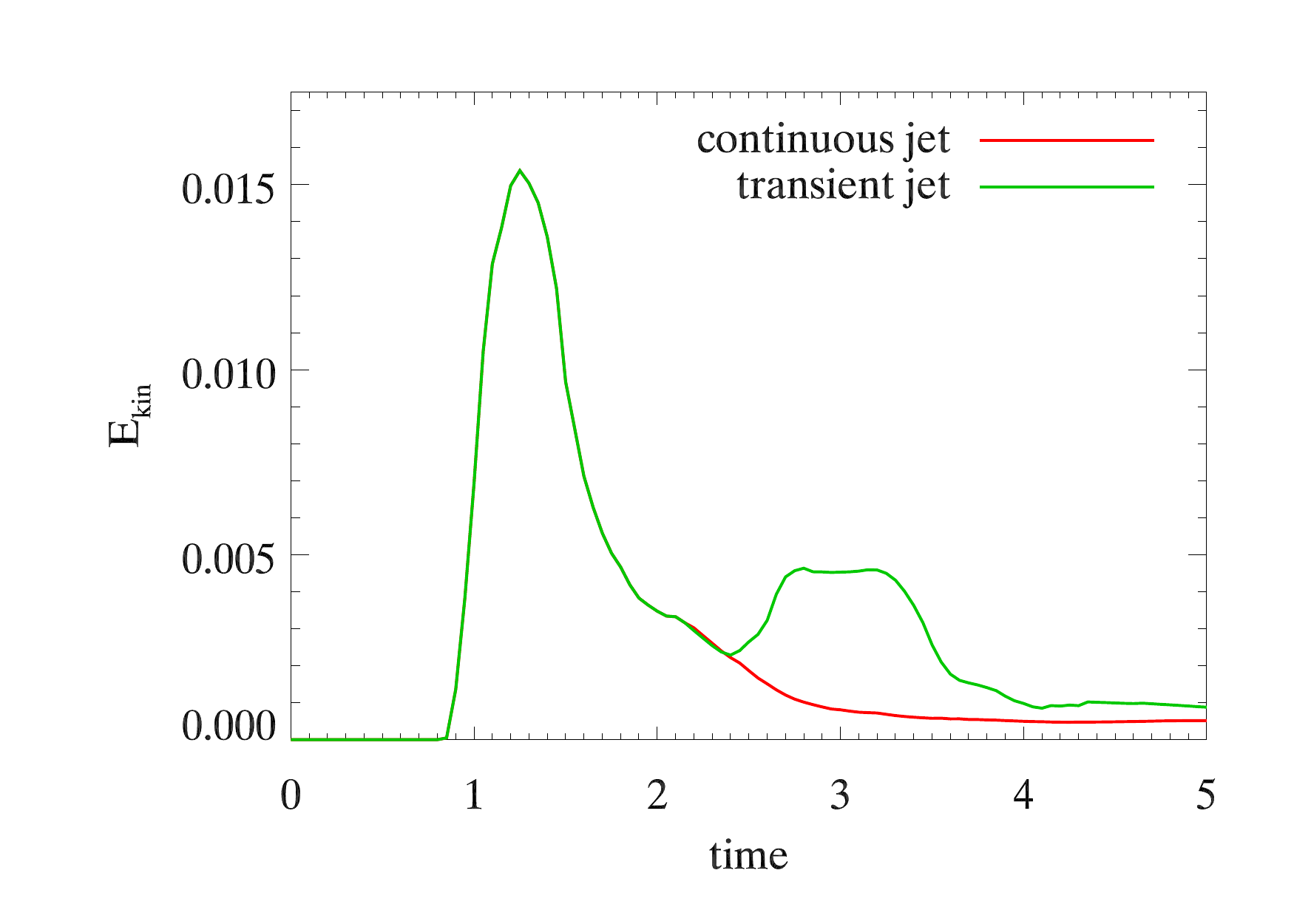}
\caption{Shows the time evolution of the energy for the runs M5c
(continuous jet) and M5t (transient jet), respectively. {\em Left
panel:} Kinetic energy per unit mass. Here, we used only contributions
from subsonic motions to calculate the energy. The slightly enhanced
energy from the transient jet during the intermediate times $1.5 < t <
3.0$ comes from the rarefaction wave travelling from the back-end of
the jet. {\em Right panel:} The total kinetic energy from a region
aside the jet (marked as boxes in the images of
Figs.~\ref{fig:jet_M5c} and \ref{fig:jet_M5t}). The bump in transient
jet case comes again from the rarefaction wave running through this
area after the jet is switched off.}
\label{fig:M5ct_sub_energy}
\eef

\befs
\showsix{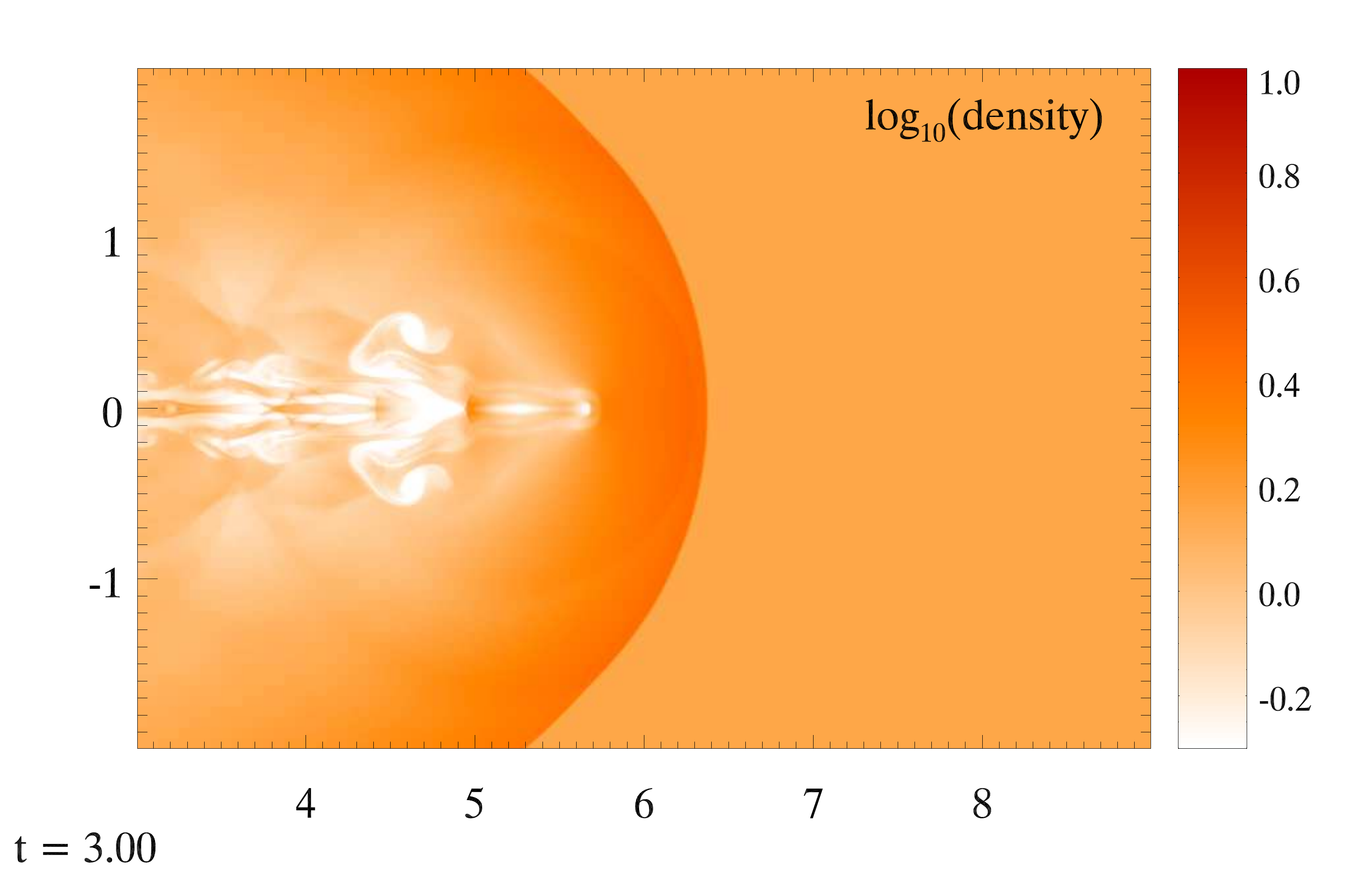}
	{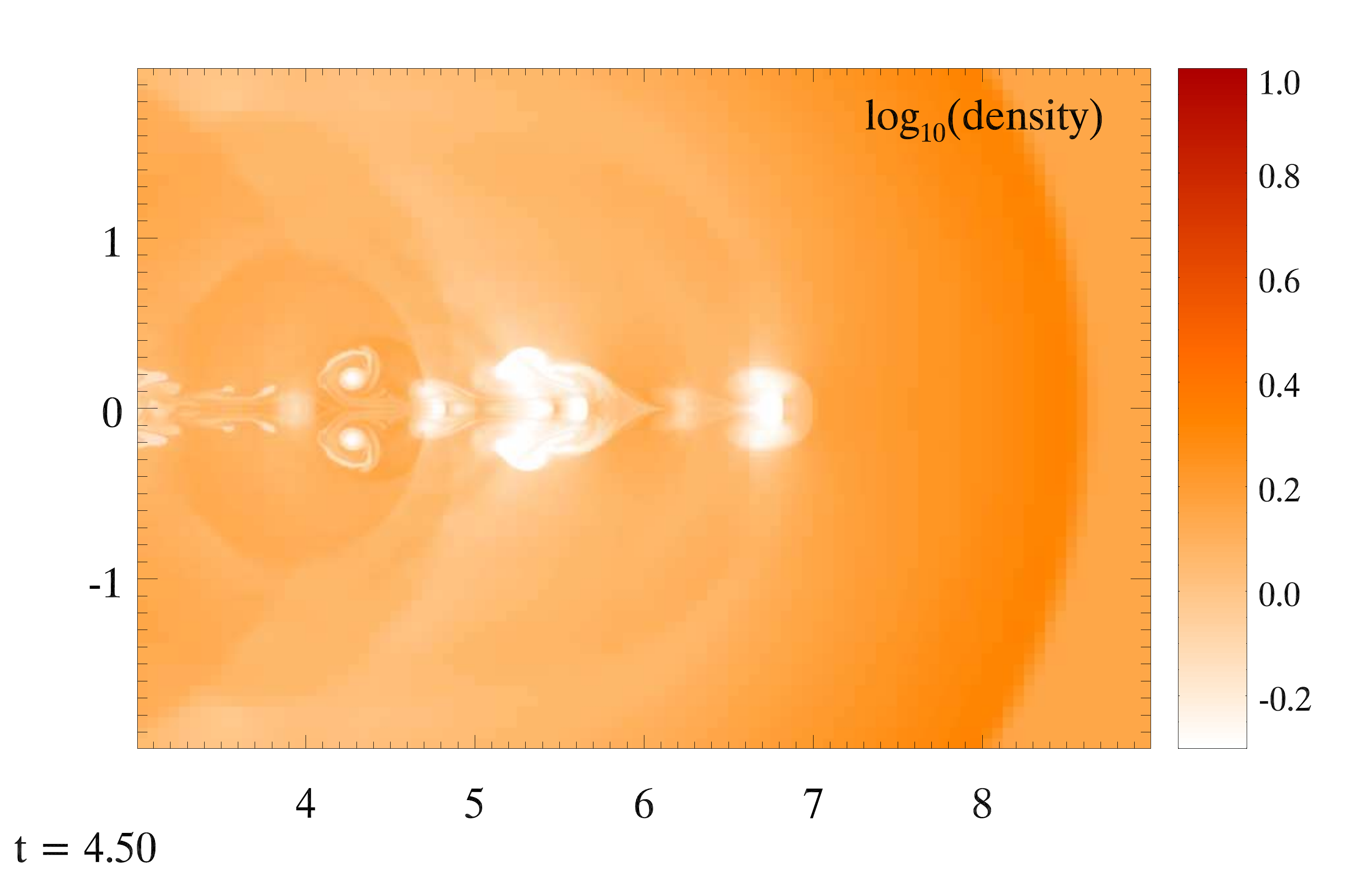}
	{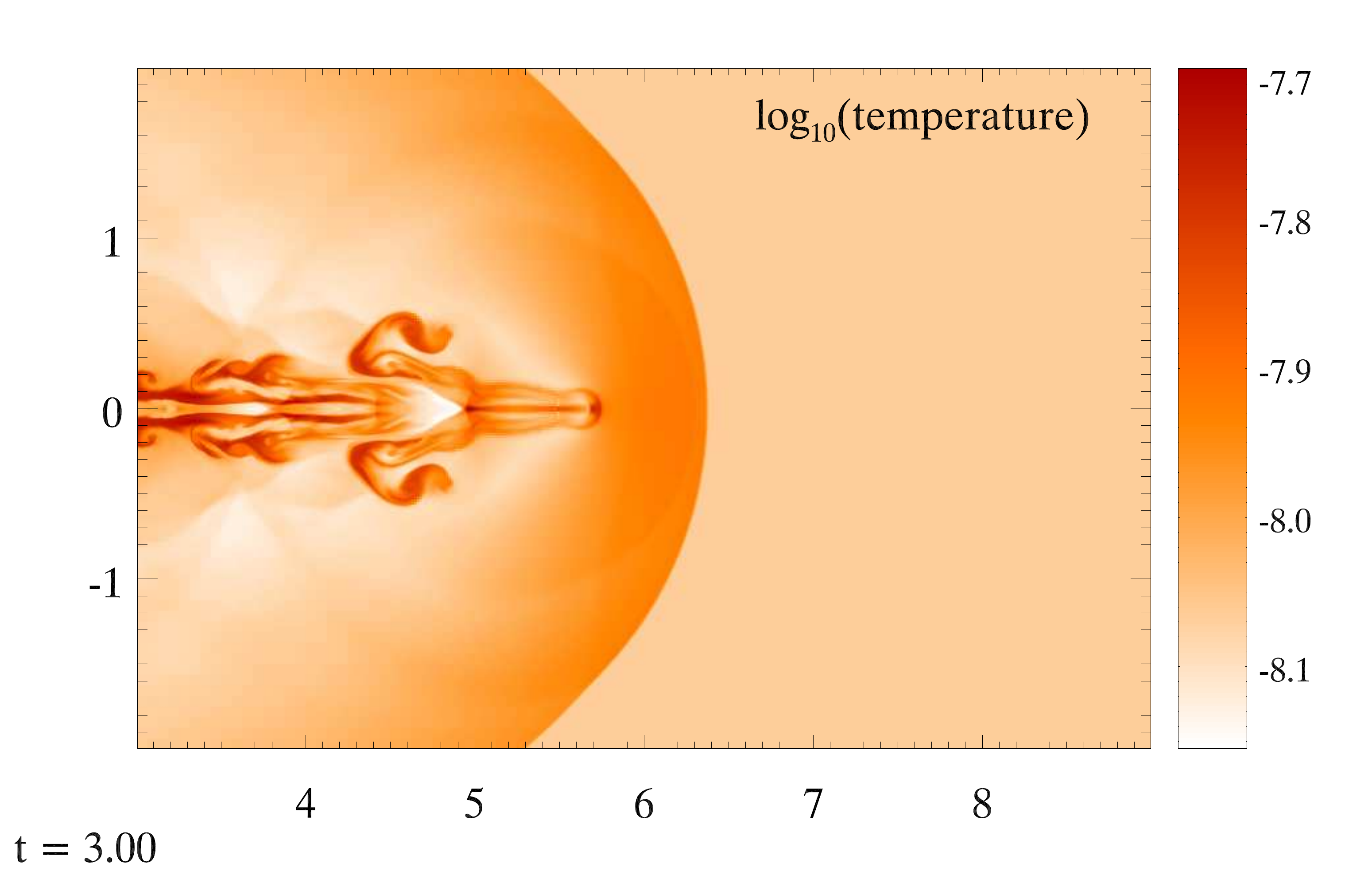}
	{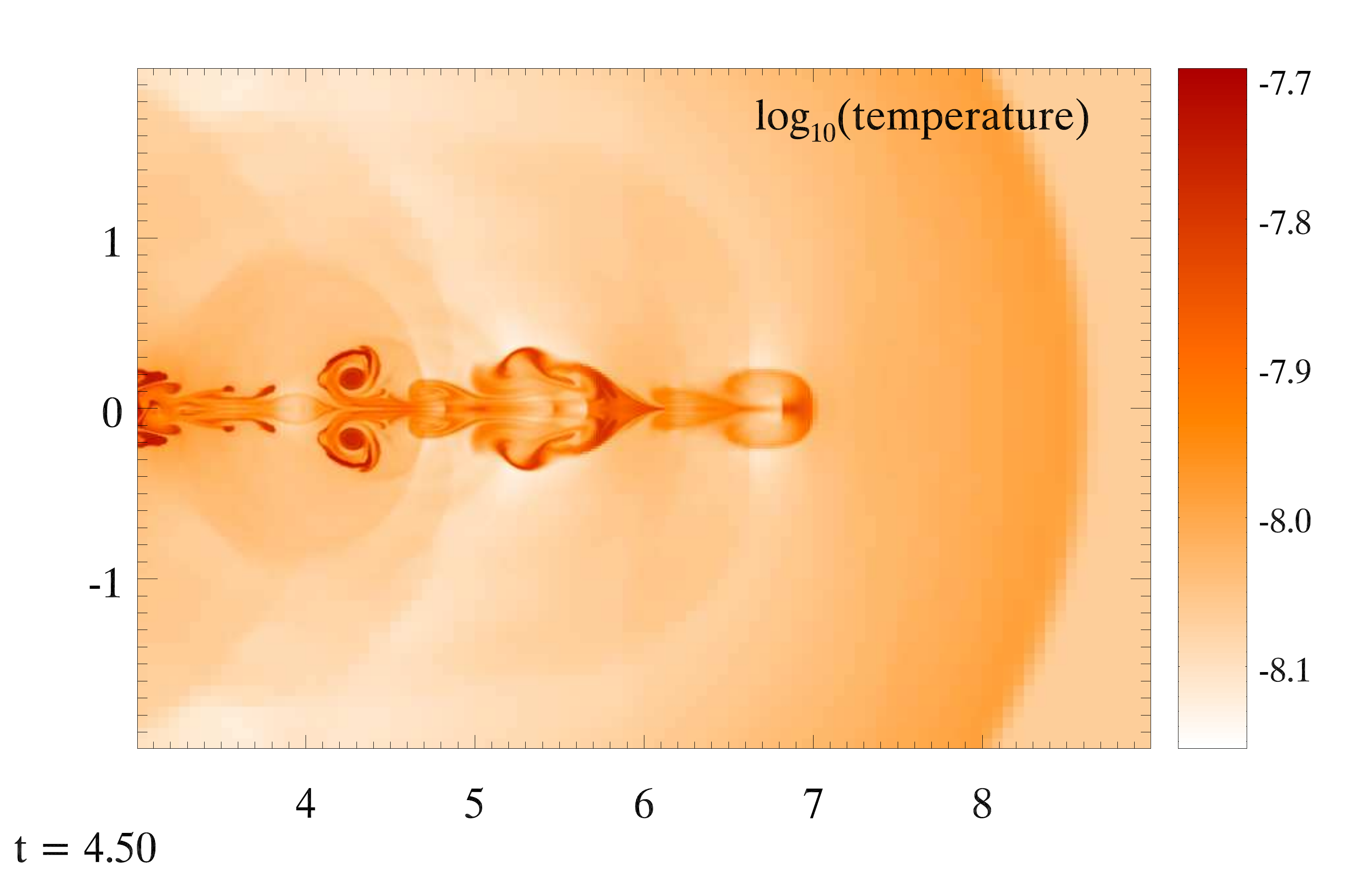}
	{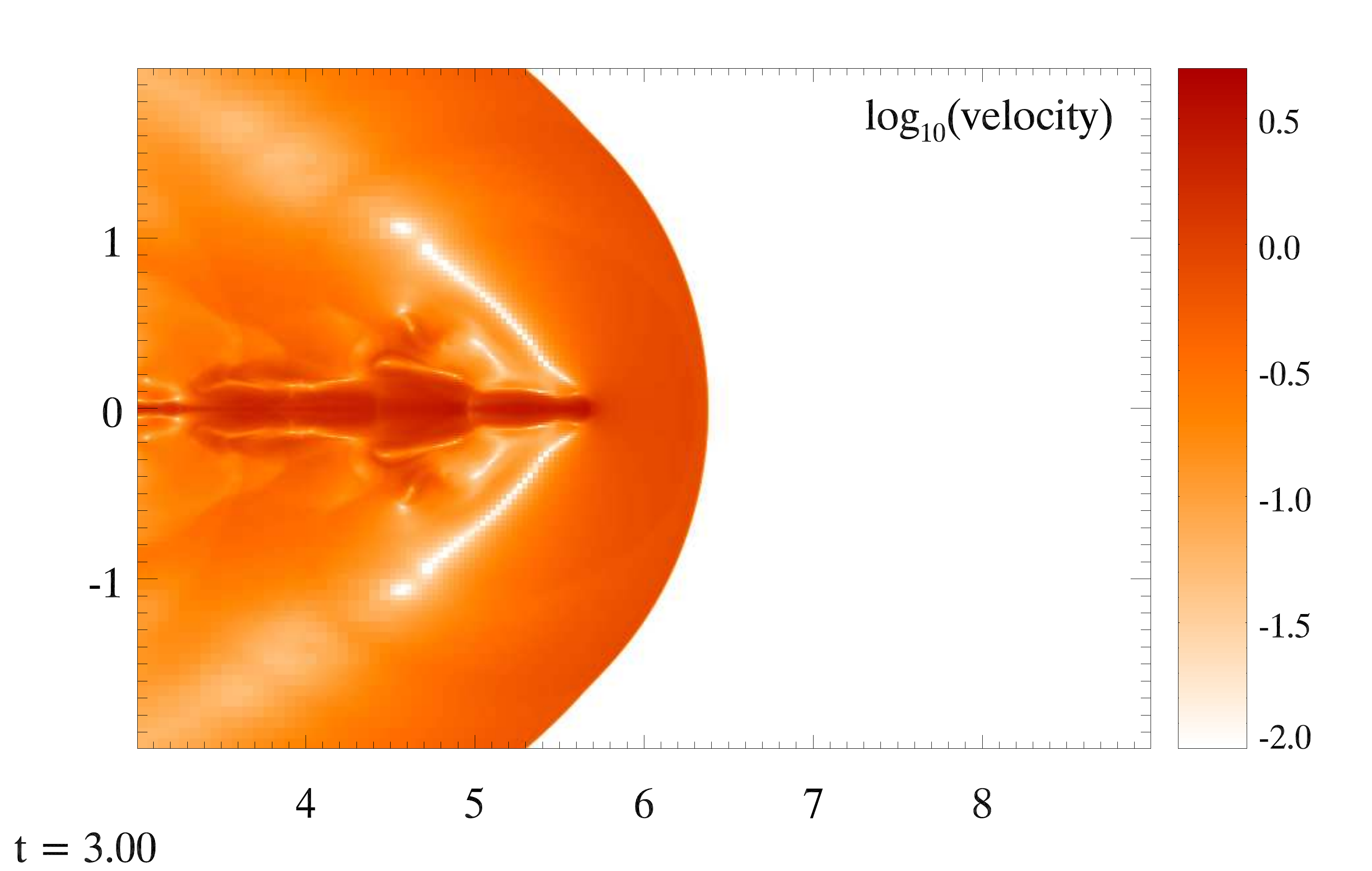}
	{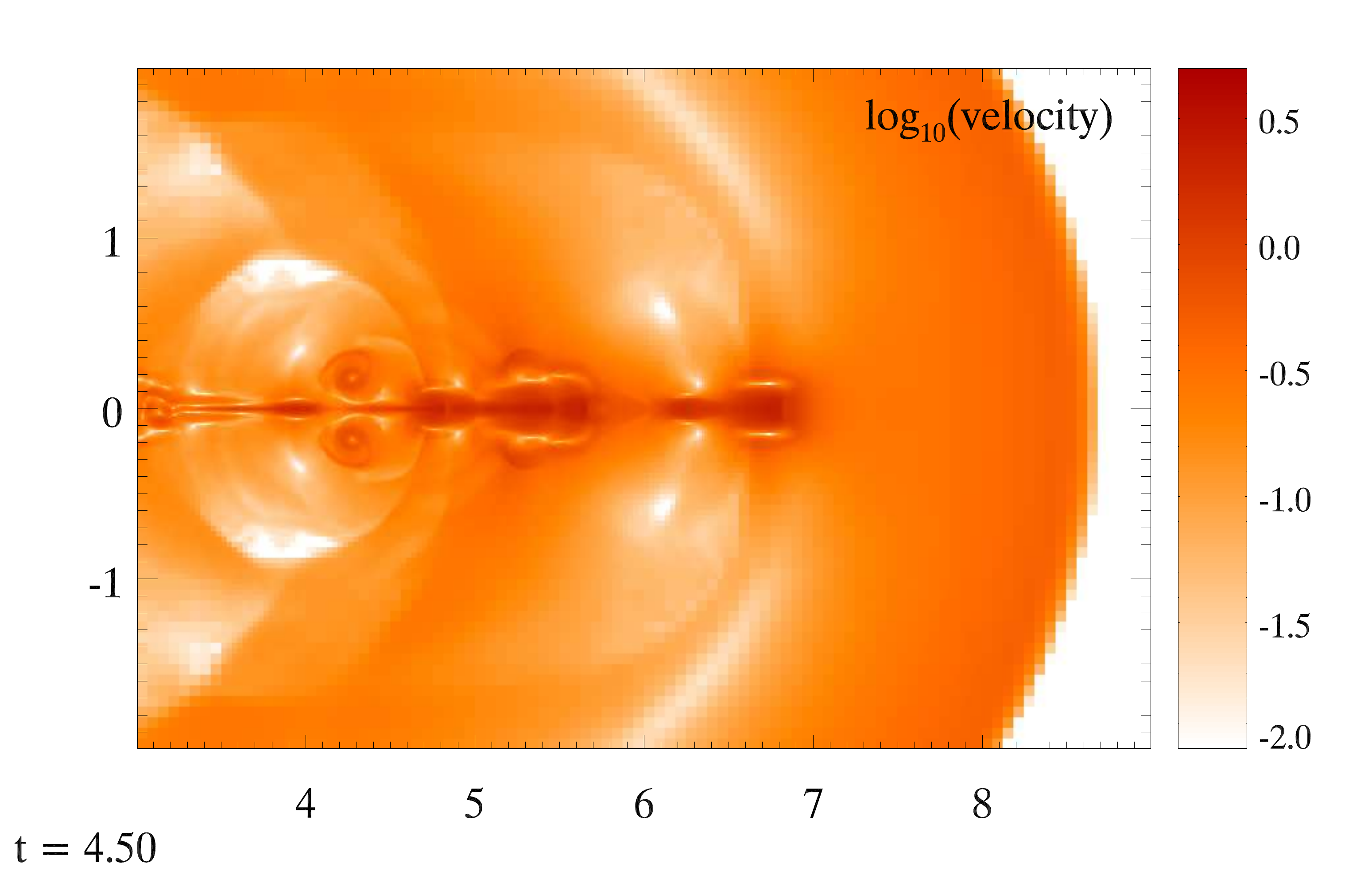}
\caption{Shows the density (top), temperature (middle) and velocity
(bottom) evolution of a transient Mach 5 jet (run M5tg1.4) at two
different times, $t = 3.0$ (left) and $t = 4.5$ (right). This
simulation run is the same than the isothermal run M5t (see
Fig.~\ref{fig:jet_M5t}) except of the equation of state. Here we use a
barotropic EOS (i.e.~$p \propto \rho^\gamma$) with a barotropic index
of $\gamma = 1.4$. Similar to the isothermal case, instabilities
develop at the edge of the jet (seen clearly in the temperature map)
but the associated velocity fluctuations are mainly subsonic
(cf.~Fig.~\ref{fig:vel_PDFs_M5tg1.4}).}
\label{fig:jet_M5tg1.4}
\eefs

\bef
\showone{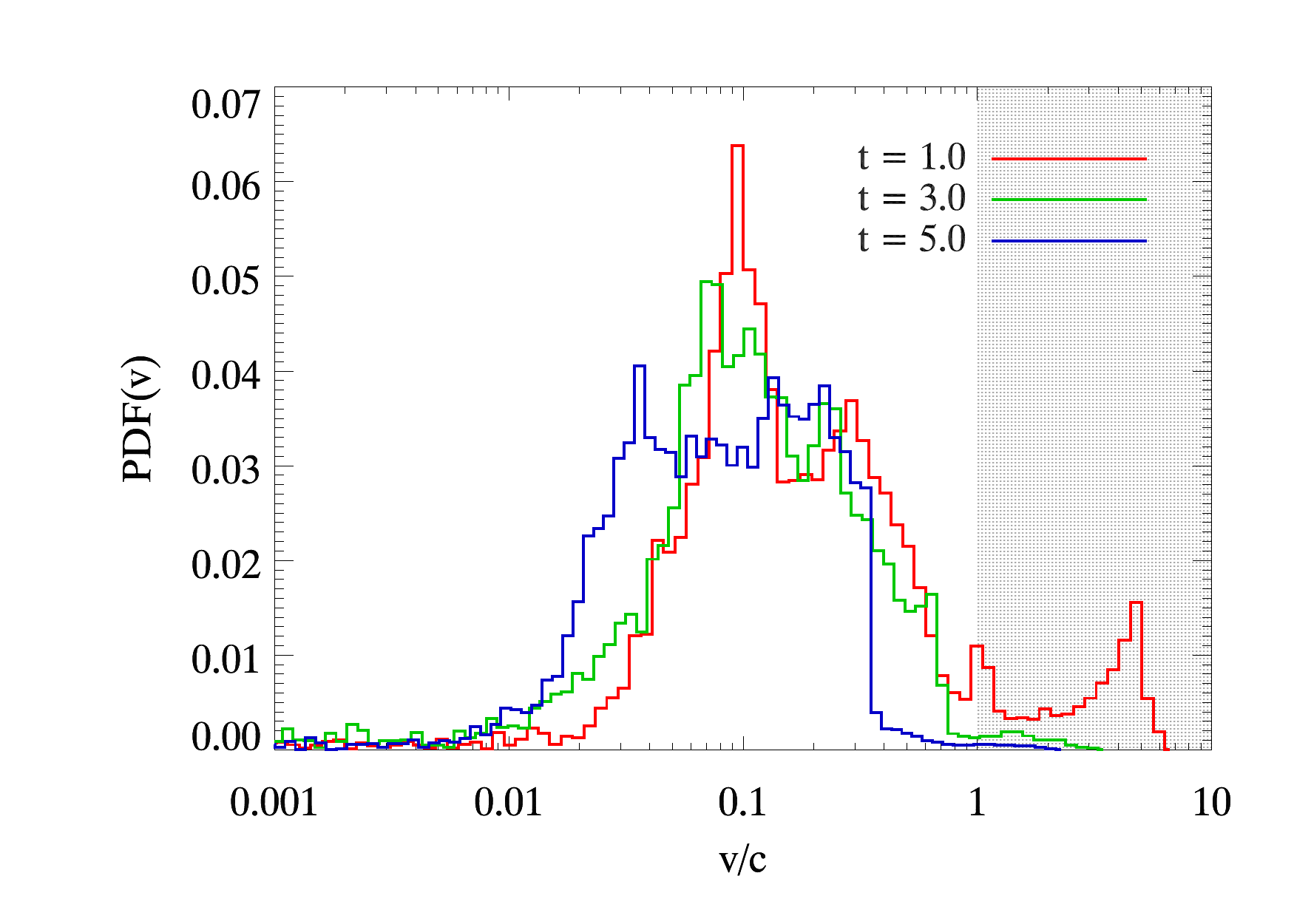}
\caption{Shows PDFs of the velocity fluctuations from the barotropic
  run M5tg1.4 at different times. This run is similar to the
  isothermal run M5t (see Fig.~\ref{fig:vel_PDFs_M5ct}) except of the
  equation of state. Here we use a barotropic index of $\gamma =
  1.4$. Note that the velocities are scaled to the {\em local} sound
  speed. See Fig.~\ref{fig:jet_M5tg1.4} for the temperature and
  velocity structure. Like in the isothermal case, supersonic velocity
  fluctuations are damped quickly and do not spread far from the jet
  itself.}
\label{fig:vel_PDFs_M5tg1.4}
\eef

\befs
\showfour{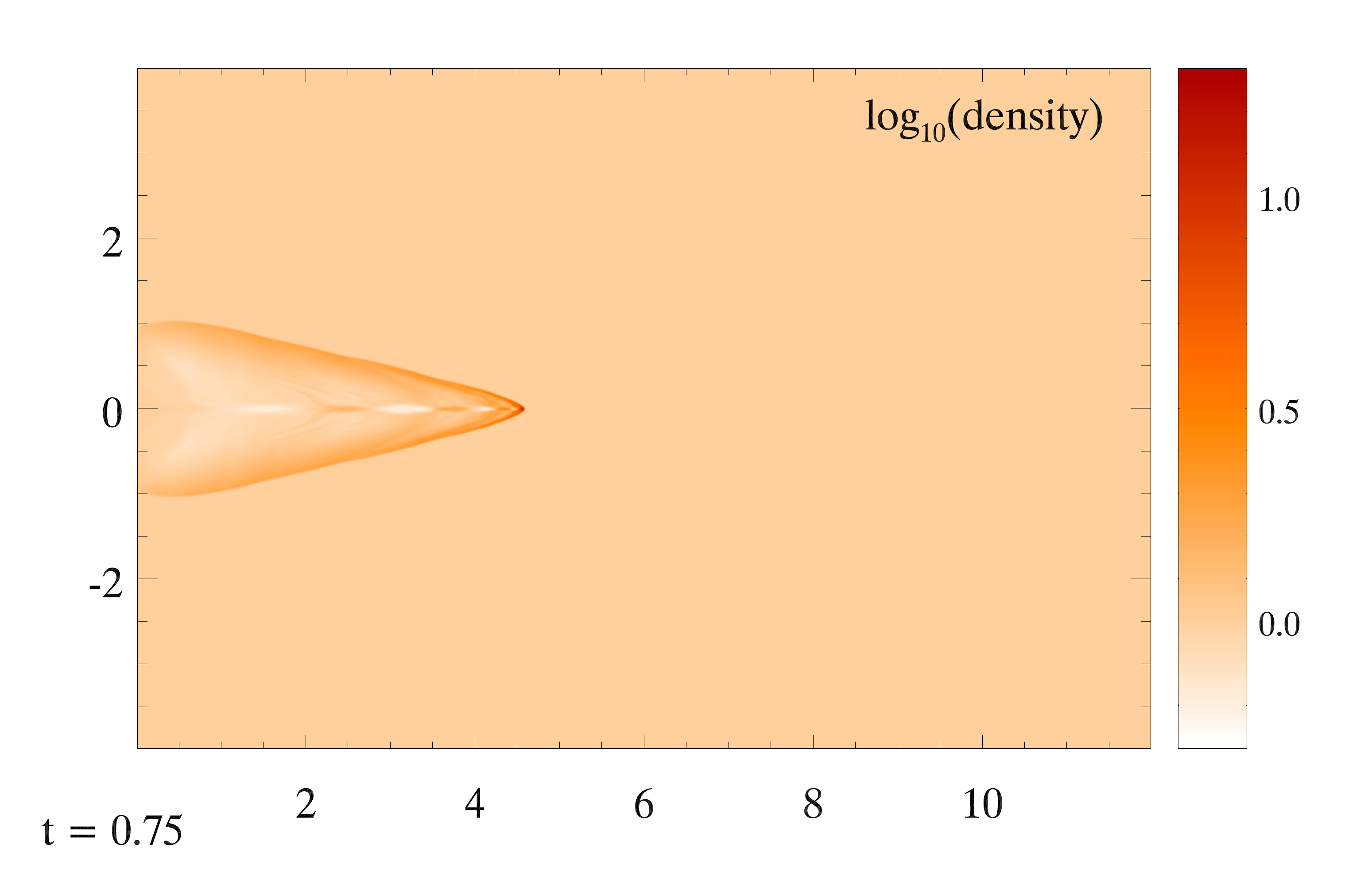}
	 {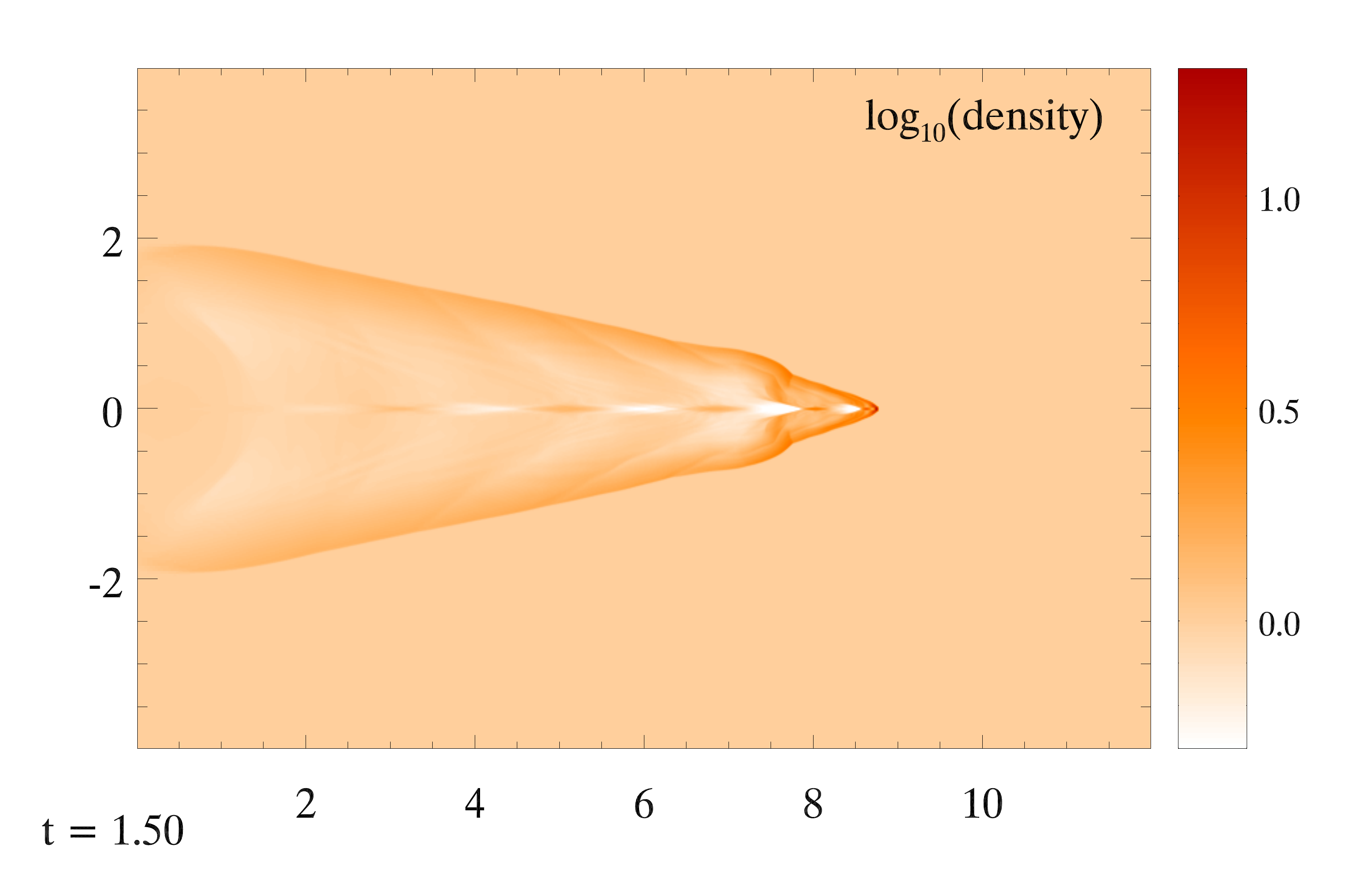}
	 {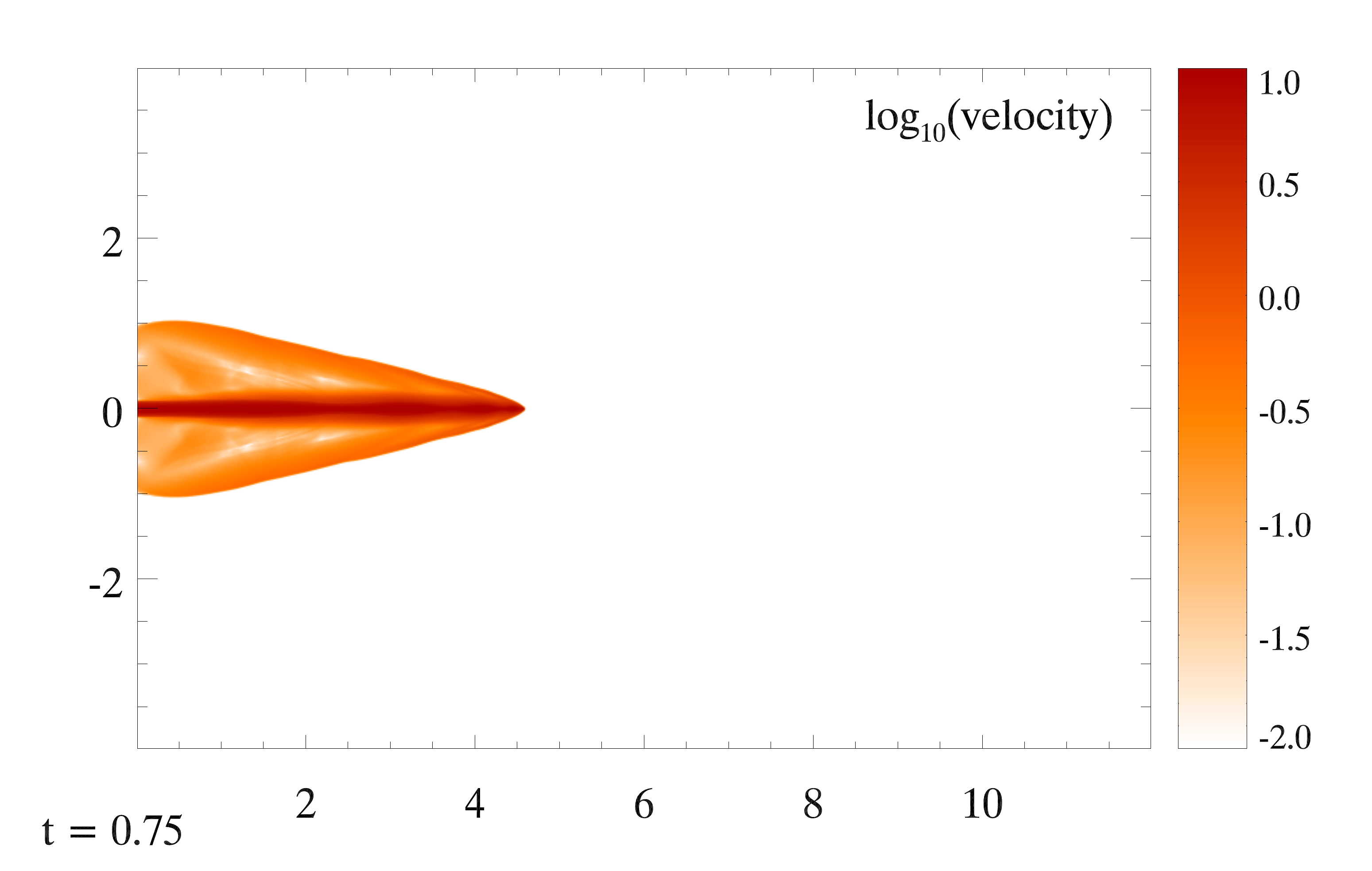}
	 {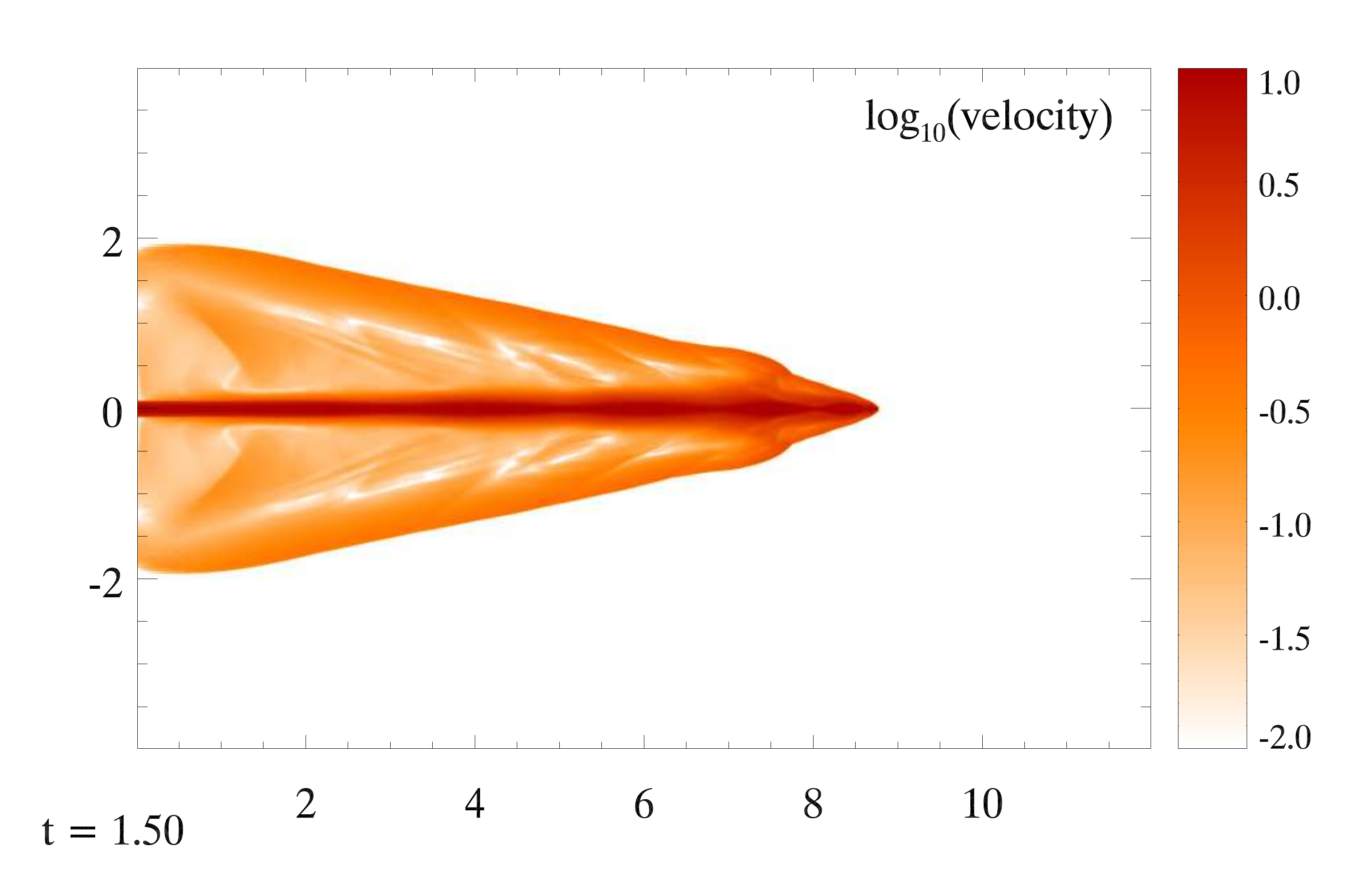}
\caption{Density (top) and velocity (bottom) evolution of a
  high-velocity Mach 10 jet (run M10c) at two different times, $t =
  0.75$ (left) and $t = 1.5$ (right). Compared to the slower Mach 5
  jet (cf.~Fig.~\ref{fig:jet_M5c}) this jet is much narrower and its
  bow shock radius is smaller, indicating that high velocity jets
  entrain less gas of their surrounding medium. Apart from very narrow
  regions around the jet edge and the tip of the bow shock, the excited
  gas motions are still mainly subsonic (see also
  Fig.~\ref{fig:vel_PDFs_M10c}).}
\label{fig:jet_M10c}
\eefs

\bef
\showone{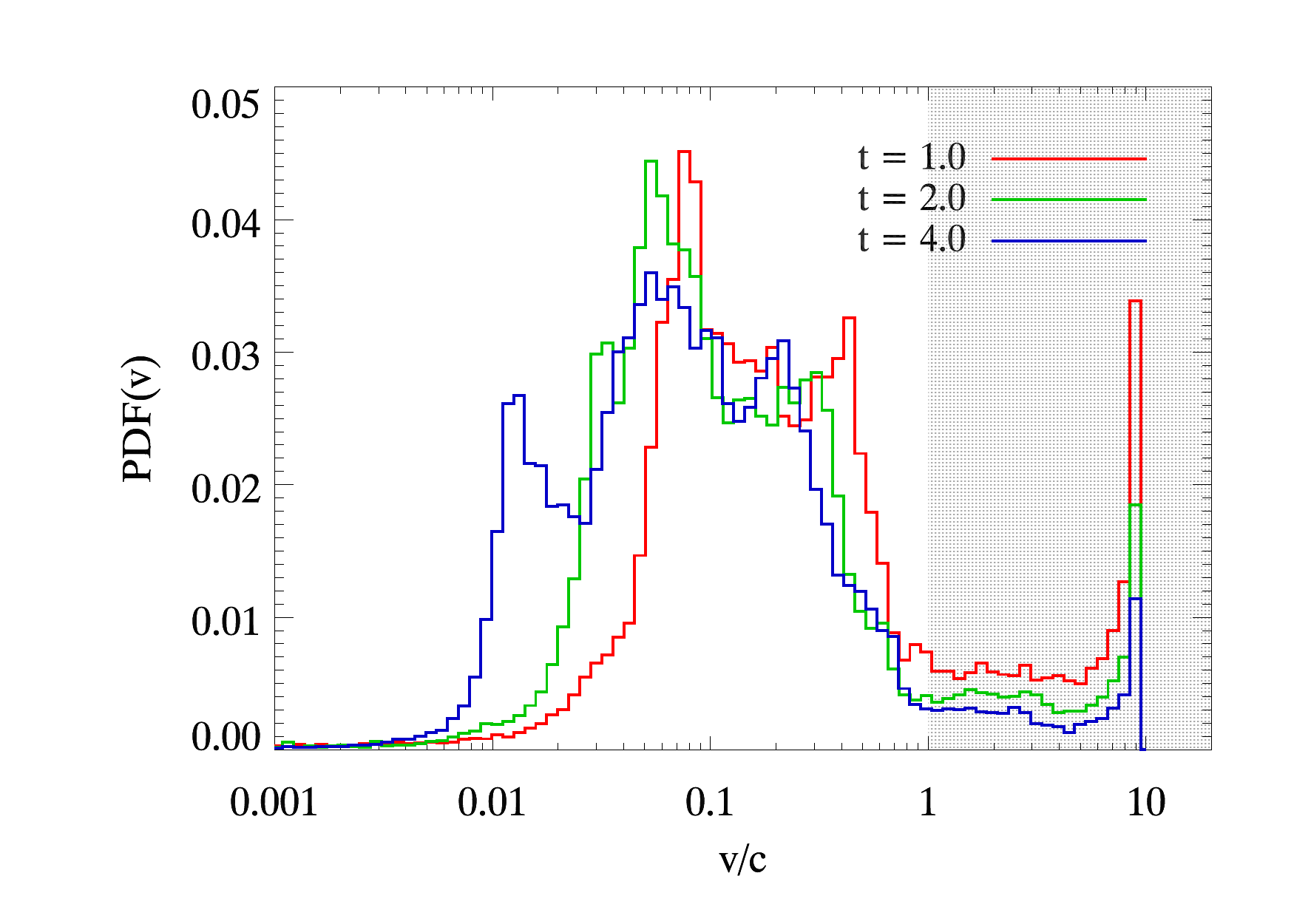}
\caption{Shows the time evolution of the probability density function of 
  velocity fluctuations in the case of the continuously driven Mach 10
  jet M10c. The amplitudes of the velocity fluctuations are given in
  units of the sound speed (i.e.~as Mach number). Essentially no
  supersonic fluctuations get excited by the jet (the peak at $v/c =
  10$ is the jet itself).}
\label{fig:vel_PDFs_M10c}
\eef

\bef
\showone{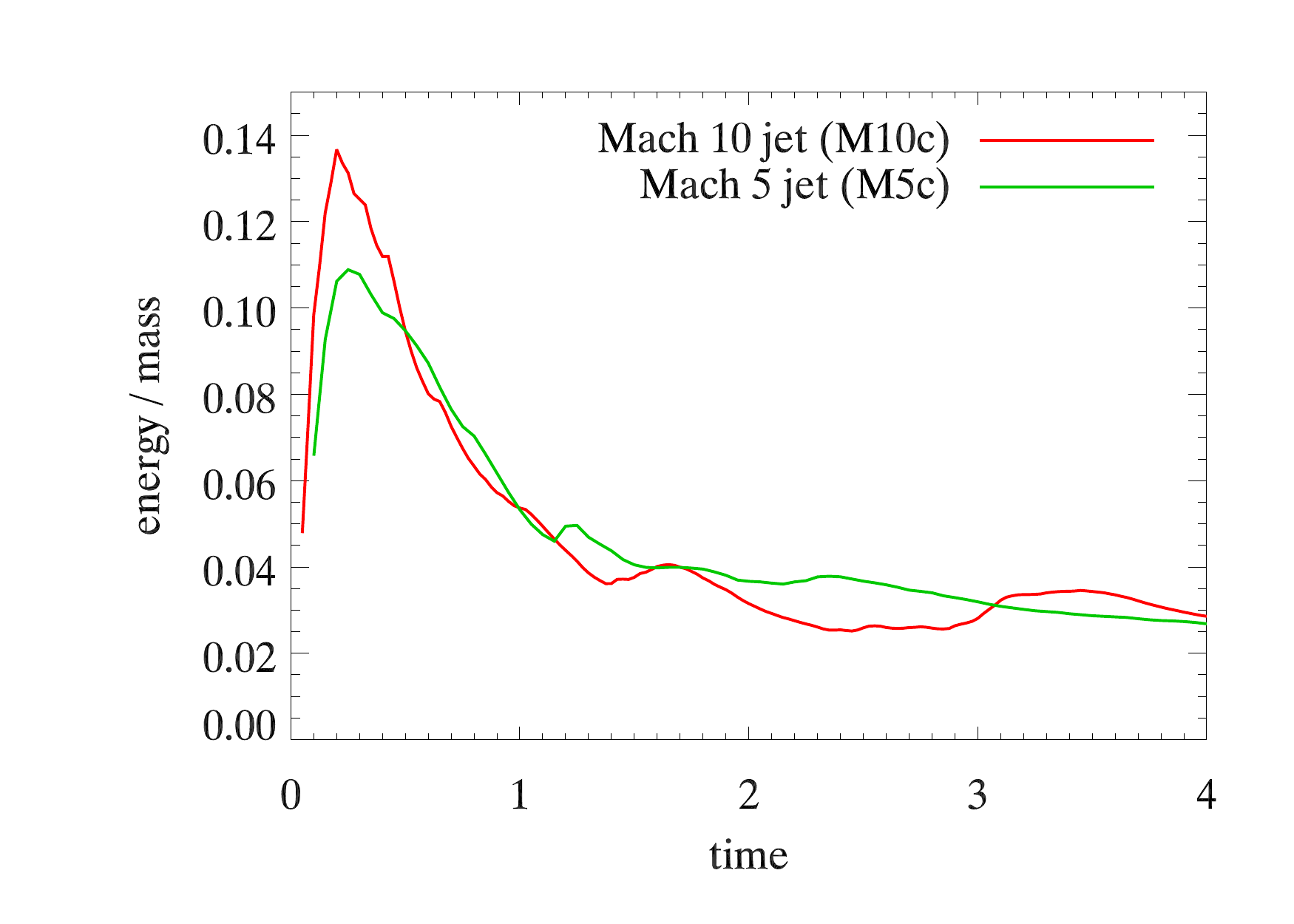}
\caption{Shows the time evolution of the kinetic energy per unit mass,
$\epsilon_{\sm{kin}}$, for the Mach 10 and Mach 5 runs M10c and M5c,
respectively. Here, the energy is calculated only for subsonic
motions. Despite the larger momentum of the Mach 10 jet it does not
excite much larger subsonic velocity fluctuations than its Mach 5
counterpart during the quiescent phase, i.e.~at time $t \simge 1.0$.}
\label{fig:M10+5_sub_energy}
\eef

We begin our investigation with the results from two dimensional
(slab) jets that are either continuously driven or have energy
injection shut off after a time $t = 1.3$.  In Fig.~\ref{fig:jet_M5c}
we show the time evolution of the density and velocity of our fiducial
run: a Mach 5 jet running into a homogeneous medium. Further
parameters of this run are: no density contrast ($\delta = 1$) and no
magnetic fields. The jet is continuously powered. This medium Mach
number jet develops knots inside the jet and visible Kelvin-Helmholtz
instabilities, including 'cats-eyes', at the edge of the jet. The
knots result from reflection waves off the jet edge where the spacing
between them is roughly $2\times \rm{jet\, radius} \times \rm{Mach\,
number}$. The bow shock has a relatively large radius and becomes
transonic close to the vicinity of the jet.

Jets from young stellar objects (YSOs) are usually driven for $10^4 -
10^5 \, \ys$ until the powering engine (disk- or stellar jet) stops
working. To study the influence of the end stage of jets we stop
driving the jet after a certain time. Fig.~\ref{fig:jet_M5t} shows the
time evolution of the density and velocity after the jet was shut off
at $t = 1.3$. Compared to the continuously driven jet (see
Fig.~\ref{fig:jet_M5c}) large velocity and density fluctuations have
already decayed, the bow shock has slowed down, and its radius
increased as expected for a slower moving jet. On the other hand, the
rarefaction wave travelling from the end of the jet excites (low
amplitude) fluctuations in the gas. This is particular important if
this pressure wave travels through a clumpy medium as we will see
later in this text.

From Fig.~\ref{fig:vel_PDFs_M5ct}, which shows the velocity PDFs for
the runs M5c and M5t, one can see that only a small fraction ($< 2\%$)
of the excited fluctuations are supersonic. The majority of the
motions are subsonic, with most around a Mach number between 0.03 and
0.1. Even in the case of a continuously driven jet, the peak of the PDF
moves towards smaller fluctuations with time, which shows that high
amplitude fluctuations decay quickly.

One of the most interesting features apparent in the PDFs is that
supersonic fluctuations close to the jet drop quickly and occupy only
very little volume until the excitation becomes subsonic (following
the PDF from the jet position at $v/c = 5$ to the left). Only subsonic
motions get excited distinctly\footnote{Note that the absolute decline
of the jet peak in the PDFs of continuously driven jets is due to our
choice of sampling. Here, the sampling volume increases as the
jet-affected regions increases with time.}. This shows that supersonic
fluctuations do not propagate far from the jet itself, and that
second-order excitations are always subsonic. This low plain in the
PDFs between the jet peak and the sonic point can be regarded as the
{\em supersonic desert}. Here, the jet is not able, although
continuously driven, to excite a large fraction of supersonic motions
in the gas. Supersonic fluctuations are highly compressive and their
energy is mainly consumed in compressing the gas instead of pushing
the gas to high velocities. The expansion of the compressed gas, in
turn, excites only {\em subsonic} fluctuations. This is the reason for
the 'supersonic desert' between the jet peak and the transonic
regime. The subsonic fluctuations are largely incompressible and can
propagate far into the medium. Given enough time, these subsonic
fluctuations will occupy most of the excited gas, so the PDF peaks
deeply in the subsonic regime.

The right panel of Fig.~\ref{fig:vel_PDFs_M5ct} (transient jet) shows
clearly the decay of supersonic motions with time. The highest
velocity fluctuations, in particular the supersonic ones, are the
fastest damped ones. This can also be seen in Fig.~\ref{fig:M5t_Ekins}
where we show the time evolution of the kinetic energy for the
supersonic and subsonic regime separately. The contributions from
supersonic motions to the kinetic energy decay much faster than the
subsonic ones. The former contributions decay as quickly as $\propto
t^{-2}$ which should be compared to the decay of subsonic fluctuations
which follow a power law $\propto t^{-1}$. Nevertheless, the total
energy decays as $\sim t^{-1.2}$ which is in agreement with previous
studies of decaying supersonic turbulence~\citep[e.g.][]{MacLow98a,
MacLow99}. Evidently, supersonic fluctuations do not excite
further supersonic motions in their neighbourhood. Instead their
momentum is mainly transfered into subsonic compressional
modes. Unlike subsonic fluctuations which spread into an ever
increasing region, the supersonic ones do not propagate into a larger
volume. This can be seen form the evolution of the affected volume in
both cases (see right panel of Fig.~\ref{fig:M5t_Ekins}).

In the transient jet case, essentially all supersonic fluctuations
decayed after one sound crossing time over the size of the
fluctuations. Again, this is expected as supersonic fluctuations decay
quickly and do not propagate far into the ambient medium. At $t = 5$
the main fluctuations peak at slightly higher amplitudes than in the
continuously driven case ($v \sim 0.05$ instead of $v \sim
0.025$). This is due to motions excited by the rarefaction wave
travelling from the back end of the jet into the ambient medium. This
feature is illustrated in Fig.~\ref{fig:M5ct_sub_energy} where we
compare the time evolution of the energy from the continuous and
transient jets, M5c and M5t, respectively. Although the transient jet
is switched off at $t = 1.3$, its energy per unit mass\footnote{Here we
show the energy per unit volume, i.e.~$\epsilon_{\sm{kin}} =
1/2\,\int\di V \, \rho\,v^2/M$ with $M = \int\di V\,\rho$, to account
for the fact that material can leave the simulation box. The
integration volume, $V$, is the regime affected by subsonic
fluctuations, i.e.~$V_{\sm{sub}} = V(0 <v<c)$.},
$\epsilon_{\sm{kin}}$, exceeds that of the continuous jet beyond
this time until $t \sim 3.0$. We therefore conclude that transient
jets are more capable of driving turbulence, albeit only in the
subsonic regime.

Although a typical PDF has one or two peaks in the subsonic regime,
these features are not due to the bow shock. We tried to identify
typical features of the jet in the PDF but, apart from the jet itself,
others do not stand out in the distribution. The bow shock region does
not occupy a large volume fraction.

In one case (run M5tg1.4) we also varied the EOS of the gas using a
barotropic relation with $p \propto \rho^{1.4}$. This gas behaviour
mimics the inefficient cooling ability in the optical thick
regime. The run M5tg1.4 should be compared to the isothermal
simulation M5t, which otherwise has the same runtime parameters. In
Fig.~\ref{fig:jet_M5tg1.4} we show density, temperature, and velocity
maps at two different times. Like in the isothermal case
Kelvin-Helmholtz instabilities develop at the edge of the jet which
are best seen in the temperature map. Otherwise, the bow shock radius
is larger in the non-isothermal case reflecting additional thermal
effects and a higher sound velocity. Nevertheless, the high amplitude
fluctuations are also damped quickly (see
Fig.~\ref{fig:vel_PDFs_M5tg1.4}) and are not found far from the jet.

The influence of the jet speed on the environment can be seen in
Fig.~\ref{fig:jet_M10c} where we show the evolution of a Mach 10 jet
(run M10c). The first feature one notices is the narrowing of the bow
shock region compared to the slower jet (run M5c). This
shows that a faster jet entrains less volume than a slower
one. Therefore we conclude that high Mach-number jets leave less
imprint on their environment than their slower counterparts, as long as
they do not become unstable. This is also corroborated by the fact
that supersonic motions do not propagate far from the
jet. Furthermore, high-velocity jets are more stable than low
Mach-number jets and are less likely to develop
instabilities. Therefore, high-velocity jets will stay collimated for
a longer time and a larger distance, which in turn reduces the volume
entrained by the jet. From the velocity PDF of this jet
(Fig.~\ref{fig:vel_PDFs_M10c}) one can see that the supersonic desert
occupies a larger velocity range in the case of a faster jet
(cf.~Fig.~\ref{fig:vel_PDFs_M5ct}) which shows again that the
supersonic velocity fluctuations are strongly suppressed compared to
the subsonic fluctuations because supersonic fluctuations do not
persist long and are not excited far from the jet itself. Rather, the
energy of supersonic modes is transfered from large to small scales
abruptly in a network of shocks~\citep{Smith00}.

Observations also support the typical velocity structure seen in our
simulations. For instance, strong velocity gradients where the speed
declines quickly in the direction off the jet axis are seen in
the HH21 jet observed by \citet{Gueth99}.

Comparing the kinetic energy evolution of the Mach 5 and Mach 10 jets
one sees from Fig.~\ref{fig:M10+5_sub_energy} that even in the
subsonic regime the higher velocity jet does not deposit much more
energy into the ambient medium. In this figure we show again the time
evolution of the kinetic energy per unit mass,
$\epsilon_{\sm{kin}}$. After an initial ramp-up phase (until $t \sim
0.3$) the energy contained in the subsonic regime starts to 'decay'
because more and more gas, i.e.~mass, is entrained (note that both
jets are continuously driven). Shortly after the peak is reached, the
energies in the fast and slower case become of similar strength
showing that the relative impact of high velocity jets is not greatly
enhanced compared to their slower counterparts.

\bef
\showone{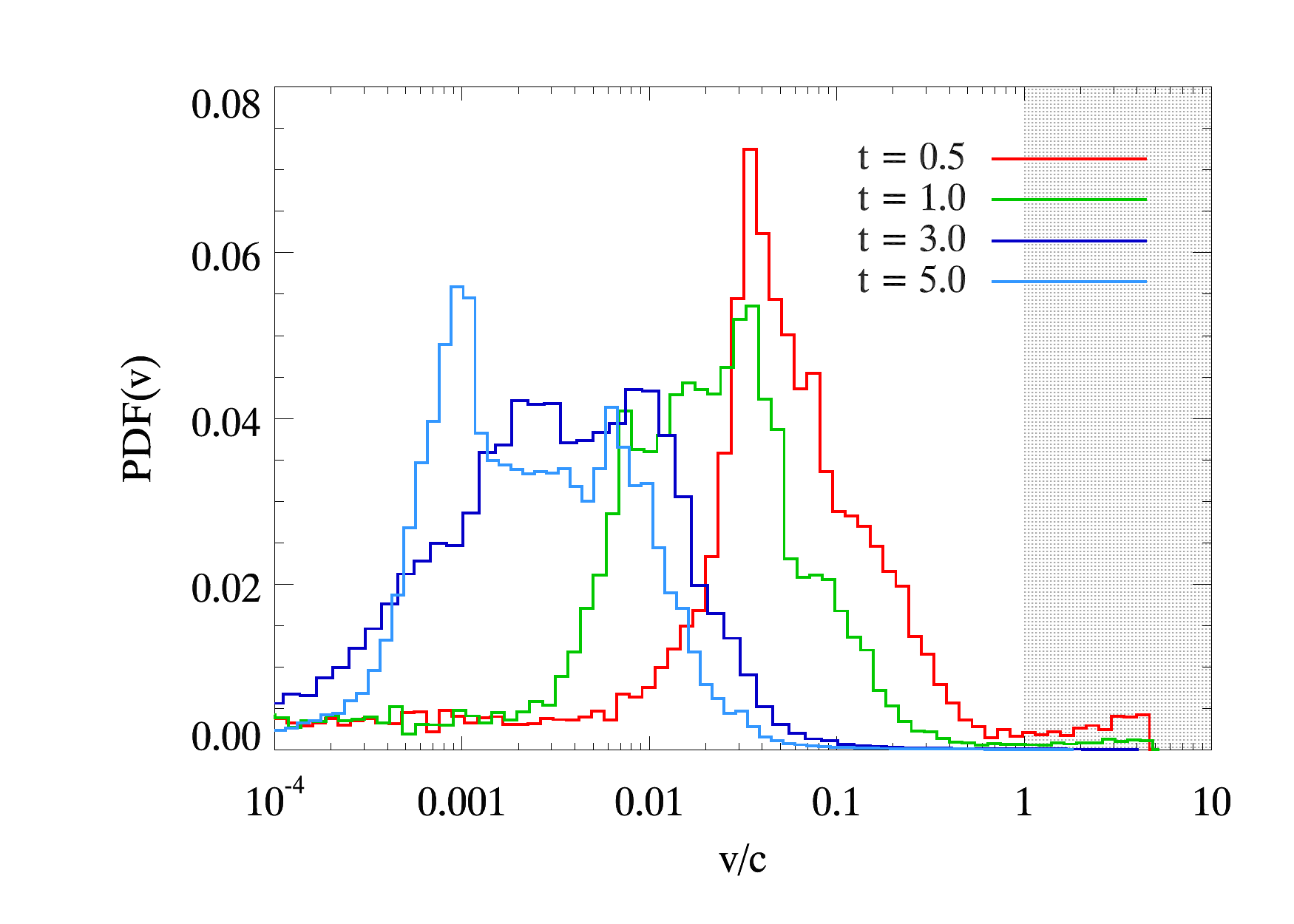}
\caption{Shows a time sequence of the velocity PDF from the
  three-dimensional run M5t3D. Compared to the 2D runs the filling
  factor of the jet ($v/c = 5$) and the supersonic regime decreased
  significantly due to the increased degrees of freedom of velocity
  excitations.}
\label{fig:vel_PDFs_M5t3D}
\eef

To compare our above presented slab jet simulations with more natural
configurations we performed a series of three-dimensional jet
simulations with similar setups. Although the study of two-dimensional
slab or axis-symmetric jets is instructive and the results represent
the gross features of jets, jets in three dimensions have more degrees
of freedom and can therefore develop more instabilities but generally
pervade a smaller fractional volume than in two dimensions.

In Fig.~\ref{fig:vel_PDFs_M5t3D} we show a time sequence of the
velocity PDFs from the 3D run M5t3D. This setup should be compared to
the two dimensional slab jet of run M5t where both have a speed of
Mach~5 and are driven until $t = 1.3$. As expected, the volume filling
factor of the jet (i.e.~the supersonic regime) in the 3D case is much
smaller than in the 2D setup. Apart from this fact the two cases
evolve quite similarly. Differences show up in particular
instabilities at the jet surface. For instance the Kelvin-Helmholtz
instabilities appear on smaller scales in the 3D case. 

Finally we mention that we did not investigate rotating jets. Jet
rotation is indicated in some sources with micro jets \citep[see
e.g.][]{Bacciotti02, Ray07} and can be fitted in principle by
magnetohydrodynamic models of jet formation in order to derive the
launching radius of the jet \citep[see e.g.][]{Anderson03, Fendt06,
Pudritz07}. Typical rotational velocities (if existent) are about a
factor 10--100 times smaller than the jet bulk speed. It is unlikely
that such a small velocity shear to the ambient medium may launch
supersonic turbulence. However, rotating jets may be affected by
additional modes of instability, and thus may give rise to a different
overall turbulence pattern. We will investigate this important aspect
in a forthcoming paper.

\subsection{Jet--Clump Interaction}

\befs
\showfour{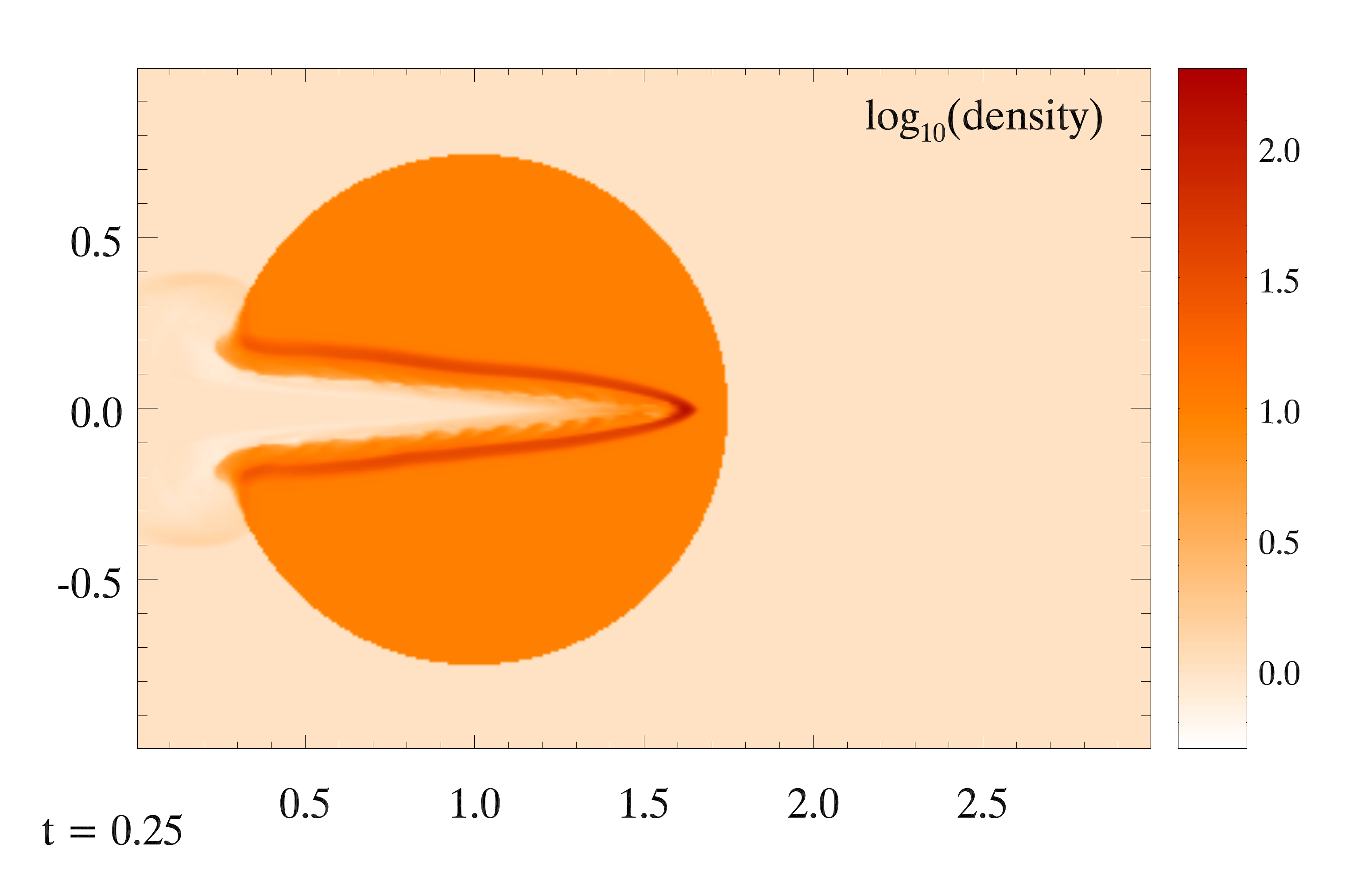}
	 {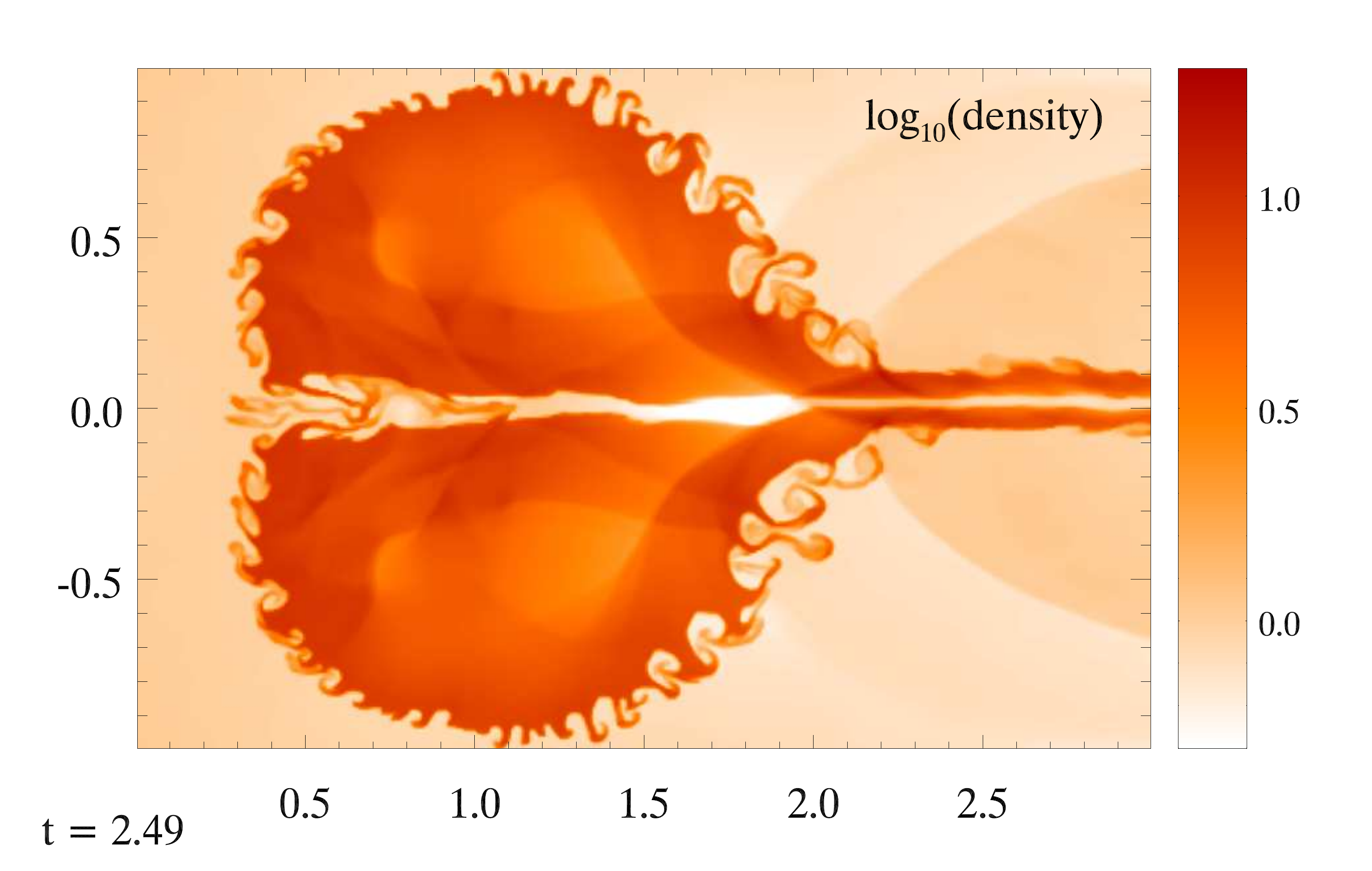}
         {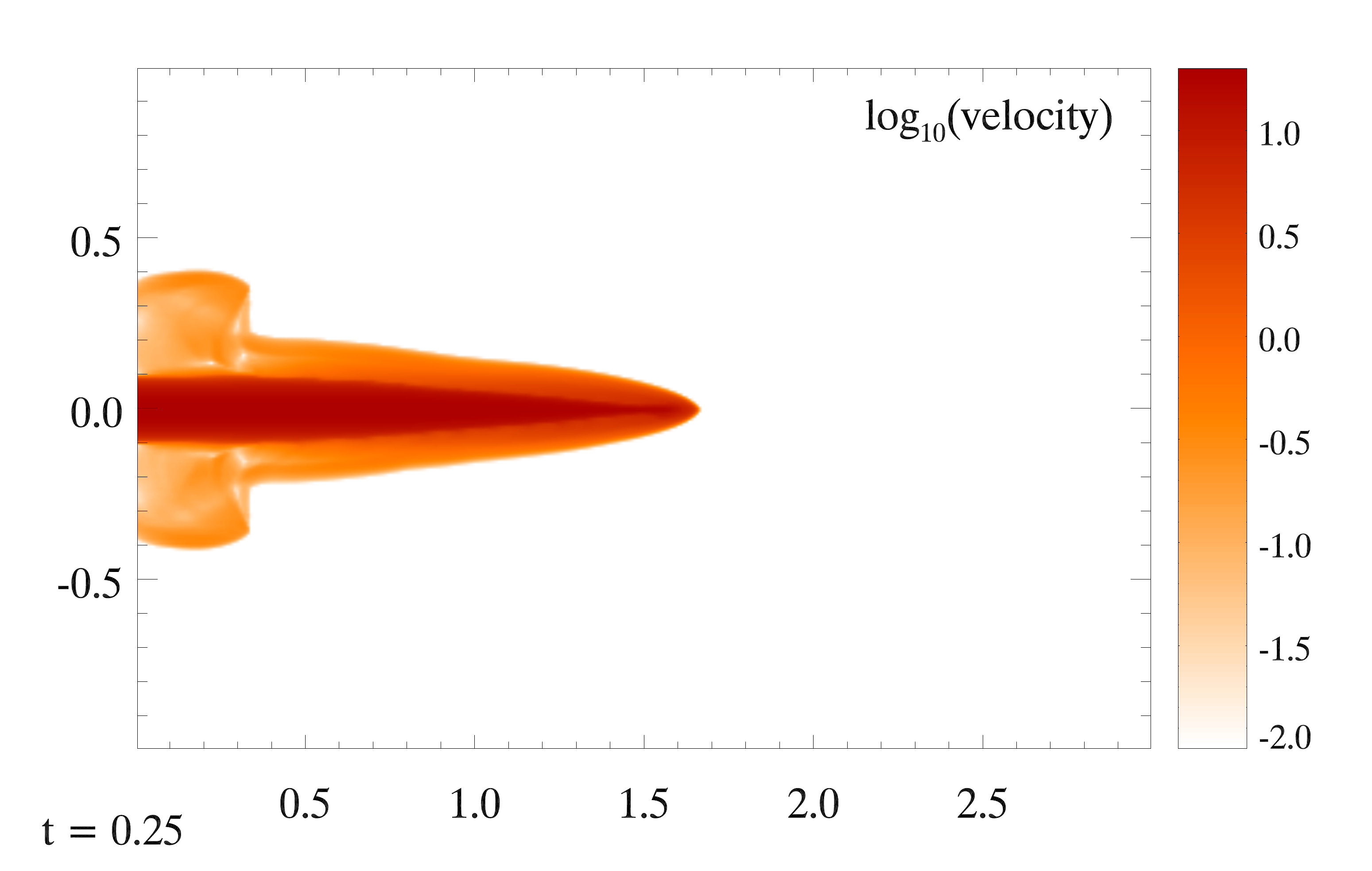}
         {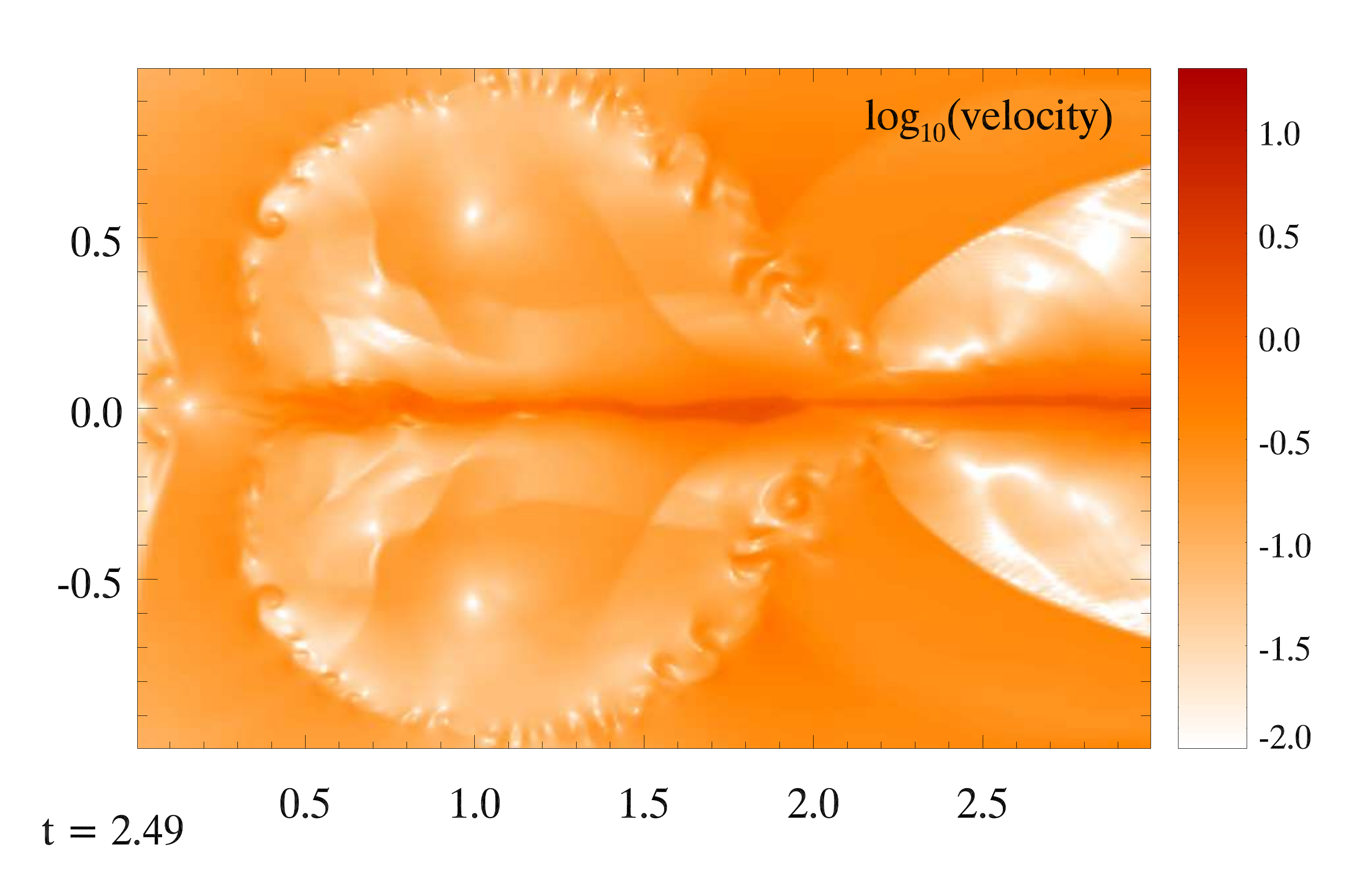}
\caption{Jet--Clump interaction. Shown are the density and velocity
distribution of a Mach 20 jet at two different times, $t = 0.25$
(left) and $t = 2.5$ (right). The earlier stage shows the situation
while the jet is still powered whereas the jet powering has ceased in
the later stage (shut off at $t = 0.6$). The imprint from the jet is
clearly visible in the clump structure even at much later stage but
the velocity fluctuations are subsonic. Note that the simulation box
is much larger than shown here (box size: $24\times 8$)}
\label{fig:jet_M20tCl}
\eefs

\bef
\showone{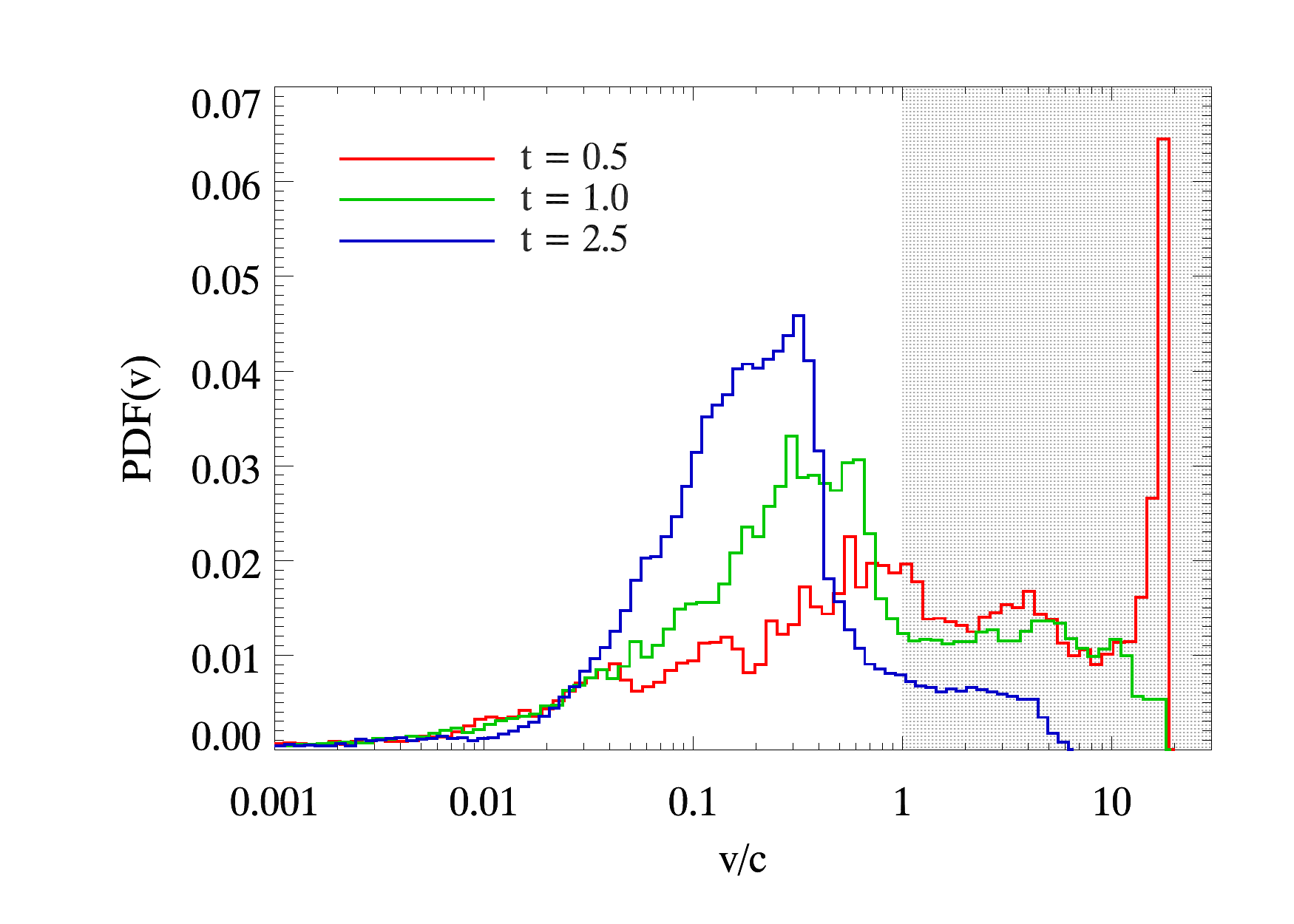}
\caption{Shows a time sequence of velocity PDFs in the case of the
  transient Mach 20 jet M20tCl interacting with a dense clump. The
  peak at $v/c = 20$ is due to the jet itself. Here again, supersonic
  velocity fluctuations are not copiously exited and are damped
  quickly after the jet driving engine ceased at $t =
  0.6$. Nevertheless, the clump is strongly disrupted by the jet and
  might even fully dissolve after some time (see Fig.~\ref{fig:jet_M20tCl})}
\label{fig:vel_PDFs_M20tCl}
\eef

The environment in which jets are launched is usually non-homogeneous,
i.e.~clumpy and often stratified. Early studies of outflows
interacting with a stratified environment (a density gradient in the
ambient media) showed that the morphology of the jet's bow shock
changes significantly in such a situation~\citep{Henney95}. More
dramatically, jets will be deflected if they hit a high-density
obstacle with a certain inclination angle~\citep[e.g.][]{Raga95,
Gouveia99}.

Here, we model a scenario of jet--clump interaction with a Mach 20 jet
running into an overdense spherical region, where the density contrast
between clump and ambient medium is $\delta_{\sm{cl}} = 10$ (here, the
jet density equals the density of the ambient medium). The ratio of
the clump radius to the jet radius is 7.5 (run M20tCl). Furthermore
the jet driving is switched off at $t =
0.6$. Fig.~\ref{fig:jet_M20tCl} shows the density and velocity of this
jet--clump interaction simulation before ($t = 0.25$) and after ($t =
2.5$) the driving engine ceased.  Initially the high Mach number jet
runs through the clump without a large influence. Even the bow shock
is very narrow inside the overdense region and leaves only a small
trace of compressed gas at the edge of the jet. The bow shock starts
to expand again in the low density region after the jet leaves the
clump (not shown in the images). Essentially a high Mach number jet
acts like a bullet penetrating a clump of material. Again, the
situation changes after the jet engine stops: a rarefaction wave
travels from the back end of jet into the ambient medium. This
pressure wave produces the main disturbance in the clump and excites
large instabilities at the clump's edge. The instabilities develop to
Rayleigh-Taylor fingers which move into the low density medium. In the
particular case here (2D slab jet, no self-gravity) the clump is
entirely dispersed by the transient jet.  This shows again that
high-velocity jets have a large impact on their environment and might
even disrupt entire cloud cores, but they are unlikely drivers of
supersonic turbulence in a large fraction of the molecular
clouds. Fig.~\ref{fig:vel_PDFs_M20tCl} shows a time sequence of the
velocity PDF in the case of this jet (M20tCl) which indicates a rapid
decay in time and amplitude of supersonic turbulence.

\subsection{High-Density Jets}

\befs
\showfour{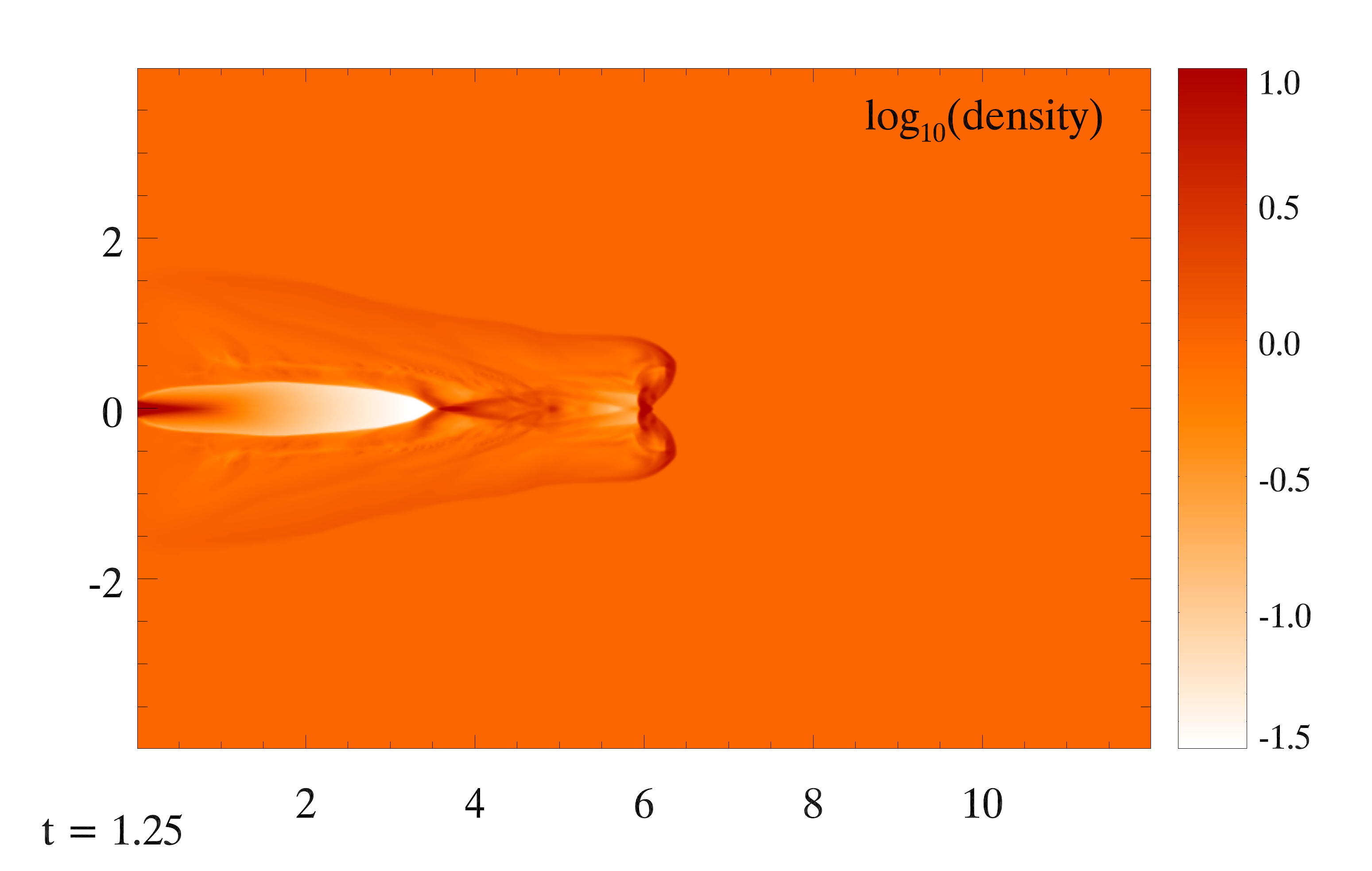}
	 {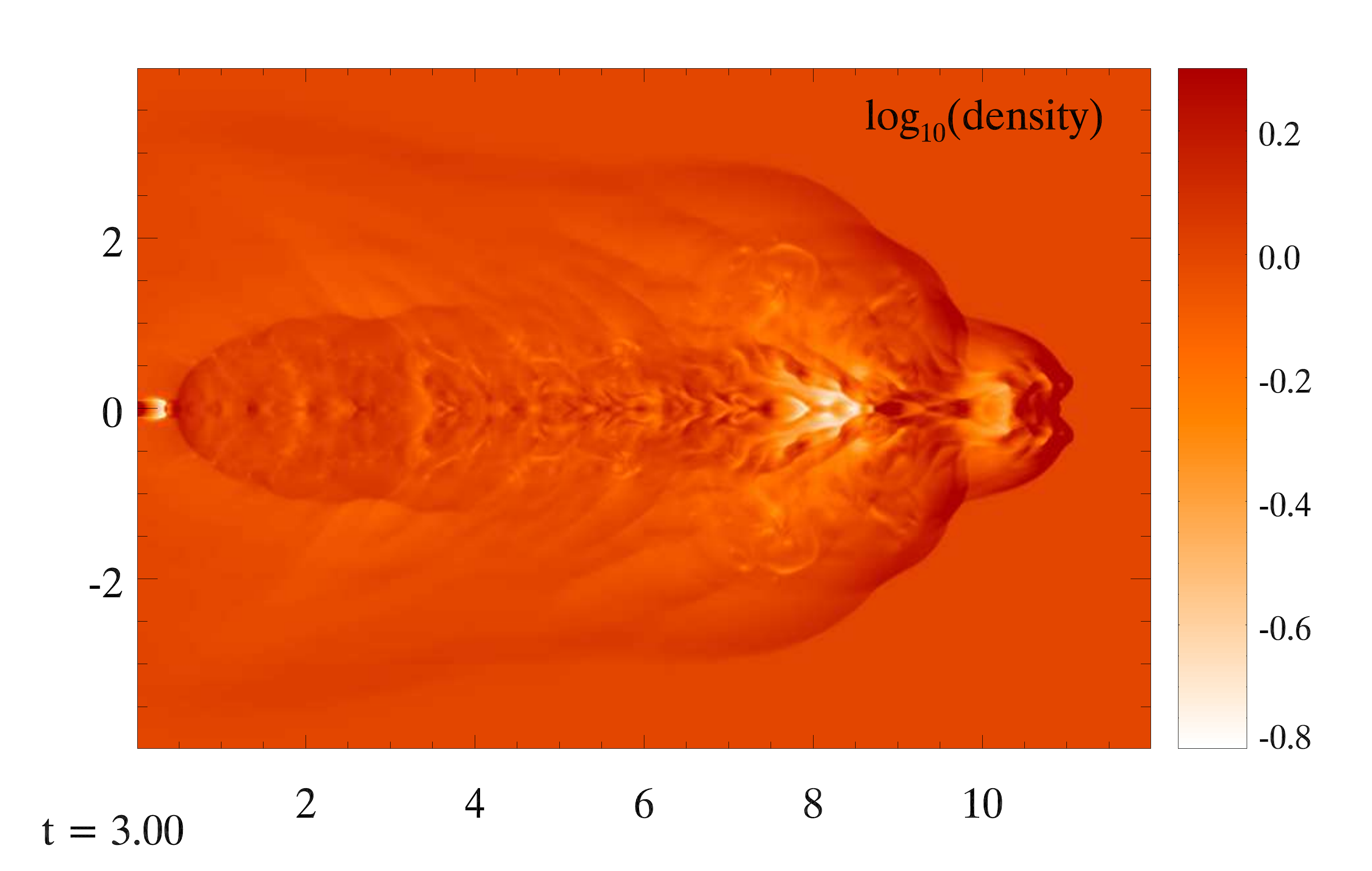}
	 {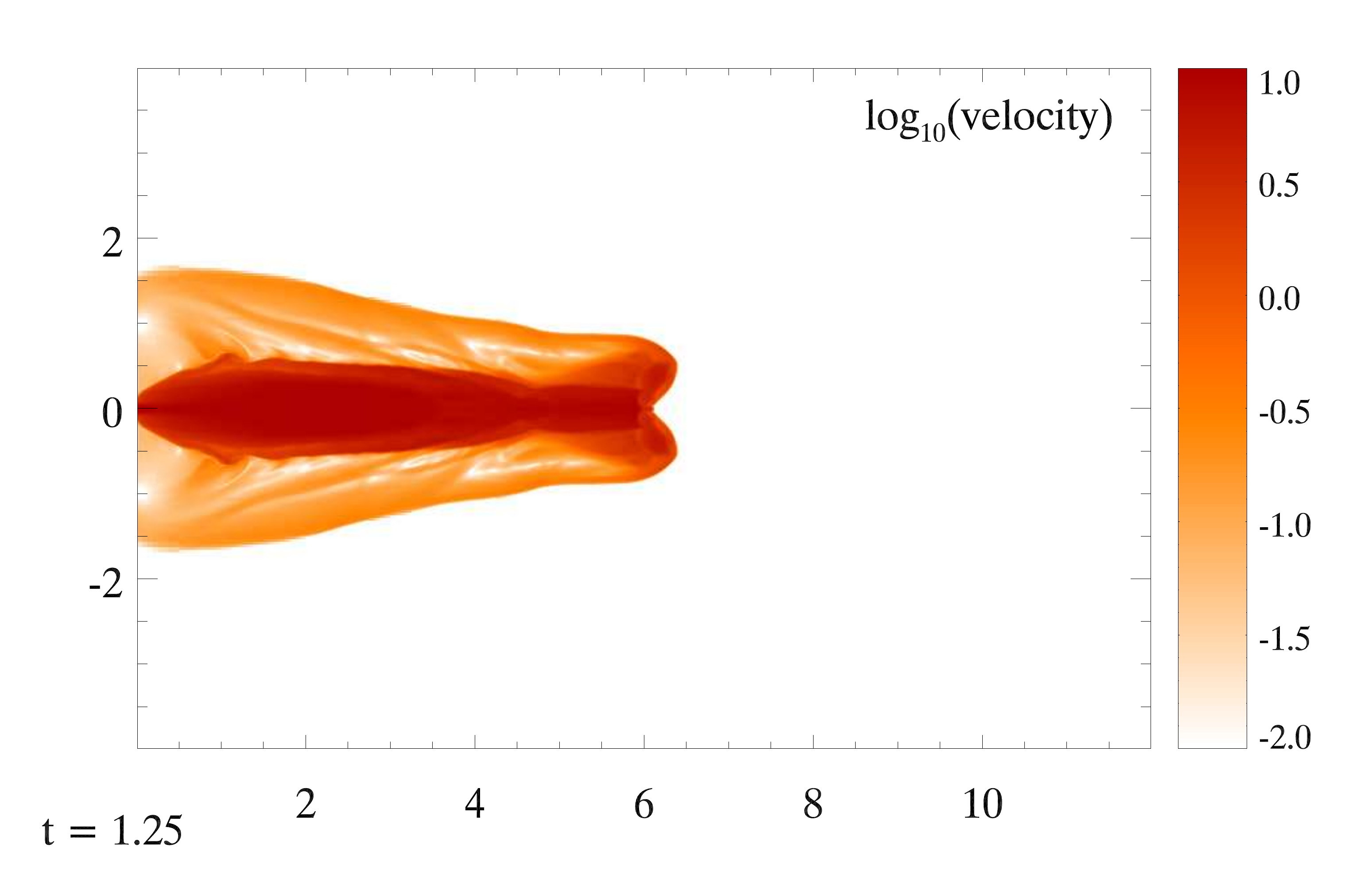}
	 {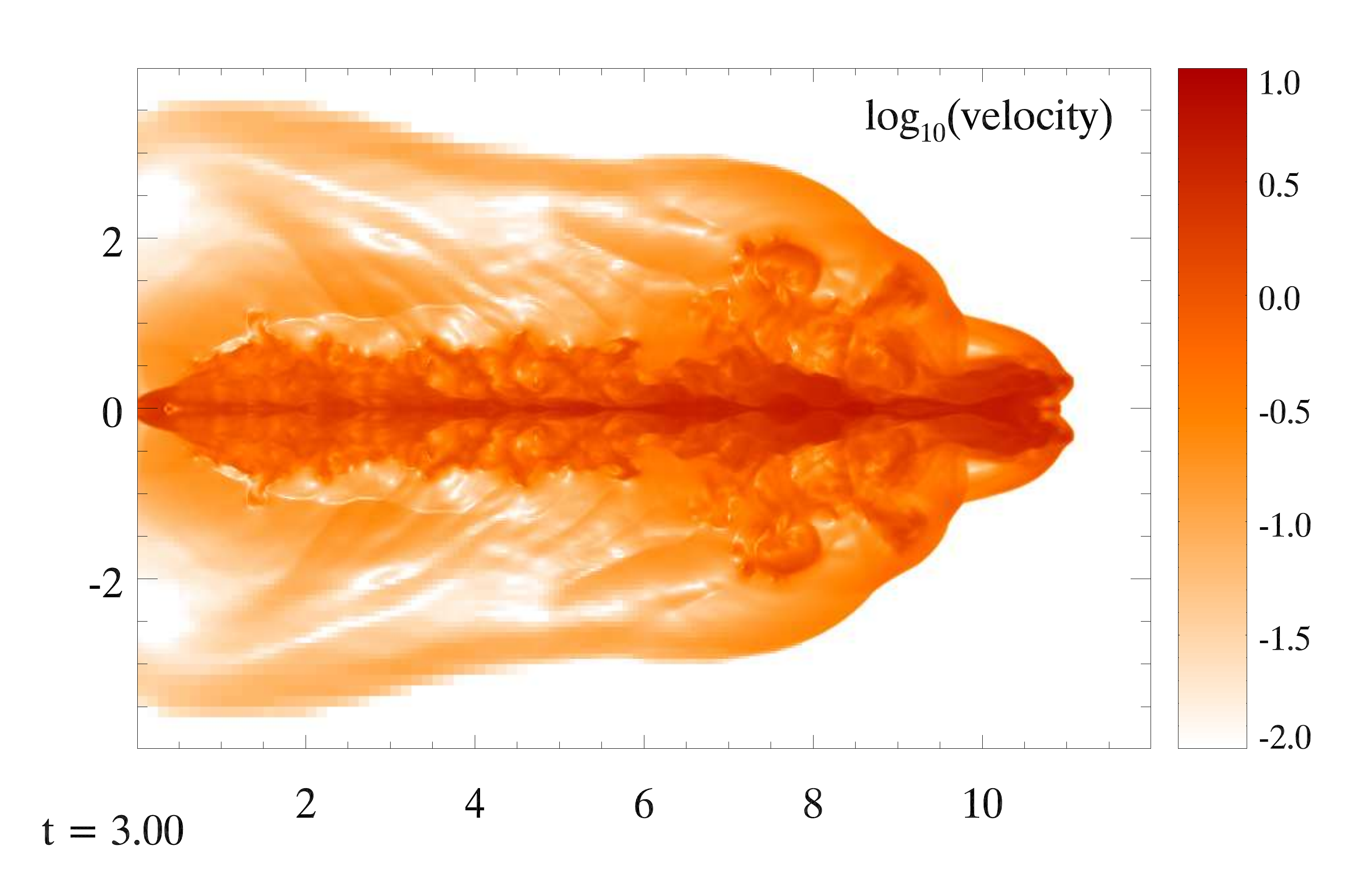}
\caption{Shows 2D cuts through the density (top) and velocity (bottom)
  of the 3D simulation M10tOd3D (Mach 10, overdense jet) at two
  different times, $t = 1.25$ (left) and $t = 3.0$ (right). The
  beamwidth is wider and the bow shock region is distinctively
  different in this case than in the case of an equal density jet
  (cf.~Fig.~\ref{fig:jet_M10c})}
\label{fig:jet_M10tOd3D}
\eefs

\bef
\showone{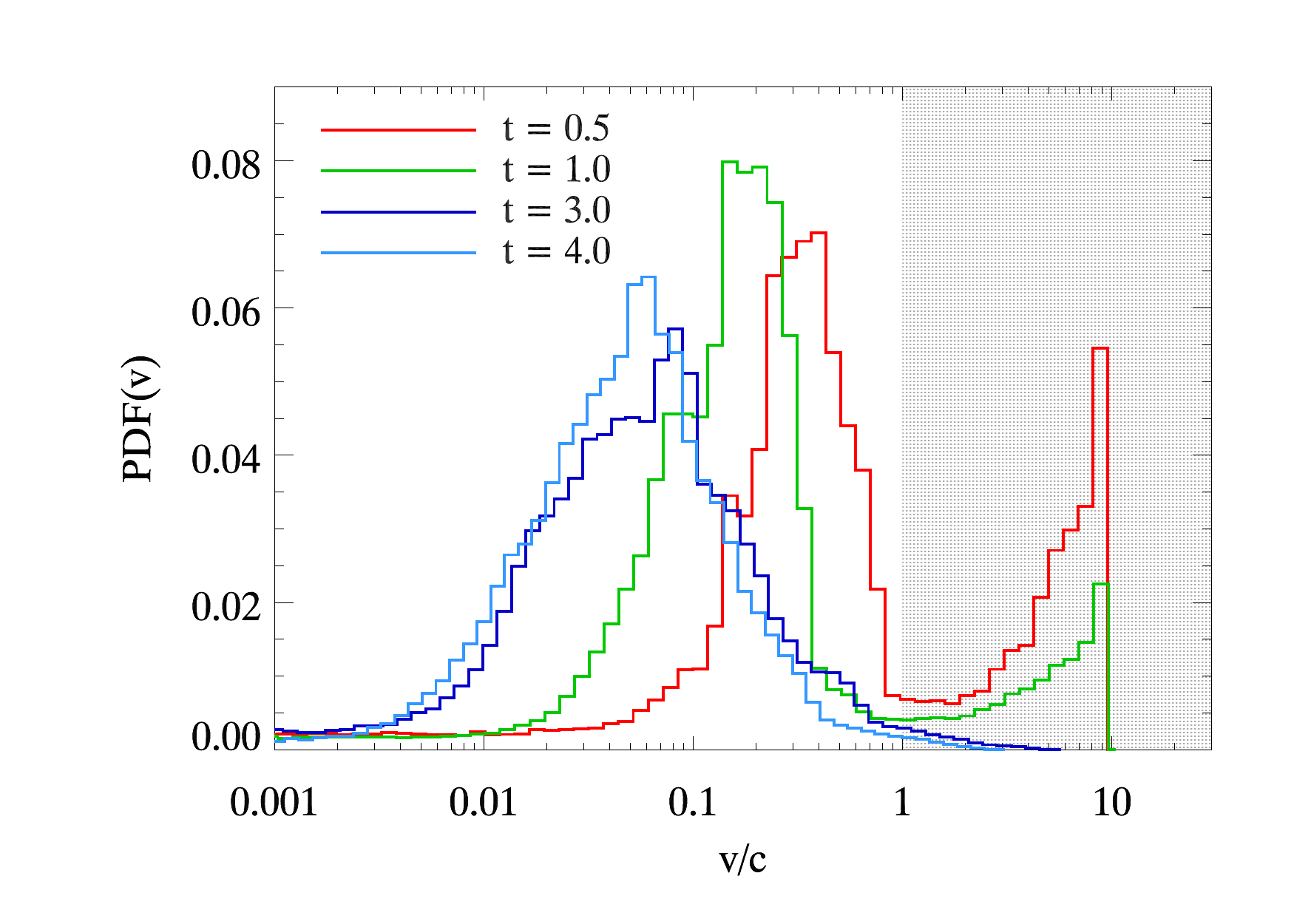}
\caption{Shows a time sequence of the velocity PDF from the three
  dimensional, overdense jet M10tOd3D before and after the driving
  engine is turned off at $t = 1.3$. A lager fraction of the excited
  fluid motions is supersonic in this overdense-jet case but the
  supersonic excitations decay quickly again after the jet driving is
  shut off at $t = 1.3$.}
\label{fig:vel_PDFs_M10tOd3D}
\eef

Jets from YSOs seem to pass through different evolutionary stages. In
particular, the youngest jests show evidence of containing high density
gas with $n(\Htwo) >10^5 \, \cm^{3}$ \citep[see review][and references
herein]{Reipurth01}. The density within these jets exceeds the mean
density of their surrounding media. Such high-density jets have the
potential to leave a larger imprint in the ambient medium than their
lower density counterparts as they carry more momentum.

In Fig.~\ref{fig:jet_M10tOd3D} we show images of the density and
velocity of an overdense ($\delta = 10$, Mach 10) jet-simulation. One
of the striking features of this jet is that the beam-width of the jet
increased by a factor of a few (lower left panel of
Fig.~\ref{fig:jet_M10tOd3D}). At the foot-point the jet has a diameter
of 0.2 in simulation units and increases to $\sim 1.0$ at a distance
of $\sim 1.0$. This broadening of high-density jets helps to pervade a
larger volume fraction and excite high amplitude velocity
fluctuations. Note that this jet is not in pressure equilibrium with
its surroundings, which means that the amount of excited motions due
to the jet expansion is an upper limit to the possible impact of
overdense jets. In Fig.~\ref{fig:vel_PDFs_M10tOd3D} we show a sequence
of velocity PDFs of this overdense jet before and after the jet's
engine ceased at $t = 1.3$. Although the supersonic wing decreases
steeply from the jet peak at $v/c = 10$ its slope is not as steep as
in the case of an equal density jet
(cf.~Fig.~\ref{fig:vel_PDFs_M10c}). As long as the overdense jet is
powered supersonic motions make up a significant fraction of the
excited gas velocities but get damped quickly again after the driving
stopped.

Another interesting feature of the overdense jet is the shape of the
bow shock. The overdense jet shows strong sideway motions due to the
higher pressure inside the jet which effectively de-collimates the jet
and increases the bow shock region forming a cone-like shape at the
tip of the jet.

\subsection{MHD Jets}

\bef
\showtwo{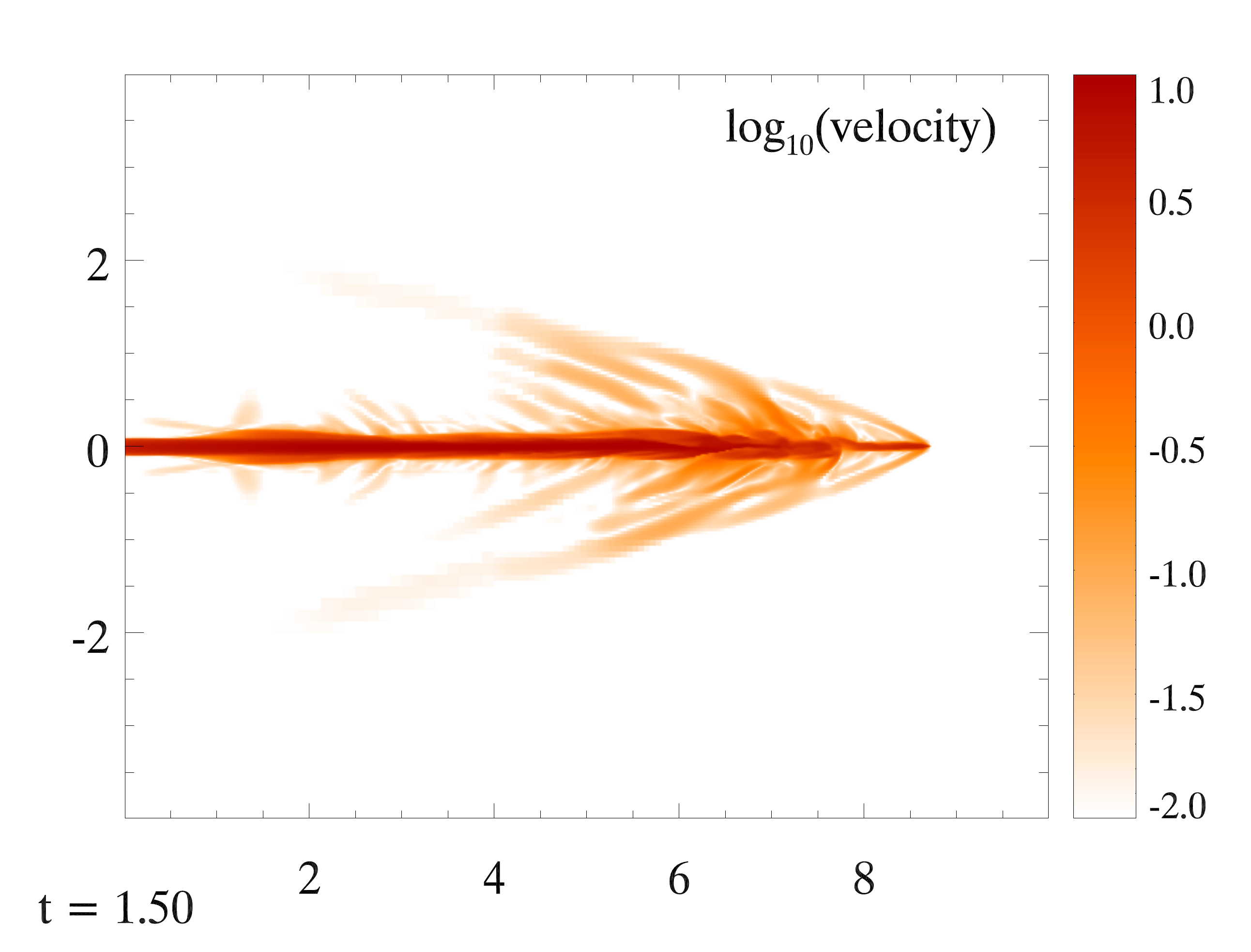}
	{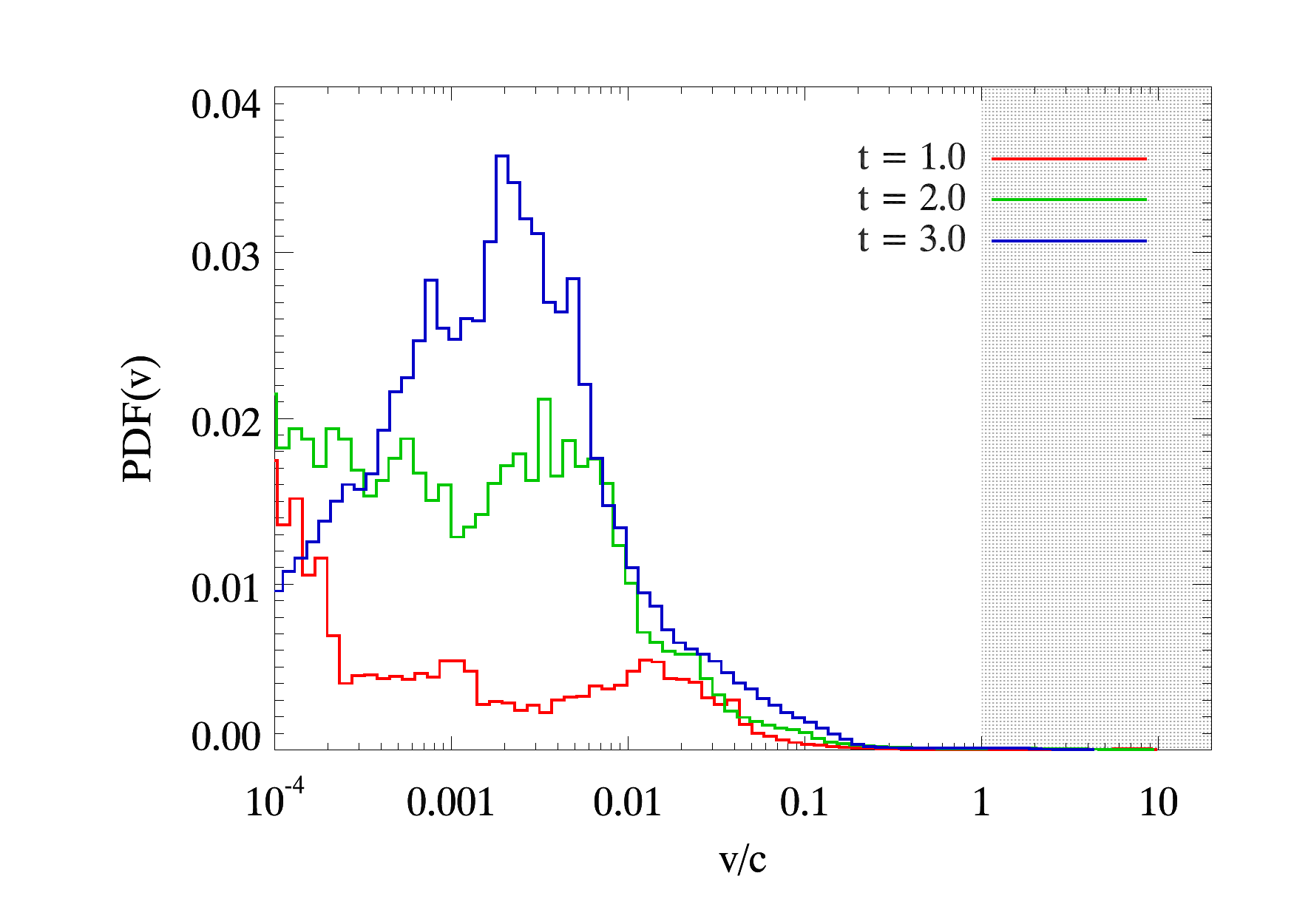}
\caption{Show the velocity structure and a time sequence of velocity
  PDFs from simulation M10tMpll3D. In this magnetised case (magnetic
  field parallel to the jet) velocity fluctuations are strongly
  suppressed compared to the non-magnetised jets (see, for example,
  Fig.~\ref{fig:jet_M10c}).}
\label{fig:jet_M10tMpll3D}
\eef

\bef
\showtwo{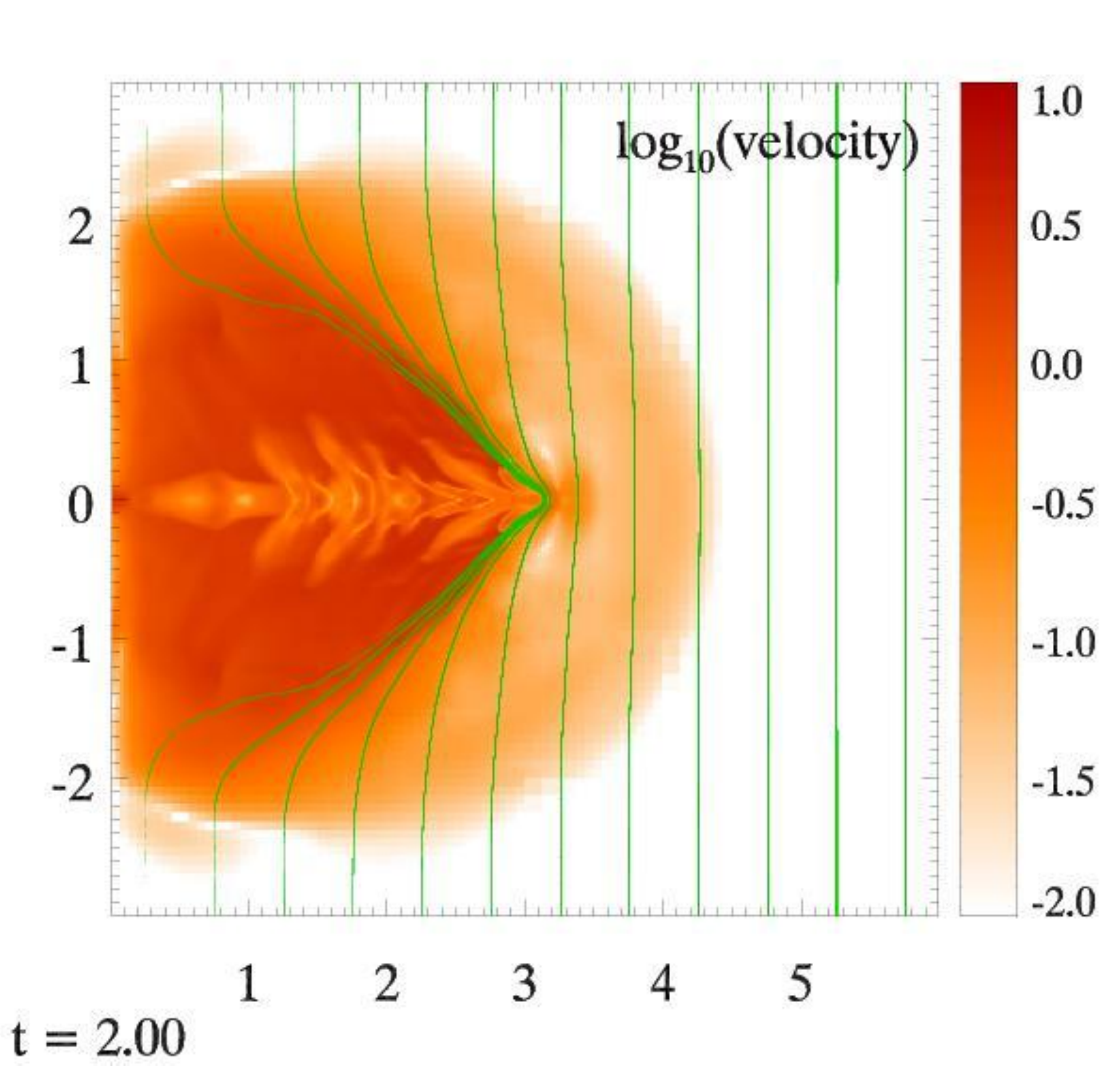}
	{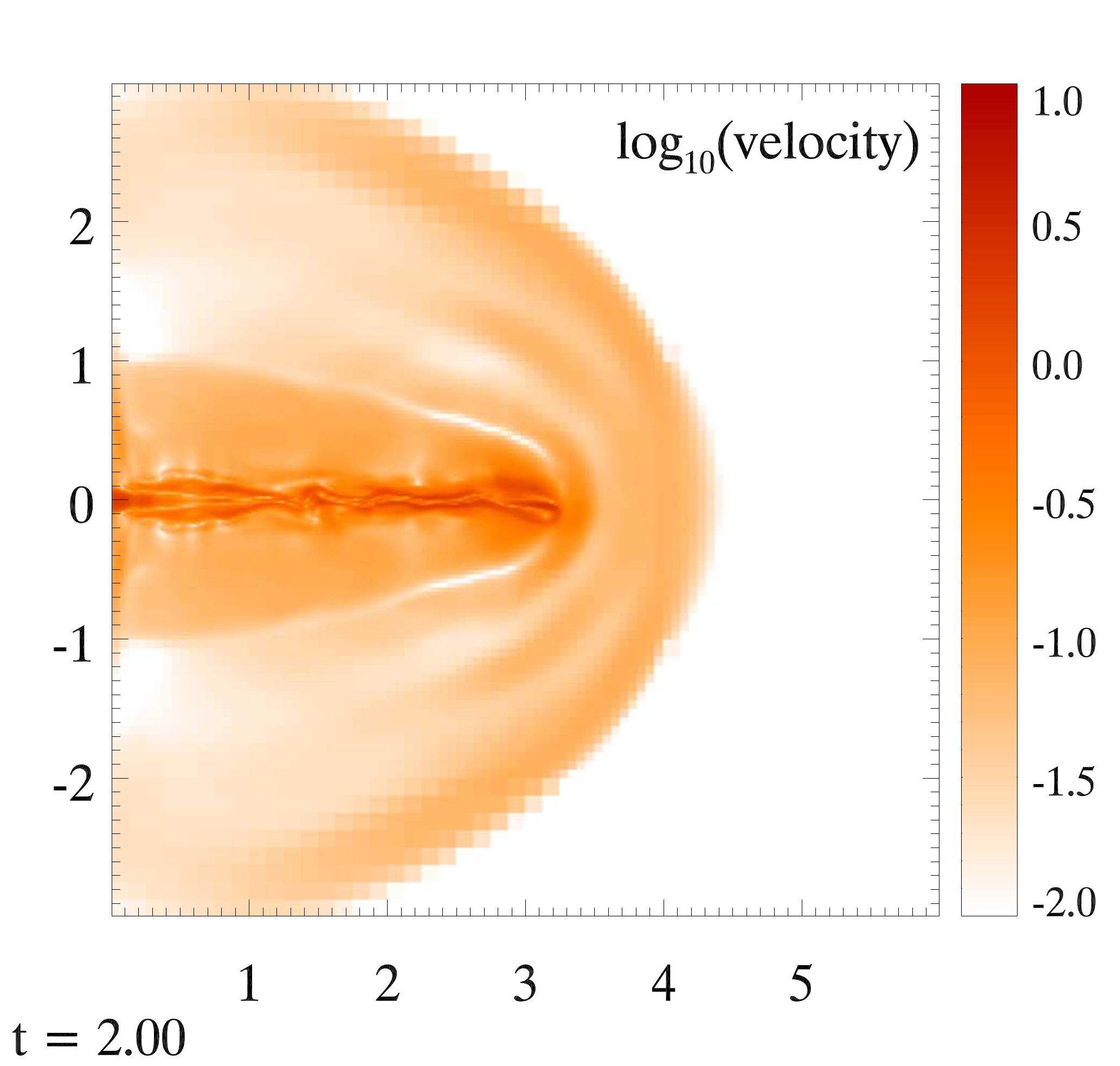}
\caption{Shows the velocity structure form simulation M10tMpe3D
  (initial homogeneous magnetic field perpendicular to the jet
  axis). The left panel shows the speed distribution in a 2D slice
  through the $xy$ plane and the right panel is an image through the
  $xz$ plane. The jet compresses and distorts the magnetic field whose
  pressure and tension accelerates gas off the jet axis as seen in the
  left panel.}
\label{fig:jet_M10tMpe3D}
\eef

\bef
\showone{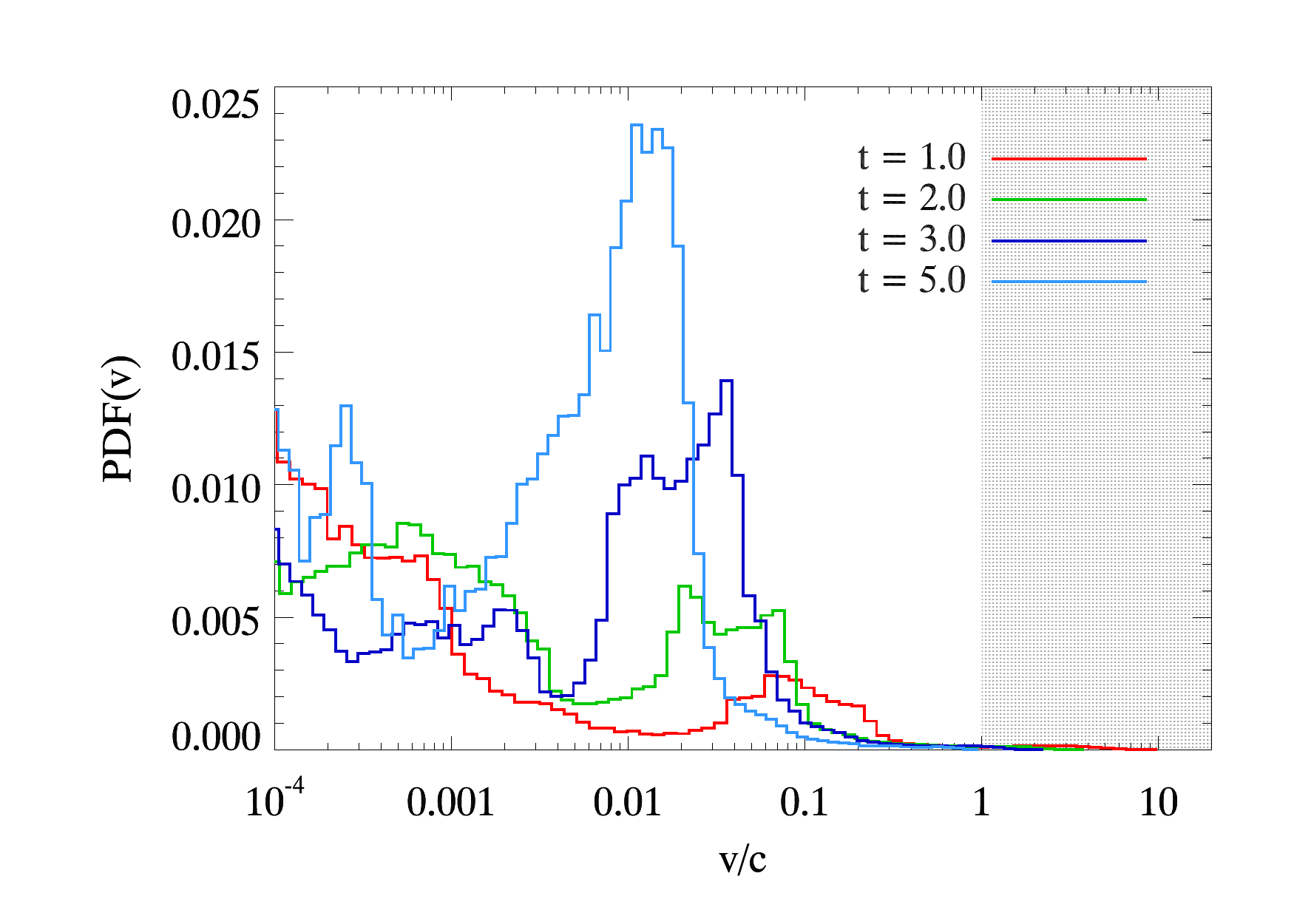}
\caption{Shows a time sequence of the velocity PDF from the MHD jet
  simulation M10tMpe3D (jet with an initially perpendicular magnetic
  field) before and after the driving engine is turned
  off at $t = 1.3$. Although fluid motions are excited along the
  magnetic field lines (see left panel of Fig.~\ref{fig:jet_M10tMpe3D})
  the majority of these velocity excitations are subsonic. Compared to
  the case of an aligned magnetic field (run M10tMpll3D) the peak of
  fluid excitations appear at higher amplitudes
  (cf.~Fig.~\ref{fig:jet_M10tMpll3D}). This high amplitude motions are
  excited by the conversion of magnetic energy into kinetic energy,
  i.e.~by the straightening of the magnetic field lines.}
\label{fig:vel_PDFs_M10tMpe3D}
\eef

Magnetic fields are observed in a variety of molecular clouds
cores~\citep[e.g.][]{Crutcher99} and may exist in all of
them. Additionally, jets are most probably magnetically launched and
collimated phenomena~\citep[e.g.][]{Blandford82, Pudritz83, Ouyed97b,
Fendt02}.  Here, we study the influence of background magnetic fields
on the jet-driven turbulence by two configurations: First we set up a
homogeneous field parallel to the jet axis (run M10tMpll3D), and
secondly, we rotate this field by 90 degrees which is then
perpendicular to the jet axis and aligned with the $y$-axis of our
simulation box (run M10tMpe3D).  The magnetic field strength in both
cases is set so that the Alfv\'en velocity, $v_{\sm{A}} = B/\sqrt{4\pi
\, \rho}$, equals the sound speed, i.e.~$v_{\sm{A}} = c$, or in other
words the plasma $\beta$-parameter is $\beta =
p_{\sm{therm}}/p_{\sm{mag}} = 2$.  As the jet is supersonic with a
sonic Mach number of $\Ma = 10$ it is also super-Alfv\'enic with an
Alfv\'en Mach number of $\Ma_{\sm{A}} = 10$.  Initally the fast
magnetosonic speed is $\sqrt{2} \, c$ which is exceeded by the jet.
Note that these numbers are defined to the external medium and measure
the jet energetics in comparison to the jet environment. The internal
dynamics of the jet is defined by the internal Mach numbers. Typical
protostellar jets are in rough energy equipartition and have internal
fast magnetosonic Mach numbers of about 3.

Interestingly, information along the jet axis is transfered by
different wave modes. In the first case where the field is aligned
with the jet, Alfv\'en waves will be able to travel along the jet. In
the second case, fast magnetosonic waves will travel across the field
in direction of the jet axis.  Of course the internal Mach numbers may
drastically change during the evolution of our simulation.  In
particular, in the second case, the parallel field will be highly
compressed by the jet (see below), increasing the field strength and
thus the Alfv\'en speed. The jet front will be in a magnetic energy
dominated region and the strong field will finally stop the jet
propagation.

In Figs~\ref{fig:jet_M10tMpll3D} -- \ref{fig:vel_PDFs_M10tMpe3D} we
summarise the results of these two simulations. From the velocity
structure and its PDF (Fig.~\ref{fig:jet_M10tMpll3D}) one can see that
the aligned magnetic field suppresses large amplitude
fluctuations. This result is obvious as transverse modes are more
difficult to excite due to the additional magnetic pressure in this
direction. Furthermore, the jet is more collimated and its bow shock
radius is smaller than in the non-magnetic case
(cf.~Fig.~\ref{fig:jet_M10c}). Any excited fluid fluctuations are also
damped more quickly in the magnetic case, but have the potential to
travel further than in the non-magnetised case. Transverse motions
distort the magnetic field lines which in turn counteract the
fluctuations leading to damped motions. Longitudinal or Alfv\'enic
excitations can travel (almost) undamped but are unlikely to excite
supersonic fluid motions~\citep[e.g.,][]{Heitsch02}.

Such a scenario can be seen in Fig.~\ref{fig:jet_M10tMpe3D} where the
magnetic field is perpendicular to the jet axis. First, the active jet
bends the magnetic field strongly while building up a large component
parallel to the jet axis. 
This parallel component has a large gradient
along the former field direction along the $y$ axis. 
The strength of this effect is of course exaggerated by our assumption
of ideal MHD. Real jets will be magnetically diffusive and thus allow
the field to penetrate the jet. However, our toy model demonstrates
how the built-up magnetic pressure gradient pushes the gas away from
the jet axis (see left panel of Fig.~\ref{fig:jet_M10tMpe3D}).  This
happens only in the direction of the initial field. Perpendicular to
it, this magnetic pressure component is negligible and does not
accelerate the gas (as seen in right panel of
Fig.~\ref{fig:jet_M10tMpe3D}).  Additionally, the tension of the
stretched magnetic field works against the jet motion and decelerates
the high velocity flow. This effect is particularly effective after
the driving engine stops and even pushes gas backwards after a
while. Obviously the jet does not propagate as far if a perpendicular
background field is present.  Note that the tension of the bent
magnetic field lines also has a force component perpendicular to the
jet and helps accelerating ambient gas away from the jet.

In summary, such a field configuration helps to excite
higher-amplitude, though still subsonic, fluctuations compared to an
aligned magnetic field. The PDFs shown in
Fig.~\ref{fig:vel_PDFs_M10tMpe3D} shows the peaks of excitations at
late times in the amplitude range $v/c \sim 0.01 - 0.1$ whereas the
peaks in the aligned case appear at $v/c \sim 0.005 - 0.01$. Note that
the Alfv\'en velocity is initially equal to the speed of sound which
in turn means that supersonic velocity fluctuations are also
super-Alfv\'enic.

\section{Summary and Conclusion}

In this work we presented a detailed study of feedback from
protostellar jets based on two and three dimensional ideal HD and MHD
simulations of
a single supersonic jets in an ambient medium. 
We focus on the particular velocity and energy structure excited
by the jet. 
Essentially, we used a statistical measure of velocity, namely the
velocity probability density function (PDF) to quantify the main
contributions to the excited motions, as subsonic or supersonic. 
The distinction between sub- and
supersonic fluctuations is a natural outcome of our investigation:
Supersonic motions decline rapidly in velocity space
(e.g.~Figs.~\ref{fig:vel_PDFs_M5ct} and \ref{fig:vel_PDFs_M10c}) and
do not propagate far from their driving source. They damp quickly
because they excite mainly compressive modes. The re-expansion of the
compressed overdensities drives mainly subsonic velocity fluctuations
that then propagate further into the ambient medium. Despite the
appearance of bow shocks and instabilities, jets do not entrain
large volumes with supersonic speeds. In particular, instabilities such
as Kelvin-Helmholtz modes at the edge of the jet, develop most
efficiently for transonic or slower velocities. High-velocity jets, on
the contrary, are bullet-like and stay very collimated, transiting the
surrounding cloud without entraining much of its gas (e.g.~see
Figs.~\ref{fig:jet_M10c} and \ref{fig:jet_M20tCl}). From the point of
view of jet-driven supersonic turbulence in molecular clouds this is a
dilemma which is difficult to circumvent. Even in the case of
overdense jets which affect more gas of the surrounding media and have
higher momenta, the supersonic motions do not propagate far from their
source as can be seen in the PDF of Fig.~\ref{fig:vel_PDFs_M10tOd3D}.

We also tested the influence of magnetic fields on the jet propagation
and jet feedback. Naturally, jets stay more collimated if the
magnetic field is aligned with the jet axis and therefore entrain less
gas. Furthermore, perpendicular motions are damped by magnetic tension
preventing a large spread of high amplitude fluctuations (see
Fig.~\ref{fig:jet_M10tMpll3D}). Perpendicular field configurations
support the propagation of such modes which are able to spread into a
large volume (see Fig.~\ref{fig:jet_M10tMpe3D}). Nevertheless, the
vast majority of these motions are still subsonic (see
Fig.~\ref{fig:vel_PDFs_M10tMpe3D}).

Based on our presented study we conclude that collimated jets from
young stellar objects are unlikely drivers of large-scale {\em
supersonic} turbulence in molecular clouds. Alternatively it could be
powered by large scale flows which might be responsible for the
formation of the cloud itself~\citep[e.g.,][]{Ballesteros99}. Energy
cascading down from the driving scale to the dissipation scale will 
then produce turbulent density and velocity structure in the inertial range 
in between \citep[][]{Lesieur97}.
If the large-scale dynamics of the interstellar medium is driven 
by gravity \citep[as suggested, e.g., by][]{LiMacLowKlessen05,LiMacLowKlessen06}
gravitational contraction would also determine to a large extent the
internal velocity structure of the cloud. Otherwise, blast waves and
expanding shells from super novae are also viable candidates to power
supersonic turbulence in molecular clouds\cite[see e.g.,][]{MacLow04}.

\acknowledgments

We thank Mordecai Mac~Low for his valuable comments on this work.
The FLASH  code was developed  in part by the  DOE-supported Alliances
Center for Astrophysical Thermonuclear Flashes (ASC) at the University
of Chicago. RB is funded by the Deutsche Forschungsgemeinschaft (DFG)
under the grant no. KL1358/4-1.


\end{document}